\documentclass[prb,preprint,floats,floatfix,12pt,superscriptaddress]{revtex4}
\usepackage{times} 
\usepackage{amssymb}
\usepackage{amsmath}
\usepackage{graphicx}     
\graphicspath{{.},{../Ps/},{../../Ps/},{../../../Obj/Sitdrop/Ps/}}
\newcommand{\no}[1]{}
\newcommand{\mylab}[1]{\label{#1}}
\renewcommand{\vec}[1]{\mathbf{#1}}
\newcommand{\vecg}[1]{\boldsymbol{#1}}
\newcommand{\tens}[1]{\mathbf{\underline{#1}}}
\newcommand{\tensg}[1]{\boldsymbol{\underline{#1}}}
\newcommand{\diff}{{d}}
\newcommand{\lambdadif}{{\lambda_\mathrm{d}}}
\newcommand{\mudif}{{\mu_\mathrm{d}}}
\newcommand{\mudift}{{\tilde{\mu}_\mathrm{d}}}
\newcommand{\Tr}{{\mathrm{T}}}

\newcommand{\Fs}{{F_{\mathrm s}}}
\newcommand{\Fstop}{{F_{\mathrm s}^{\mathrm{top}}}}
\newcommand{\Fsbot}{{F_{\mathrm s}^{\mathrm{bot}}}}
\newcommand{\fsurf}{{f_{\mathrm s}}}
\newcommand{\xsurf}{{{\vec{x}}_\mathrm{s}}}
\newcommand{\Fb}{{F_{\mathrm b}}}

\newcommand{\etaco}{{\eta_{\mathrm {co}}}}
\newcommand{\etanc}{{\eta_{\mathrm {di}}}}

\begin{document}
\bibliographystyle{../../../../Literatur/Bibdir/Bibstyles/pof}
\title{Decomposition driven interface evolution for layers of
binary mixtures: \\ {I}. Model derivation and stratified base states}
\author{Uwe Thiele}
\email{thiele@pks.mpg.de}
\homepage{http://www.uwethiele.de}
\altaffiliation[Present address:]{School of Mathematics, Loughborough University,
Loughborough, Leicestershire, LE11 3TU, UK}
\affiliation{Institut f\"ur Physik, Universit\"at Augsburg, D-86135 Augsburg, Germany}
\affiliation{Max-Planck-Institut f\"ur Physik komplexer Systeme, 
N{\"o}thnitzer Str.\ 38, D-01187 Dresden, Germany}
\author{Santiago Madruga}
\email{santiago@pks.mpg.de, smadruga@gmail.com}
\affiliation{Max-Planck-Institut
f\"ur Physik komplexer Systeme, N{\"o}thnitzer Str.\ 38, D-01187 Dresden, Germany}
\author{Lubor Frastia}
\email{lfrastia@gmail.com}
\affiliation{Department of Chemical Engineering, Technion -- Israel Institute of
Technology, Haifa 32000, Israel}
%
%
\begin{abstract}
A dynamical model is proposed to describe the coupled decomposition and
profile evolution of a free surface film of a binary mixture. An example is a
thin film of a polymer blend on a solid substrate undergoing simultaneous
phase separation and dewetting.  The model is based on model-H describing the
coupled transport of the mass of one component (convective Cahn-Hilliard
equation) and momentum (Navier-Stokes-Korteweg equations) supplemented by
appropriate boundary conditions at the solid substrate and the free surface.

General transport equations are derived using phenomenological
non-equilibrium thermodynamics for a general non-isothermal setting
taking into account Soret and Dufour effects and interfacial
viscosity for the internal diffuse interface between the two
components.  Focusing on an isothermal setting the resulting model is
compared to literature results and its base states
corresponding to homogeneous or vertically stratified flat layers are
analysed.
\end{abstract}

%
\maketitle
%
%
%
%
\section{Introduction} \mylab{intro}
%
Driven by applications in coating technology, micro and nano
structuring of soft matter layers, and, in general, the development of
(multi-)functional surfaces the understanding of thin films of simple
and complex fluids is of growing importance. Recent years have seen on
the one hand major advances in experimental techniques of preparation
and analysis and on the other hand intense developments of the
theoretical description of the statics and dynamics of homogeneous and
structured films.\cite{deGe85,TDS88,ODB97,GeKr03,Muel03b,Thie03,WeRu04,Seem05,KaTh07}  The
dynamics of the structuring often represents examples for micro- and
nano-fluidic flows, a present focus of interest in its own right.
\cite{SSA04,SqQu05}

For thin one-layer free surface films of one-component simple or
polymeric liquids experimental results
\cite{Reit92,ShRe96,Beck03,Seem05} and theoretical understanding
\cite{RuJa73,Mitl93,TVN01,BeNe01,KKS01,SRd02,Beck03}
are well developed.
However, experiments increasingly focus on complex situations like the
evolution of multilayer films of partially miscible \cite{KKY02} or
immiscible \cite{GeKr03} liquids, complex fluids like polymer blends
that might undergo dewetting {\it or/and} decomposition
\cite{YKK99,KKY02,OKYK04} or solutions of polymers, nanoparticles, colloids or
polymer blends with interacting convective motion, phase separation,
evaporation/condensation and evolving rheology.
\cite{Deeg97,Mert98,TMP98,PiFr99,Weh99,Deeg00,YNKA02,GRDK02,MTB02,%
RRGB03,MBM04,Borm05,Borm05b,GPBR05,HeJo05,Muel06}  Theoretical
descriptions exist, however, only for a small part of the
experimentally known complex scenario and phenomena involving free
surface thin films. Recent advances include a fully nonlinear thin
film description in long wave approximation for two layers of
immiscible liquids under air
\cite{PBMT04,FiGo05,PBMT05,BGS05,PBMT06} and between two plates,
\cite{MPBT05} the analysis of the dewetting behaviour on chemically or topographically 
heterogeneous substrates,\cite{KaSh01,KaSh02,BKTB02,TBBB03}
the study of the dynamics of depinning of a driven drop
on a heterogeneous substrate,\cite{ThKn06} the description of films with surface active 
nanoparticles,\cite{WCM03}
the inclusion of evaporation/condensation in the
thin film description.\cite{BBD88,SLT98,Dano98,OrBa99,KKS01,WCM03,Pism04}

Thin films of polymer blends are one of the 'simplest' complex systems
listed above and extensive experimental results can be found in the
literature.
\cite{BrBr92,KKY02,GeKr03,Krau93,SKF94,HOS95,SWBS03,GeKr97,JHBK96,Kari98,StKl96,%
WaCo00b,WaCo03,YKK99}  
Described effects include the dependency of
the evolving structure (stratified bi- or tri-layer structure
that is laterally homogeneous; purely lateral phase separation, checkerboard
structure) on substrate properties,\cite{Jone91,BrBr92,GeKr97,GeKr98}
surface roughening or film morphology changes during phase separation,
\cite{JHBK96,WaCo00,WaCo00b,WaCo03} surface directed spinodal decomposition,
\cite{Jone91} subsequent vertical phase separation and dewetting,
\cite{YKK99,KKY02} and surface phase inversion.\cite{SKF94}  The
influence of heterogeneous substrates was also studied.\cite{Kari98,NEDK99}
However, as detailed below to our knowledge there exists no theoretical 
description of the involved processes that takes into account the evolving 
composition of the mixture 
{\it and} the evolving surface profile of the film. 

The aim of the present paper is to present such a description based on
the underlying transport equations.  To cover the coupled time
evolution of the film thickness and concentration profiles one has to
supplement the coupled transport equations for momentum and
concentration by appropriate boundary conditions at the free surface
and at the solid substrate.

Two groups of studies are present in the literature that address part
of the involved questions. On the one hand, the classical
Cahn-Hilliard model \cite{CaHi58} describing purely diffusive
decomposition of a binary mixture was studied for films in a gap
between two solid plates.\cite{FMD97,FMD98,Kenz01}  At the plates boundary conditions
prescribe zero diffusive flow through the plates, energetic preference
of the plates for one component and an enforced (or reduced) demixing
at the plates. However, such a model can in principle not account for
an evolving surface deflection in decomposing films as observed in
phase-separating polymer blends on homogeneous
\cite{JHBK96,WaCo00,WaCo00b,WaCo03} or patterned \cite{Kari98,NEDK99}
substrates. An evolving film profile is {\it by definition}
related to a convective flow of the mixture.

On the other hand, the coupling of momentum and concentration transport for
weakly miscible fluids (i.e.\ decomposing mixtures) is well
studied for bulk systems using the so-called model-H.
\cite{HoHa77,AMW98,JaVi96,VMM99,VMM99b}
It couples the convective Cahn-Hilliard equation and the Navier-Stokes
equations ammended by a concentration dependent stress tensor, the so
called Korteweg stresses.\cite{Kort1901} For a survey of the history
see Ref.~\onlinecite{Jose90}.  A variant of model-H is also used for the
dynamics of momentum and density of a single component near a
liquid-gas phase transition in isothermal
\cite{AMW98,PiPo00,Pism01,Rohd05} or non-isothermal conditions.
\cite{Anta96,JaVi96,BoBe03} Two-phase liquids and binary mixtures between two
solid plates are also investigated.\cite{Anta95,Tana01,BoBe03}
A similar model for miscible liquids was studied, for instance, in
Refs.\,\onlinecite{Jose90} and \onlinecite{GJPR91}. There the convective Cahn-Hilliard
equation is replaced by a 'normal' convective diffusion equation.
Nowadays, model-H is also applied to multiphase flows in closed micro
channels of different geometries (straight quadratic channel,
\cite{KJB03} T-junction. \cite{deMe06})

To describe a film of a mixture under air, model-H is used to describe
the dynamics within the film, i.e.\ modelling the creation and
evolution of 'internal' diffuse interfaces within the film. The bulk
model has to be completed by boundary conditions at the solid
substrate (discussed in Refs.~\onlinecite{FiDi97,Bind98,FNB95}) {\it and} by
boundary conditions at the free surface. The latter represents an
'external' sharp interface. The interaction of the internal diffuse
interfaces with the external sharp one via a solutal Marangoni effect
results in an additional driving force of the evolution.

Other related work involves an ad-hoc lubrication approximation model
coupling evolution equations for the film thickness and the mean
concentration in the film.\cite{Clar04,Clar05} We believe that such a
model might be correct in the limit of weak vertical variation of
concentration but is not able to describe vertically stratified films
and their evolution.  An alternative approach uses microscopic
discrete models like lattice gas (for results on a binary alloy see
Ref.~\onlinecite{PlGo97}) or Molecular Dynamics.\cite{Furu97} This leads,
however, often to a strong restriction in length- and time-scales that
can be studied.

We present our work as a sequence of papers.  The present first part
derives and discusses the basic transport equations and analyses
steady base states.  The accompanying second part performs a detailed
stability analysis with respect to transversal instability modes for
the various qualitatively different base states. Thereby the
consequences of convective transport are studied in detail and the
sequence of patterning processes is predicted.  A planned further
sequel will focus on the nonlinear evolution.

The present paper is structured as follows. In Section~\ref{dermh} we
derive the coupled transport equations for momentum, density and
temperature in the framework of phenomenological non-equilibrium
thermodynamics. After discussing the physical interpretation of the
individual contributions representing, for instance, an internal Soret
effect (i.e.\ Soret effect for the internal diffuse interface) and
internal interface viscosity, the model is
simplified assuming an isothermal setting, vanishing interface
viscosity and internal energies resulting from a setting close to the
critical point of demixing. The resulting model-H is compared to
versions found in the literature discussing the issue of 
defining pressure and chemical
potential.  Section~\ref{sec-bc} introduces boundary conditions at the
rigid solid substrate and the free interface. It is explained in
detail why the incorporation of convective flow is a necessary
precondition to describe evolving surface deflections. Next,
Sections~\ref{sec-nondim} and~\ref{sec-en} introduce the
non-dimensionalization and local energies, respectively.  Homogeneous
and vertically stratified, transversally homogeneous 
steady state solutions are analysed in
Section~\ref{sec-base}. The final Section~\ref{sec-conc}
summarizes, compares to the literature and gives an outlook on the
sequel. The Appendix~\ref{deribc}
uses variational calculus to independently 
derive the boundary conditions in the static limiting case.
%
\section{Derivation of extended model-H}
\mylab{dermh}
%
First we present a derivation of an extended model-H that accounts for all
cross couplings of the transport equations for momentum, concentration and
temperature. This includes Soret- and Dufour-effects 
with nonlinear coefficients and interface
viscosity for the diffuse interface. It follows in spirit the derivations of
the Navier-Stokes equations given in Refs.\,\onlinecite{Batc00,LaLi87f}. 
The online version shows new terms related to the concentration field in red.

\subsection{General transport equations} \mylab{chap6.3.2}
\subsubsection{Conserved quantities}
Starting point are the transport, conservation and balance laws for
the relevant phenomenological thermodynamic entities. In general, we
have for a conserved scalar or vector field $a(\vec{x},t)$ the
transport equation
\begin{equation}
\frac{\partial}{\partial t}a\,+\,\vec{\nabla}\cdot\vec{j'}_a\,=\,0
\mylab{fk04000}
\end{equation}
where $\vec{j'}_a$ is a general flux density that is a vector or
second order tensor. Note that a dotless product corresponds always
to a tensor (or outer) product, whereas a dot `$\cdot$' product
is an inner product (resulting in a tensor of the order $n-2$, where $n$
is the order of the respective tensor product).
The contribution by convective transport with the velocity $\vec{v}$ is
expressed explicitly by $\vec{j'}_a = \vec{j}_a + a\vec{v}$ where
$\vec{j}_a$ is the
diffusive flux caused by (several) microscopic mechanisms.

By definition the total mass density $\rho(\vec{x},t)$ is transported
by convection only, i.e.\ the mass density flux is
\begin{equation}
\vec{j'_\rho}\equiv\rho\vec{v}=\vec{g}
\mylab{fk041}
\end{equation}
corresponding to the momentum density $\vec{g}$. The transport equation for 
the density (continuity equation) is
\begin{equation}
\frac{\partial}{\partial t}\rho\,+\,\vec{\nabla}\cdot\vec{g}\,=\,0
\mylab{fk040}
\end{equation}
The density of the momentum $\vec{g}$ as well as the density of the 
total energy $\epsilon$
are transported by convective and diffusive fluxes, i.e.\
\begin{eqnarray}
\frac{\partial}{\partial t}\vec{g}\,+\,\nabla\cdot\tens{j}'_\vec{g}&=&0
\mylab{fk042}\\
\frac{\partial}{\partial t}\epsilon\,+\,\vec{\nabla}\cdot\vec{j}_{\epsilon}'&=&0
\mylab{fk043}
\end{eqnarray}
where $\tens{j}'_\vec{g}$ is the tensor of the momentum flux density
and $\vec{j}_{\epsilon}'$ is the energy flux density.
Note that all densities are per volume.

By explicitly denoting the transport by convection as before,
$\vec{j'_\epsilon}=\epsilon\vec{v}+\vec{j_\epsilon}$ and
$\tens{j}'_\vec{g}=\vec{v} \vec{g} + \tensg{\sigma}$, where
$\tensg{\sigma}$ is the usual symbol for the diffusive momentum flux
$\tens{j}_\vec{g}$ (sometimes also called pressure tensor corresponding to
the negative of the stress tensor). We will use underlined symbols to denote
tensors of 2nd or higher order.  Eqs.\,(\ref{fk043}) and (\ref{fk042})
result in
\begin{eqnarray}
\frac{\partial}{\partial t}\epsilon
+\vec{\nabla}\cdot(\epsilon\vec{v})
+\vec{\nabla}\cdot\vec{j_\epsilon}
&=&0
\mylab{fk046}\\
\mbox{and}\qquad \frac{\partial}{\partial t}\vec{g}
+\vec{\nabla}\cdot(\vec{v} \vec{g})
+ \vec{\nabla}\cdot\tensg{\sigma}
&=&0\mylab{fk045}
\end{eqnarray}
respectively.

Introducing the material time derivative 
$D/D t= \partial/\partial t\,+\,\vec{v}\cdot\nabla$ one obtains for 
the velocity field [(\ref{fk045}) and (\ref{fk040})]
\begin{equation}
\rho\frac{D \vec{v}}{D t}\,+\,\nabla\cdot\tensg{\sigma}\,=\,0
\mylab{transvel}
\end{equation}
For a binary mixture of fluids a transport 
equation for the mass density of one of the components has to be added beside the
one for the total density $\rho$.  
Choosing $\rho_1$ we have
\begin{equation}
\frac{\partial}{\partial t}\rho_1+\vec{\nabla}\cdot\vec{j}'_{\rho_1}\,=\,0,
\mylab{dens1b}
\end{equation}
i.e.\
\begin{equation}
\frac{\partial}{\partial t}\rho_1+\vec{\nabla}\cdot(\rho_1\vec{v}) 
+ \vec{\nabla}\cdot\vec{j}_{\rho_1}=0.
\mylab{dens2}
\end{equation}
The density of the other component is $\rho_2=\rho-\rho_1$.
These are all the conserved quantities. The conservation of angular momentum 
is guaranteed
by the symmetry properties of the stress tensor (see below).\cite{Batc00}
Note that all densities used throughout the paper are volume densities.
Next we discuss the transport
equations for non-conserved quantities.

\subsubsection{Non-conserved quantities}
The transport equations for non-conserved quantities contain additional 
source terms, i.e.\ for a general field $a$ one writes
\begin{equation}
\frac{\partial a}{\partial t} + \vec{\nabla}\cdot\vec{j}_{a}'\,=\,Q_a
\mylab{fk045b}
\end{equation}
where $Q_a$ is a possibly space- and time-dependent source density. Relevant
non-conserved quantities are the densities of the 
internal energy $u$ and of the entropy $s$.

For systems with small gradients of concentration and/or temperature
the energies do only depend on the local fields. For strong gradients, 
however, this statement does not hold any more and the energy of a system 
will depend also on field gradients. In the present case we consider strong
density gradients related to diffuse interfaces between 
different phases. The underlying assumption is that for a demixing system
gradients in $\rho_1$ might be much larger than all other gradients. The latter 
enter the theory as parametric dependencies on space only.

We define the internal energy $u(\rho,\rho_1,s,t)$ 
as the thermodynamic equilibrium value for a local
fluid element, i.e.\ it shall not depend on gradients. 
The conserved total energy $\epsilon$,
however, shall include gradient terms in $\rho_1$. 
The relation between the two is
\begin{equation}
\epsilon \,=\, u + \frac{\rho}{2}\,v^2 + \frac{\xi}{2} 
(\nabla \rho_1)^2.
\mylab{internal1}
\end{equation}

Note that the unit of $\xi$ is
$[\xi]=\text{m}^7/(\text{kg}\,\text{s}^2)$. The energy densities have
units $[\epsilon]=[u]=\text{Nm}/\text{m}^3=\text{kg}/(\text{m s}^2)$.
The transport equation for the internal energy is
\begin{equation}
\frac{\partial u}{\partial t} + \vec{\nabla}\cdot\vec{j}_{u}'\,=\,Q_u
\mylab{fk045c}
\end{equation}
with
\begin{equation}
\vec{j}_{u}'=u\vec{v}+\vec{j}_{u}.
\mylab{fk045cc}
\end{equation}
For the irreversible processes in question
entropy is not conserved. The transport equation for its
volume density is
\begin{equation}
\frac{\partial s}{\partial t} 
+ \vec{\nabla}\cdot\vec{j'}_s = Q_s = \frac{R}{T}
\mylab{entrop1}
\end{equation}
with $\vec{j'}_s = s \vec{v} + \vec{j}_s$ the total entropy flux
density. We write the source density $Q_s$ in the usual form $R/T$ where
$R$ is the so-called dissipation function and $T$ the temperature.\cite{Batc00}
\subsection{Determination of thermodynamic forces}\mylab{chap7.3.2neu}
The flux densities $\tensg{\sigma}$,
$\vec{j}_{\epsilon}$, $\vec{j}_{u}$, $\vec{j}_{\rho_1}$, $\vec{j}_s$ and source terms $R/T$, $Q_u$
remain to be determined.
The specific transport equation for the internal energy is obtained
from the transport equation of the total energy Eq.\,(\ref{fk046}) using
Eqs.~(\ref{fk040}), (\ref{fk045}), (\ref{dens2}), and (\ref{internal1}). It reads
\begin{eqnarray}
\frac{D}{D t}u + u \nabla\cdot\vec{v} + \nabla\cdot \vec{j}_u
&&\,=\,\left\{-\tensg{\sigma}+\xi(\nabla\rho_1)(\nabla\rho_1)
-\xi\left[\tfrac{1}{2}(\nabla\rho_1)^2+\rho_1\Delta\rho_1\right]\tens{I}\right\}:\nabla\vec{v}
\nonumber\\
&&-\xi\,(\Delta\rho_1)\nabla\cdot \vec{j}_{\rho_1}
\mylab{eq-7.3.2-1}
\end{eqnarray}
with
\begin{equation}
\vec{j}_u=\vec{j}_{\epsilon} - \vec{v}\cdot\tensg{\sigma} 
-\xi (\nabla\rho_1) \left[\rho_1 (\nabla\cdot\vec{v})
+ \nabla\cdot \vec{j}_{\rho_1}\right].
\mylab{zus1}
\end{equation}
The symbol `:' stands for a double inner product, i.e.\
$\tens{a}:\tens{b}=\sum_{ij}a_{ij}b_{ji}$.

The time evolution of the entropy is deduced using a local form of 
Gibbs relation for each fluid element, i.e.\ from the assumption that 
small fluid elements are in thermodynamic equilibrium. 
Gibbs relation for a local fluid element of volume $V$ writes
\begin{equation}
dU = T\,dS -p\,dV + \hat{\mu}_1\,dN_1 + \hat{\mu}_2\,dN_2
\mylab{zus2}
\end{equation}
where $U$, $S$, $p$, $\hat\mu_1$, $\hat{\mu}_2$, $N_1$, $N_2$ stand for
internal energy, entropy, pressure, chemical potentials of component 1 and
2, and particle numbers of component 1 and 2.
The chemical potentials (with hat) are related to particle numbers.
The relation (\ref{zus2}) is transformed expressing extensive variables by the corresponding
densities using
$U=u V$, $N_i=n_i V=\rho_i\,N_a/M_i\, V$, $N=N_1+N_2$,
$\rho_2=\rho-\rho_1$, and $S=s V$, where $N$ is the total particle number,
$\rho$ the density of a mixture, $M_i$ the molar mass of component $i$, and $N_a$ is
the Avogadro number. One obtains
\begin{equation}
du = T\,ds +\mu_2\,d\rho + \mudif\,d\rho_1
+ (-u + \mu_2\rho  + \mudif\rho_1 +Ts-p)dV/V,
\mylab{zus3}
\end{equation}
The chemical potential $\mu_2=\hat{\mu}_2 N_a/ M_2$ of
component 2 and the difference of the chemical potentials of
components 1 and 2 $\mudif=\hat{\mu}_1 N_a/ M_1-\hat{\mu}_2 N_a/ M_2$
are related to volume densities and have units
$[\mu_2]=[\mudif]=\text{m}^2/\text{s}^2$.

Relation (\ref{zus3}) is valid for arbitrary local volume
$V$, i.e.\ one obtains the local Gibbs relation
\begin{equation}
d u=T\,d s + \mu_2\,d\rho+ \mudif\,d\rho_1
\mylab{eq-7.3.2-4}
\end{equation}
and the local Gibbs-Duhem relation
\begin{equation}
p=-u + T s + \mu_2\rho + \mudif\rho_1.
\mylab{eq-7.3.2-9}
\end{equation}
Here, we observe that, within the framework of volume density quantities,
$p$ behaves as a thermodynamic potential that is related with $u$
by the Legendre transform (\ref{eq-7.3.2-9}). Furthermore,
\begin{equation}
T=\left(\frac{\partial u}{\partial s}\right)_{\rho,\rho_1}\qquad
\mu_2=\left(\frac{\partial u}{\partial\rho}\right)_{s,\rho_1}\qquad
\mbox{and}\qquad
\mudif=\left(\frac{\partial u}{\partial\rho_1}\right)_{s,\rho}.
\mylab{Tmusdefs}
\end{equation}
Eq.\,(\ref{eq-7.3.2-4}) is divided by a small time span $dt$ that is,
however, large as compared to typical microscopic time scales yielding
\begin{equation}
\frac{du}{dt}=T\frac{ds}{dt}  + \mu_2\frac{d\rho}{dt}
+\mudif\frac{d\rho_1}{dt}.
\mylab{eq-7.3.2-5}
\end{equation}
This relation is valid in all local volume elements that might be
convected by the flow, i.e.\ the derivatives $d/dt$ correspond to
Lagrangian or material time derivatives denoted above $D/Dt$. Using
equations (\ref{fk040}), (\ref{dens2}), and (\ref{eq-7.3.2-1}) one
transforms (\ref{eq-7.3.2-5}) into the wanted form of
Eq.\,(\ref{entrop1})
\begin{eqnarray}
&&\hspace*{-.5cm}\frac{\partial s}{\partial t}+\vec{\nabla}\cdot\left[
s\vec{v}+\frac{\vec{j}_u}{T} 
 + \frac{{\vec{j}_{\rho_1}}}{T}(\xi\Delta\rho_1 - \mudif)
\right]
=\nonumber\\
&&\frac{1}{T}\left\{-\tensg{\sigma}+\xi(\nabla\rho_1)(\nabla\rho_1)+\left(p
-\xi\rho_1\Delta\rho_1\,-\,\frac{\xi}{2}(\nabla\rho_1)^2\right)\,
\tens{I}\right\}:(\vec{\nabla}\vec{v})
\nonumber\\
&& \,+\,\vec{j}_u\cdot\vec{\nabla}\left(\frac{1}{T}\right)
+ \vec{j}_{\rho_1}\cdot\nabla\left(\frac{\xi\Delta\rho_1 -
\mudif}{T}\right)
\mylab{entrop2}
\end{eqnarray}
with the pressure $p$ and the flux of internal energy $\vec{j}_u$ given by (\ref{eq-7.3.2-9}) and (\ref{zus1}), 
respectively. Here it is already possible to see 
the structure of the dissipative contribution to the pressure or stress tensor 
(the part in $\{ \}$ on the r.h.s.).
The reversible entropy transport (cp.\ Eqs.\,(\ref{entrop1}) and (\ref{entrop2})), i.e.\ 
the entropy flux
\begin{equation}
\vec{j}'_s\,=\,s\vec{v}+\frac{\vec{j}_u}{T} 
+ \frac{{\vec{j}_{\rho_1}}}{T}(\xi\Delta\rho_1 - \mudif)
\mylab{entrflux}
\end{equation}
contains the convective transport, the transport via the heat flux and the 
transport via the diffusive flux of species one.
Comparing Eqs.\,(\ref{entrop1}) and (\ref{entrop2}) allows to identify 
the source term for the entropy. It is related to irreversible processes
and written in terms of the dissipation function
\begin{eqnarray}
R\,&=&\,\left\{-\tensg{\sigma}+\xi(\nabla\rho_1)(\nabla\rho_1)+\left(p
-\xi\rho_1\Delta\rho_1\,-\,\frac{\xi}{2}(\nabla\rho_1)^2\right)\,
\tens{I}\right\}:(\vec{\nabla}\vec{v})\nonumber\\
\,&&\,+T\vec{j}_u\cdot\vec{\nabla}\left(\frac{1}{T}\right)
+ T\,\vec{j}_{\rho_1}\cdot\nabla\left(\frac{\xi\Delta\rho_1 -
\mudif}{T}\right).
\mylab{eq-7.3.2-11}
\end{eqnarray}
We can directly deduce the reversible part $\tensg{\sigma}^r$ of the
pressure tensor $\tensg{\sigma}=\tensg{\sigma}^r + \tensg{\sigma}^d$,
because only the dissipative part $\tensg{\sigma}^d$ contributes to
the dissipation function, i.e.\
\begin{equation}
\tensg{\sigma}^r\,=\,\xi(\nabla\rho_1)(\nabla\rho_1)+\left(p
-\xi\rho_1\Delta\rho_1\,-\,\frac{\xi}{2}(\nabla\rho_1)^2\right)\,
\tens{I}.
\mylab{pressrev}
\end{equation}
Note that negative of $\tensg{\sigma}^r-p\tens{I}$ is known as the capillary or
Korteweg stress tensor in the literature.\cite{Kort1901,JaVi96,AMW98,PiPo00}
The dissipative part $\tensg{\sigma}^d$
is also called viscose pressure tensor or friction tensor.
The dissipation function has the structure $R=\sum_\alpha\,
\vec{j}_\alpha\cdot\vec{f}_\alpha$, where the $\vec{j}_\alpha$ and
$\vec{f}_\alpha$ are general thermodynamic fluxes and forces,
respectively, that might be tensors.  Correspondingly the `$\cdot$'
stands here for a 'complete' inner product (scalar product).

We have the fluxes $\vec{j}_u$, $\vec{j}_{\rho_1}$, 
and $-\tensg{\sigma}^d$ with the corresponding forces
\begin{eqnarray}
\vec{f}_u\,&=&\,T\,\nabla\left(\frac{1}{T}\right)
\mylab{tdforce1}\\
\vec{f}_{\rho_1}&=&T\,\nabla\left(\frac{\xi\Delta\rho_1 -
\mudif}{T}\right)
\mylab{tdforce2}\\
\tens{f}_\vec{g}&=&\vec{\nabla}\vec{v}.
\mylab{tdforce3}
\end{eqnarray}
In the last step of the derivation, the thermodynamic fluxes have to
be determined.  Following Onsager, we first make the basic ansatz of
linear non-equilibrium thermodynamics, i.e.\ we postulate a linear
dependence of the fluxes on {\it all} the forces (if symmetry
permits), i.e.\
\begin{equation}
\vec{j}_\beta\,=\,\sum_\alpha \tens{L}_{\beta\alpha}\cdot\vec{f}_\alpha
\mylab{tdflux1}
\end{equation}
with $\tens{L}_{\alpha\beta}=\tens{L}_{\beta\alpha}$ (Onsager relation, resulting from microscopic 
reversibility).
Specifically, we get for the fluxes
\begin{eqnarray}
\vec{j}_u \,&=&\, T\,\tens{L}_{uu}\cdot\vec{\nabla}\left(\frac{1}{T}\right) 
+ T\,\tens{L}_{u\rho_1}\cdot\nabla\left(\frac{\xi\Delta\rho_1 - \mudif}{T}\right)
\mylab{tdflux2}\\
\vec{j}_{\rho_1} \,&=&\, T\,\tens{L}_{\rho_1 u} \cdot\vec{\nabla}\left(\frac{1}{T}\right)
\,+\, T\,\tens{L}_{\rho_1 \rho_1}\cdot\nabla\left(\frac{\xi\Delta\rho_1 - \mudif}{T}\right)
\mylab{tdflux3}\\
\tensg{\sigma}^d \,&=&\, -\tens{L}_{\vec{g} \vec{g}}: \nabla\vec{v}
\mylab{tdflux4}
\end{eqnarray}
Note that there is no linear coupling between the momentum flux and
the thermodynamic forces corresponding to temperature and
concentration gradients. However, when discussing the total energy for
systems with large gradients in the density $\rho_1$ we included
quadratic terms in the density gradient.
For consistency, a nonlinear term, quadratic in
the forces $\vec{f}_{\rho_1}$, should be added to relation (\ref{tdflux4})
resulting in
\begin{equation}
\tensg{\sigma}^d \,=\, -\tens{L}_{\vec{g} \vec{g}}: \nabla\vec{v}
- T^2\,\tens{Q}_{\vec{g}\rho_1} :
\left(\nabla\,\frac{\xi\Delta\rho_1 - \mudif}{T}\right)\,
\left(\nabla\,\frac{\xi\Delta\rho_1 - \mudif}{T}\right).
\mylab{tdflux4b}
\end{equation}
The additional term is related to irreversible aspects of the dynamics
of the diffuse interface and can be seen as a generalization of the term
$\tens{S}_\ast$ proposed in the conclusion of Ref.~\onlinecite{Anta95} that
reads
\begin{equation}
\tens{S}_\ast = \beta(\nabla\zeta)(\nabla\zeta),
\end{equation}
where $\beta$ is an undetermined empirical coefficient and
$\zeta$ is the difference between the chemical potentials of components~1 and~2
for an inhomogeneous equilibrium which in our terms is
$\zeta = \mudif-\xi\Delta\rho_1$ as is discussed below, after
Eq.\ (\ref{mh-And1}) and also in Ref.~\onlinecite{AMW98}.
$\tens{Q}_{\vec{g}\rho_1}$ corresponds to a tensor of interfacial
viscosities.
Related issues are discussed for sharp interface
theories in Ref.~\onlinecite{EBW91}.

The $\tens{L}_{\alpha\beta}$ are tensors of various orders:
$\tens{L}_{uu}$ is of order 2, whereas $\tens{L}_{\vec{g} \vec{g}}$
and $\tens{Q}_{\vec{g}\rho_1}$ are of order 4 (i.e.\ in the general
case there are $3^4=81$ components: viscosities). Assumption of an
isotropic medium significantly reduces the number of coefficients.\cite{Batc00}

Considering small interfacial viscosities only, we neglect the
corresponding terms and finally get
\begin{eqnarray}
\vec{j}_u \,&=&\, - \frac{k_1}{T} \nabla T 
+ k_2\,T\,\nabla\left(\frac{\xi\Delta\rho_1 - \mudif}{T}\right)
\mylab{tdflux5}\\
\vec{j}_{\rho_1} \,&=&\,-\frac{k_2}{T} \nabla T
\,+\, k_3\,T\,\nabla\left(\frac{\xi\Delta\rho_1 - \mudif}{T}\right)
\mylab{tdflux6}\\
\tensg{\sigma}^d \,&=&\, -\zeta\tens{I}(\nabla\cdot\vec{v})-\eta\left[
\nabla\vec{v} + (\nabla\vec{v})^\Tr
-\frac{2}{3}\tens{I}(\nabla\cdot\vec{v})\right]
\mylab{tdflux7}
\end{eqnarray}
where $\zeta$ and $\eta$ are the dynamic and shear bulk viscosity,
respectively, and the kinetic coefficients have the units
$[k_1]=\text{kg}\,\text{m}/\text{s}^3$,
$[\eta]=[\zeta]=[k_2]=\text{kg}/\text{m}\,\text{s}$ and
$[k_3]=\text{kg}\,\text{s}/\text{m}^3$.
The governing equations are now obtained by introducing the fluxes
(\ref{tdflux5}) to (\ref{tdflux7}) into the corresponding transport
equations.  Introducing (\ref{tdflux6}) into (\ref{dens2}) we get for
the transport of $\rho_1$
\begin{equation}
\frac{\partial}{\partial t}\rho_1+\vec{\nabla}\cdot(\rho_1\vec{v}) 
+ \vec{\nabla}\cdot  \left[-\frac{k_2}{T} \nabla T
\,+\, k_3\,T\,\nabla\left(\frac{\xi\Delta\rho_1 - \mudif}{T}\right)\right] = 0
\mylab{dens3}
\end{equation}
and for momentum transport feeding (\ref{tdflux7}) and
(\ref{pressrev}) into (\ref{transvel}) we have
\begin{eqnarray}
&&\rho\frac{\partial \vec{v}}{\partial t}\,+\,\rho\vec{v}\cdot\nabla\vec{v}\,=\,
-\nabla\cdot\left\{
\xi(\nabla\rho_1)(\nabla\rho_1)+\left(p
\,-\,\xi\rho_1\Delta\rho_1\,-\,\frac{\xi}{2}(\nabla\rho_1)^2\right)\,\tens{I}
\right.\nonumber\\
&&\left.-\zeta\tens{I}(\nabla\cdot\vec{v})-\eta\left[
\nabla\vec{v} + (\nabla\vec{v})^\Tr
-\frac{2}{3}\tens{I}(\nabla\cdot\vec{v})\right]
\right\}
\mylab{mom3}
\end{eqnarray}
Assuming that $\mu_2$ and $\mudif$ are given, to obtain a closed set of
equations we still need an equation for the evolution of the
temperature field. We obtain it by multiplying the transport equation
for entropy (\ref{entrop2}) with $T$, and expressing the entropy via
the Gibbs relation in a similar way as Batchelor (p.\,35ff and
p.\,136ff of Ref.~\onlinecite{Batc00}).

After reordering Eq.\,(\ref{entrop2}) yields
\begin{equation}
T\frac{Ds}{Dt} + Ts\vec{\nabla}\cdot\vec{v} =
-T\vec{\nabla}\cdot\left[\frac{\vec{j}_u}{T} 
+ \frac{{\vec{j}_{\rho_1}}}{T}(\xi\Delta\rho_1 - \mudif)
\right] + R.
\mylab{entrop3}
\end{equation}
All terms on the r.h.s.\ are already known. We next determine the
l.h.s.\ terms. Consistently with Gibbs relation (\ref{zus2}) one can
express the entropy as a function of $T, V, N_1$ and $N_2$. Using the
second law of thermodynamics we write
\begin{equation}
dQ = T dS = C_V dT + T\left(\frac{\partial S}{\partial V}\right)_{T,N_i}
dV + T\left(\frac{\partial S}{\partial N_1}\right)_{T,V,N_2} dN_1
 + T\left(\frac{\partial S}{\partial N_2}\right)_{T,V,N_1} dN_2,
\mylab{dheat1}
\end{equation}
where $Q$ stands for the heat supplied to the system. Fixing $N_1$ and $N_2$
the
last two terms vanish. Expressing volume $V$ in terms of $T, p, N_1$, and
$N_2$, assuming fixed $N_1$ and $N_2$ yields
\begin{equation}
dV = \left(\frac{\partial V}{\partial T}\right)_{p,N_i} dT
 + \left(\frac{\partial V}{\partial p}\right)_{T,N_i} dp
\nonumber
\end{equation}
Substituting into Eq.\,(\ref{dheat1}) we obtain (assuming fixed $N_1$ and
$N_2$)
\begin{align}
T dS &= \left[C_V
+ T\left(\frac{\partial S}{\partial V}\right)_{T,N_i}
\left(\frac{\partial V}{\partial T}\right)_{p,N_i}\right] dT
+ T\left(\frac{\partial S}{\partial V}\right)_{T,N_i}
\left(\frac{\partial V}{\partial p}\right)_{T,N_i} dp
\nonumber\\
&= C_p dT + T\left(\frac{\partial S}{\partial p}\right)_{T,N_i} dp.
\mylab{dheat2}
\end{align}
Introducing the thermal expansion coefficient
$\beta=\frac{1}{V}\left(\frac{\partial V}{\partial T}\right)_{p,N,N_1}$
and comparing the respective prefactors of $dT$ in the two lines of
Eq.\,(\ref{dheat2}) we rewrite Eq.\,(\ref{dheat1}) as
\begin{equation}
dQ = T dS = C_V dT + \frac{C_p-C_V}{\beta V}
dV + T\left(\frac{\partial S}{\partial N_1}\right)_{T,V,N_2} dN_1
 + T\left(\frac{\partial S}{\partial N_2}\right)_{T,V,N_1} dN_2.
\mylab{dheat3}
\end{equation}
Assuming that the material constants $C_p, C_V$ and $\beta$ are known we next
focus on the last two terms of (\ref{dheat3}). Consistently
with Eq.~(\ref{zus2}), we write for the Helmholtz free energy, $F=U-TS$,
of a local fluid element $V$ in thermodynamic equilibrium
\begin{equation}
dF = d(U-TS) = -S dT -p dV + \hat{\mu}_1 dN_1
 + \hat{\mu}_2 dN_2.
\mylab{dfreeenerg1}
\end{equation}
Partial differentiation of (\ref{dfreeenerg1}) with respect to
$N_1$ and $T$ gives
\begin{eqnarray}
-\left(\frac{\partial S}{\partial N_1}\right)_{T,V,N_2}
 &=& \left(\frac{\partial}{\partial N_1}
\left(\frac{\partial F}{\partial T}\right)_{V,N_i}\right)_{T,V,N_2}
\qquad\text{and}\nonumber\\
\left(\frac{\partial\hat{\mu}_1}{\partial T}\right)_{V,N_i}
 &=& \left(\frac{\partial}{\partial T}
\left(\frac{\partial F}{\partial N_1}\right)_{T,V,N_2}\right)_{V,N_i},
\end{eqnarray}
respectively. Identifying the mixed 2nd derivatives we obtain the so called
Maxwell relation for $S$ and $\hat\mu_1$
\begin{equation}
\left(\frac{\partial S}{\partial N_1}\right)_{T,V,N_2}=
-\left(\frac{\partial\hat{\mu}_1}{\partial T}\right)_{V,N_i}
\mylab{maxwellSmu}
\end{equation}
Analogously, we obtain the Maxwell relation for $S$ and $\hat\mu_2$. Using
them we rewrite (\ref{dheat3}) as
\begin{equation}
T dS = C_V dT + \frac{C_p-C_V}{\beta V}
dV - T\left(\frac{\partial\hat{\mu}_1}{\partial T}\right)_{V,N_i} dN_1
 - T\left(\frac{\partial\hat{\mu}_2}{\partial T}\right)_{V,N_i} dN_2.
\mylab{dheat4}
\end{equation}
Note that there exist other ways to express the heat. The one chosen
here is advantageous because the equation explicitly contains $dV$.
This easily allows  to consider the incompressible case by setting $dV=0$
(see below Section~\ref{mhf}).  The dependencies of the chemical
potentials on temperature will be also given.

Next, let us rewrite
local Gibbs and Gibbs-Duhem relations, Eqs.~(\ref{eq-7.3.2-4}) and
(\ref{eq-7.3.2-9}), in terms of the density of the Helmholtz free
energy, $f(\rho,\rho_1,T,t) = u - Ts$
\begin{align}
df &= -s dT +\mu_2 d\rho + \mudif d\rho_1,
\mylab{dfreeenerg1dens}\\
p &= -f +\mu_2 \rho + \mudif \rho_1.
\mylab{gibbsduhemf}
\end{align}
using the above introduced procedure for obtaining local
Gibbs~(\ref{eq-7.3.2-4}) and Gibbs-Duhem~(\ref{eq-7.3.2-9}) relations,
we derive from Eq.\,(\ref{dheat4}), the relations for the volume
densities of the extensive quantities
\begin{align}
T ds &= c_V dT - T\left(\frac{\partial\mu_2}{\partial
T}\right)_{\rho,\rho_1}d\rho
- T\left(\frac{\partial\mudif}{\partial T}\right)_{\rho,\rho_1}d\rho_1
\qquad\mbox{and}
\mylab{dheat5}\\
Ts & = \frac{c_p-c_V}{\beta}
 -T\left(\frac{\partial\mu_2}{\partial T}\right)_{\rho,\rho_1}\rho
 -T\left(\frac{\partial\mudif}{\partial T}\right)_{\rho,\rho_1}\rho_1,
\mylab{dheat6}
\end{align}
respectively. As above, $\mu_2=\hat{\mu}_2 N_a/ M_2$ and 
$\mudif=\hat{\mu}_1 N_a/ M_1-\hat{\mu}_2 N_a/ M_2$ are related to
densities not to particle number.
Note that we also changed the notation for partial derivative
with respect to $T$ regarding the chemical potentials $\mu_2,\mudif$
as defined by (\ref{dfreeenerg1dens}).

Dividing finally Eq.~(\ref{dheat5}) by $dt$, identifying $d/dt$ with the 
material derivative $D/Dt$, substituting into Eq.\,(\ref{entrop3}), using
Eqs.\,(\ref{fk040}) and (\ref{dens2}) and reordering we obtain the transport
equation for the temperature field
\begin{equation}
c_V\frac{DT}{Dt} \,+\, \frac{c_p-c_V}{\beta}\,\vec{\nabla}\cdot\vec{v}
\,+\, T\,\left(\frac{\partial\mudif}{\partial T}\right)_{\rho,\rho_1}\,
\vec{\nabla}\cdot\vec{j}_{\rho_1}
\,=\,
-T\vec{\nabla}\cdot\left[\frac{\vec{j}_u}{T} 
 + \frac{{\vec{j}_{\rho_1}}}{T}(\xi\Delta\rho_1 - \mudif)
\right] + R.
\mylab{temptransp}
\end{equation}
%
\subsection{Model-H -- bulk equations}
\mylab{mhf}
%
Next, we simplify the coupled equations for temperature, momentum and
volume density of component 1 by assuming a fluid with constant density 
$\rho$
(which implies incompressibility $\nabla\cdot\vec{v}=0$), in an
isothermal setting (constant $T$). Further, we express the density
$\rho_1$ in terms of a mass concentration $c_1=\rho_1/\rho$ and obtain
from Eq.\,(\ref{dens3}) the convective Cahn-Hilliard equation
\begin{equation}
\frac{\partial c_1}{\partial t} + \vec{v}\cdot\nabla c_1
+ \vec{\nabla}\cdot  \left[
M_1\nabla\left(\sigma_{c_1}\Delta c_1 - \mudift \right)\right] \,=\,
0,
\mylab{dens4}
\end{equation}
where we introduced ${\mudift}=\rho \mudif$,
$\sigma_{c_1}=\rho^2\xi$ and $M_1=k_3/\rho^2$. Note that $c_1$ is
dimensionless and $[\sigma_{c_1}]=\text{m}\,\text{kg}/\text{s}^2$,
$[\mudift]=\text{kg}/\text{m}\,\text{s}^2$ and
$[M_1]=\text{s}\,\text{m}^3/\text{kg}$.

The momentum equation \,(\ref{mom3}) reduces to
\begin{equation}
\rho\frac{\partial \vec{v}}{\partial t}\,+\,\rho\vec{v}\cdot\nabla\vec{v}\,=\,
-\nabla\cdot\left\{
\sigma_{c_1}(\nabla c_1)(\nabla c_1)+\left[p
\,-\,\sigma_{c_1} c_1\Delta c_1\,-\,\frac{\sigma_{c_1}}{2}(\nabla c_1)^2\right]\,\tens{I}
\right\}
+ \eta \Delta\vec{v},
\mylab{mom4}
\end{equation}
where the mechanical pressure is given by the local Gibbs-Duhem
relation~(\ref{eq-7.3.2-9}), i.e.\ $p = -u + \mu_2\rho + T s +
\mudift c_1$. We emphasize that $p$ as the mechanical pressure
for a homogeneous material in the thermodynamic equilibrium is a
locally defined variable that should not depend on any gradient or
derivative.
However, for simplicity we introduce an effective pressure
$p_{\mathrm{eff}_1}$
that incorporates all terms in the square brackets in equation~(\ref{mom4}).
Equations~(\ref{dens4})
and (\ref{mom4}) are normally called model-H.\cite{AMW98}

In the literature model-H is presented in various forms. Especially,
the momentum equation is written in different ways. Most differences
arise from different definitions of the pressure $p_{\mathrm{eff}_1}$
introduced in the last paragraph.  In the following we indicate how to
translate the different formulations and point out 'irreducible'
differences.

The review by Anderson et al.\ \cite{AMW98} gives as transport equation
for the momentum in a binary mixture (their Eq.\,(17b) with (16a) and (19))
in our notation)
\begin{equation}
\rho\left[\frac{\partial}{\partial t}\vec{v}\,+\,\vec{v}\cdot\nabla\vec{v}\right]
\,=\, -\nabla\cdot\left[(p_\mathrm{A}-\frac{\sigma_{c_1}}{2} (\nabla c_1)^2)\tens{I}
+\sigma_{c_1} (\nabla c_1)(\nabla c_1)\right] + \eta\Delta\vec{v} 
\mylab{mh-And1}
\end{equation}
However, their equation\,(20) for their chemical potential $\mu_c$
indicates that they do not follow their Eq.\,(10), but already absorbed
additional terms into their $\mu_c$. Our formulation coincides with theirs
identifying their $p_\mathrm{A}-\sigma_{c_1} (\nabla c_1)^2/2$ and our
$p_{\mathrm{eff}_1}$. The difference in the formulation arises because
Anderson et al.\ use $\mu_c$ in place of our
$\mudift-\sigma_{c_1} \Delta c_1=\bar{\mu}$ in the
thermodynamic pressure definition, i.e.\ their $\mu_c=\bar{\mu}$ is the
chemical potential for an inhomogeneous equilibrium. Then also their
Eq.\ (21) corresponds to our (\ref{dens4}).

Jasnov and Vi\~nals \cite{JaVi96} present two forms for the momentum equation 
\begin{equation}
\rho\left[\frac{\partial}{\partial t}\vec{v}\,+\,\vec{v}\cdot\nabla\vec{v}\right]
\,=\, - \nabla p_\mathrm{JV} + \eta\Delta\vec{v} + \bar{\mu} \nabla c_1
\mylab{mh-JaVi1}
\end{equation}
[their Eq.\,(2)] and 
\begin{equation}
\rho\left[\frac{\partial}{\partial t}\vec{v}\,+\,\vec{v}\cdot\nabla\vec{v}\right]
\,=\, - \nabla \tilde{p}_\mathrm{JV} + \eta\Delta\vec{v} - c_1 \nabla \bar{\mu}.
\mylab{mh-JaVi2}
\end{equation}
[their Eq.\,(2) with the replacement described in the last paragraph
of their appendix]. They also use $\bar{\mu}$ instead of $\mudift$.  We
introduce different symbols $p_\mathrm{JV}$ and $\tilde{p}_\mathrm{JV}$ for the
respective pressures.
The second form can be obtained from ours taking into account
$\nabla\cdot\left[p_{\mathrm{eff}_1}\tens{I}+\sigma_{c_1} (\nabla c_1)
(\nabla c_1)\right]=c_1\nabla\bar{\mu}+\nabla p$. The first
form just follows from integration by parts and redefining the
pressure again: $\tilde{p}_\mathrm{JV}= p_\mathrm{JV} - c_1 \bar{\mu}$.

The form of model-H presented in the review 
by Hohenberg and Halperin \cite{HoHa77} gives a momentum equation (their
Eq.\,(5.1b)) that agrees on the first view with the second form of 
Jasnov and Vi\~nals~(\ref{mh-JaVi2}). However, they dropped
the pressure term, i.e.\ in the limit of constant concentration 
their model does not reduce to the Navier-Stokes equations.

Finally, we rewrite model-H in terms of the
difference of concentrations $c=c_1-c_2=2c_1-1$. Introducing
new parameters $\sigma_c=\sigma_{c_1}/4$ and $M=4M_1$ and specifying the
chemical potential $\mudift = 2 \partial_c f(c)$, where $f(c)$ is the
concentration dependent part of the local free energy, results in 
\begin{equation}
\partial_t  c + \vec{v}\cdot\nabla c \,=\, 
- \nabla \cdot\left\{ M \nabla \left[\sigma_c \Delta c - \partial_c f(c)
\right]\right\}.
\mylab{mh-eqc-our}
\end{equation}
and
\begin{equation}
\rho\frac{\partial \vec{v}}{\partial t}\,+\,\rho\vec{v}\cdot\nabla\vec{v}\,=\,
-\nabla\cdot\left\{\sigma_c(\nabla c)(\nabla c) + p_{\mathrm{eff}}\,\tens{I}
\right\} \,+\, \eta \Delta\vec{v}
\mylab{mh-mom-our}
\end{equation}
where
\begin{equation}
p_{\mathrm{eff}}\,=\,p\,-\,\sigma_c (c+1)\,\Delta c\,-\,\frac{\sigma_c}{2}(\nabla c)^2.
\mylab{pco-our}
\end{equation}
Fixing $f(c)$ to be a symmetric double well potential,
Eq.\,(\ref{mh-eqc-our}) corresponds to the convective Cahn-Hilliard
equation studied, for instance, in Ref.~\cite{GNDZ01,WORD03}.  The energy
will be further discussed in Section\,\ref{sec-en}.
Because of its importance for the boundary conditions (see
Section~\ref{sec-bc}) we also give the stress tensor
\begin{equation}
\tensg{\tau} \,=\,-p_{\mathrm{eff}}\tens{I}\,-\,\sigma_c (\nabla c)(\nabla c)
\,+\,\eta\left(\nabla \vec{v} + (\nabla \vec{v})^\Tr\right), 
\mylab{mh-tau-our}
\end{equation}
where $(p-p_{\mathrm{eff}})\tens{I}-\sigma_c (\nabla c)(\nabla c)$ represents
the Korteweg stress.\cite{Kort1901,Jose90}
The pressure $p_{\mathrm{eff}}$ can be calculated from the Poisson equation
\begin{equation}
\Delta p_{\mathrm{eff}} \,=\, -\sigma_c (\nabla\nabla):[(\nabla c)(\nabla c)]
\,-\,\rho(\nabla \vec{v}):(\nabla \vec{v}).
\mylab{mh-pres-our}
\end{equation}

In the literature the various formulations of model-H are mainly used 
to describe the behavior of bulk flows.
\cite{AMW98,JaVi96,VMM99,VMM99b}  Systems confined between rigid plates
are considered in some cases \cite{Anta95,JaVi96,BoBe03} assuming (i)
the diffuse interface is far away from the plates, and (ii) the walls
are neutral with respect to the two components. However, the role of
energetically biased plates and the evolution of a free surface of the
binary mixture have to be understood in their interaction with the
bulk flow to be able to describe an evolving free surface film on a
solid support. The necessary boundary conditions are discussed next.
%
\section{Boundary conditions} \mylab{sec-bc}
%
\subsection{Concentration} \mylab{sec-bc-conc}
%
For the concentration field the boundary conditions were discussed in
connection with a purely diffusive transport for a system confined by
rigid plates.\cite{FMD97,FMD98,Kenz01} Assuming the velocity is zero
at the rigid substrate (no-slip condition, see
Section~\ref{sec-bc-vel}) the conditions for the full model-H are
similar. We have zero diffusive flux through the substrate ($z=0$)
\begin{equation} 
\partial_z [\sigma_c \Delta c - \partial_c f(c)] \,=\,0
\mylab{bc-conc1}
\end{equation} 
and obtain in the general case an evolution equation for the concentration
(see Appendix~\ref{deribc})
\begin{equation}
\partial_t  c +\vec{v}\cdot\nabla c\,=\, - M^- 
[-\sigma_c \partial_z c - \sigma^-  \Delta_\parallel c 
\,+\,\partial_c f^-(c)] 
\mylab{bc-conc2b}
\end{equation}
where $\Delta_\parallel=\nabla_\parallel\cdot\nabla_\parallel$ and 
$\nabla_\parallel= (\partial_x,\partial_y)$. Here, however, we will
focus on a surface energy that (i) does not depend on concentration gradients
($\sigma^-=0$) and (ii) relaxes instantaneously to its equilibrium value 
($M^-\rightarrow\infty$).

At the free surface [$z=h(x,y,t)$] one has the condition of zero diffusive flux 
through the moving surface, i.e.\ $\vec{n}\cdot\vec{j}_{\rho_1}=0$
with $\vec{j}_{\rho_1}$ as defined in Eqs.\,(\ref{dens2}) and
(\ref{tdflux6}). 
\begin{equation}
\vec{n}=\frac{(-\partial_x h,-\partial_y h, 1)}{\bigl(1+(\partial_xh)^2+(\partial_yh)^2\bigr)^{1/2}}
\end{equation}
is the normal vector of the free surface.
The change from the total flux $\vec{j}'_{\rho_1}$ [Eq.\,(\ref{dens1b})]
to $\vec{j}_{\rho_1} = \vec{j}'_{\rho_1} - \rho_1 \vec{v}$ exactly accounts
for the transformation into the frame moving locally with the surface.
One gets
\begin{equation}
\vec{n}\cdot\nabla\left(\sigma_c \Delta c - \partial_c f(c) \right) \,=\,0.
\mylab{bc-der2}
\end{equation}
The second condition is in the general case again an evolution equation for the
concentration field on the boundary as derived in Appendix~\ref{deribc}: 
The evolution equation is valid in the local comoving frame, i.e.\
\begin{equation}
\partial_t  c + \vec{v}\cdot\nabla c\,=\, - M^+ 
[\sigma_c(\vec{n}\cdot\nabla)c \,-\,\sigma^+ \Delta_s c \,+\,\partial_c
f^+(c)],
\mylab{bc-conc2}
\end{equation}
where $\Delta_s=\nabla_s\cdot\nabla_s$ and the surface nabla operator is
defined as $\nabla_s=(\tens{I}-\vec{n}\vec{n})\cdot\nabla$.
In the following we assume as above $\sigma^+=0$ and $M^+\rightarrow\infty$.
We will drop the respective terms after the non-dimensionalization in
Section~\ref{sec-nondim}. 
Note, however, that the units of the surface parameters
differ from the ones of the corresponding bulk parameters:
$[M^\pm]=\text{s}/\text{kg}$,
$[\sigma_s^\pm]=\text{kg}\,\text{m}^2/\text{s}^2$, $[\partial_c
f^\pm]=\text{N}/\text{m}=\text{kg}/\text{s}^2$.
%
\subsection{Velocity} \mylab{sec-bc-vel}
%
The boundary conditions for the velocity fields are the 
no-slip and no-penetration condition at the solid substrate ($z=0$)
\begin{equation}
\vec{v}=0,
\mylab{bc-der0}
\end{equation}
and the force equilibrium at the free surface ($z=h$)
\begin{equation} 
(\tensg{\tau}-\tensg{\tau}_{air})\cdot\vec{n} \,=\,
-\gamma(c)\,\vec{n}\,\nabla\cdot\vec{n}
\,+\,\nabla_s\gamma(c)
\mylab{th-hddf5}
\end{equation}
Note that $\nabla\cdot\vec{n}$
corresponds to the curvature of the free surface.
We assume that the ambient air does not transmit any force
($\tensg{\tau}_{air}=0$). 
The first term on the r.h.s.\ of Eq.\,(\ref{th-hddf5}) corresponds to the 
Laplace or curvature pressure whereas the second one represents a Marangoni 
force tangential to the interface and results from the variation  
of the surface tension along the surface caused normally by a solutal or thermal
Marangoni effect. As shown in the Appendix~\ref{deribc}
 these terms can be derived from
a minimization procedure.

For a pure Navier-Stokes problem the Marangoni term is often modeled
as a linear dependence of the surface tension on concentration or
temperature.  Here, however, one has to use a condition in accordance
with the interface energies introduced when discussing the boundary
conditions for the concentration field, i.e.\ at
Eq.\,(\ref{bc-conc2}). For $\sigma^+=0$ and $M^+\rightarrow\infty$
[see Eq.\,(\ref{bc-conc2})] the surface tension $\gamma(c)$
corresponds to $f^+(c)$ plus a constant (reference tension $\gamma_0$,
see below Section~\ref{sec-en}).  For $\sigma^+\neq0$ the surface
tension depends as well on concentration gradients
$\gamma=\gamma(c,(\nabla_s c)^2)$, a concept that has not yet been
followed in the literature.  Considering a finite $M^+$ would
correspond to a $\gamma(c,t)$, i.e.\ to a dynamical surface tension
characterized by a relaxation time towards its equilibrium value.
Both complications will not be considered further in the present
paper.

The boundary condition (\ref{th-hddf5}) is of vectorial character, 
i.e.\ three scalar conditions are derived by projecting it
onto $\vec{n}$, $\vec{t}_1$, and $\vec{t}_2$, respectively, where
\begin{equation}
\vec{t}_1=\frac{(1,0,\partial_x h)}{\bigl(1+(\partial_x h)^2\bigr)^{1/2}},
\quad\vec{t}_2=\frac{(0,1,\partial_y h)}{\bigl(1+(\partial_y h)^2\bigr)^{1/2}}
\mylab{t1t2}
\end{equation}
are the (non orthogonal) tangent vectors. The resulting scalar conditions
\begin{eqnarray}
&& -\sigma_c\,(\vec{n}\cdot\nabla c)^2 - p_{\mathrm{eff}}
\,+\,2\eta\,\vec{n}\cdot\left(\nabla \vec{v}\right)\cdot\vec{n}
\,=\,-\gamma(c)\,\nabla\cdot\vec{n}
\mylab{th-hddf7}\\
&& -\sigma_c\, (\vec{t}_1\cdot\nabla c)(\vec{n}\cdot\nabla c)
\,+\,\eta\,\vec{t}_1\cdot\left(\nabla \vec{v} + (\nabla \vec{v})^\Tr\right)\cdot\vec{n}
\,=\,\vec{t_1}\cdot\nabla\gamma(c) 
\mylab{th-hddf6}\\
&& -\sigma_c\, (\vec{t}_2\cdot\nabla c)(\vec{n}\cdot\nabla c)
\,+\,\eta\,\vec{t}_2\cdot\left(\nabla \vec{v} + (\nabla \vec{v})^\Tr\right)\cdot\vec{n}
\,=\,\vec{t_2}\cdot\nabla\gamma(c)
\mylab{th-hddf6b}
\end{eqnarray}
correspond to equilibria of  normal and
tangential forces, respectively.

At the free surface one has furthermore the 
kinematic condition, i.e.\ the prescription that the surface follows the flow field
\begin{equation} 
\partial_t h \,=\, \vec{n}\cdot\vec{v} \,\sqrt{1+(\nabla_\parallel h)^2}
\mylab{th-hddf4}
\end{equation} 
which can be written in a more compact form as
\begin{equation} 
(\partial_t \vec{h})\cdot\vec{n} = \vec{v}\cdot\vec{n}
\mylab{th-hddf4b}
\end{equation} 
where vector $\vec{h} = h(x,y,t)\,\vec{e}_z$ is tracking the free surface.
%
\section{Non-dimensionalization} 
\mylab{sec-nondim}
%
Next, we non-dimensionalize the bulk equations and the boundary equations 
in 2 steps: (i) introduction
of abstract scales for velocity, pressure, length, concentration and energy density
that leads us to a set of dimensionless numbers; (ii) introduction of problem
specific length and velocity scales to obtain a minimal set of dimensionless
numbers valid for the problems without external driving studied here.
\subsection{Abstract scales}
Introducing scales
\begin{equation}
\begin{array}{ccccc}
\text{dimensionless} &\quad& \text{Scale} &\quad& \text{dimensional}\\
t'&&\tau_v=l/U&&t= \tau_v t'\\
\vec{x}'&&l&&\vec{x}=l\vec{x}'\\
\vec{v}'&&U&&\vec{v}=U\vec{v}'\\
p'&&P&&p=P p'\\
c'&& C &&c= C c'\\
f'(c')&& E &&f(c)= E f'(c')\\
f'^\pm(c')&& E^\pm &&f^\pm(c)= E^\pm f'^\pm(c')
\end{array}
\mylab{scales1}
\end{equation}
one obtains after dropping the primes the
dimensionless bulk equations
\begin{equation}
\partial_t  c + \vec{v}\cdot\nabla c \,=\, 
- \text{Ts}\,\nabla \cdot\left\{ \nabla \left[\text{Ko}  \Delta c -
\partial_c f(c)
\right]\right\}.
\mylab{mh-eqc-sca}
\end{equation}
and
\begin{equation}
\text{Ps}\left[
\frac{\partial \vec{v}}{\partial t}\,+\,\vec{v}\cdot\nabla\vec{v}\right]\,=\,
-\nabla\cdot\left\{
\text{Ko'} (\nabla c)(\nabla c) + p_{\mathrm{eff}}\,\tens{I}
\right\} + \frac{\text{Ps}}{\text{Re}}\Delta\vec{v}
\mylab{mh-mom-sca}
\end{equation}
where
\begin{equation}
p_{\mathrm{eff}}\,=\,p\,-\,\text{Ko'}(c+1)\,\Delta c
\,-\,\tfrac{1}{2}\text{Ko'}(\nabla c)^2.
\mylab{pco-sca}
\end{equation}
We defined the dimensionless numbers
\begin{eqnarray}
\text{Reynolds number}&\quad& \text{Re}\,=\,
\frac{U l\rho}{\eta}\nonumber\\
\text{Korteweg number 1}&\quad&\text{Ko}\,=\,\frac{\sigma_c\,C^2}{l^2 E}
\nonumber\\
\text{Korteweg number 2}&\quad&\text{Ko'}\,=\,\frac{\sigma_c\, C^2}{l^2 P}
\nonumber\\
\text{Time scale ratio}&\quad&\text{Ts}\,=\,\frac{M E}{U l\,C^2}
\nonumber\\
\text{Pressure scale ratio}&\quad&\text{Ps}\,=\,\frac{\rho U^2}{P}
\mylab{scale2}
\end{eqnarray}
We propose the name 'Korteweg number'  because both of them are related to the 
Korteweg stresses. 
The Korteweg numbers can be seen as  'bulk Marangoni numbers'.
For the determination of the energy scale $E$ see Section\,\ref{sec-en}.

The scaled boundary conditions for the concentration field at both
interfaces are the no-flux condition
\begin{equation}
0\,=\, 
\vec{n}\cdot\nabla \left[\text{Ko}  \Delta c - \partial_c f(c)
\right].
\mylab{bc-sca0}
\end{equation}
and the evolution equations for the concentration at the surface
\begin{equation}
\partial_t  c + \vec{v}\cdot \nabla c \,=\, - \text{Ts}^\pm\,
[\text{Ko}\,\vec{n}\cdot\nabla c
\,-\, \text{Ko}^\pm\, \Delta_s c 
\,+\, \text{En}^\pm  \partial_c f^\pm(c)].
\mylab{bc-sca}
\end{equation}
For the substrate one sets $\vec{n}=(0,0,-1)$.

At the free surface the conditions for the normal and tangential forces are
\begin{eqnarray}
&& -\text{Ko'}\,(\vec{n}\cdot\nabla c)^2 - p_{\mathrm{eff}}
\,+\,2\frac{\text{Ps}}{\text{Re}}\,\vec{n}\cdot\left(\nabla \vec{v}\right)\cdot\vec{n}
\,=\,-\text{S}\,\gamma(c)\,\nabla\cdot\vec{n}
\mylab{bc-sca3}\\
&& -\text{Ko'}\, (\vec{t}_1\cdot\nabla c)(\vec{n}\cdot\nabla c)
\,+\,\frac{\text{Ps}}{\text{Re}}\,\vec{t}_1\cdot\left(\nabla \vec{v} 
+ (\nabla \vec{v})^\Tr\right)\cdot\vec{n}
\,=\,\text{S}\,\vec{t_1}\cdot\nabla\gamma(c)
\mylab{bc-sca2}\\
&& -\text{Ko'}\, (\vec{t}_2\cdot\nabla c)(\vec{n}\cdot\nabla c)
\,+\,\frac{\text{Ps}}{\text{Re}}\,\vec{t}_2\cdot\left(\nabla \vec{v} 
+ (\nabla \vec{v})^\Tr\right)\cdot\vec{n}
\,=\,\text{S}\,\vec{t_2}\cdot\nabla\gamma(c)
\mylab{bc-sca2b}
\end{eqnarray}
respectively, where $\gamma$ is the dimensionless surface tension
referred below in Section~\ref{sec-en} as $\gamma'$. The
dimensionless numbers are either given above or listed next
\begin{eqnarray}
\text{Boundary Korteweg number}&\quad& \text{Ko}^\pm\,=\, 
\frac{\sigma^\pm C^2}{l^3 E}
\nonumber\\
\text{Boundary time scale ratio}&\quad& \text{Ts}^\pm\,=\,
\frac{E l^2 M^\pm}{U C^2}
\nonumber\\
\text{Boundary energy number}&\quad& \text{En}^\pm\,=\,
\frac{E^\pm}{l E}
\nonumber\\
\text{Surface tension number}&\quad&\text{S}\,=\,\frac{\gamma_0}{l P}
\mylab{scale4}
\end{eqnarray}
\subsection{Specific scales}
For relaxational settings, i.e.\ systems without external driving forces one 
might specify scales based on the 'internal' diffusive or convective
transport. Assuming very viscous liquids and taking into account that all
structure formation will be driven by the decomposition process, it is
convenient to base all scales on the diffusive processes only.  
Fixing Ko${}=1$ and Ts${}=1$ length and velocity scales become
\begin{equation}
l=\sqrt{\frac{\sigma_c}{E}}\,C\qquad\text{and}\qquad U=\frac{M E}{l\,C^2},
\mylab{scale2b}
\end{equation}
respectively. Choosing a pressure scale based on the energy density scale
\begin{equation}
P =  E
\mylab{scale2c}
\end{equation}
identifies the two Korteweg numbers, i.e.\ Ko'$=$Ko$=1$. 
The specific forms of Reynolds and Pressure numbers are then 
\begin{equation}
\text{Re}=\frac{M E \rho}{\eta\,C^2}
\qquad\text{and}\qquad 
\text{Ps}=\frac{\rho M^2 E}{l^2 C^4}=\frac{\rho M^2 E^2}{\sigma_c C^6},
\mylab{scale2bb}
\end{equation}
respectively, i.e.\ \text{Ps}/\text{Re}$=\eta M/l^2C^2 =\eta M E/\sigma_c C^4$.
The dimensionless numbers related to the boundaries become
\begin{equation}
\text{Ts}^\pm=\frac{l^3 M^\pm}{M}=\frac{M^\pm \sigma_c^{3/2}\,C^3}{M E^{3/2}}
\qquad\text{and}\qquad 
\text{Ko}^\pm=\frac{\sigma^\pm}{l \sigma_c}
=\frac{\sigma^\pm E^{1/2}}{C\sigma_c^{3/2}}.
\mylab{scale2d}
\end{equation}
Note that $\text{Ts}^\pm\rightarrow\infty$ for $M^\pm\rightarrow\infty$ and 
$\text{Ko}^\pm=0$ for $\sigma^\pm=0$ (see discussion in Section~\ref{sec-bc}).
\begin{equation}
\text{En}^\pm=\frac{E^\pm}{l E}
=\frac{E^\pm}{\sigma_c^{1/2} E^{1/2} C}
\qquad\text{and}\qquad 
\text{S}=\frac{\gamma_0}{l E}
=\frac{\gamma_0}{\sigma_c^{1/2} E^{1/2} C}
\mylab{scale2e}
\end{equation}
The relation of 
$\text{S}$ and the 'classical' Marangoni number is discussed in the next section.
%
\section{Local energies} \mylab{sec-en}
%
For the local bulk and surface energies we use a simple polynomial, an
approximation valid near the critical point. However, it is straightforward
to introduce other expressions derived using Flory-Huggins
or more advanced theories.\cite{GeKr03}
For the bulk energy a symmetric quartic potential is used
\begin{equation}
f(c)\,=\,f_0\,-\,\frac{a(T)}{2} c^2\,+\,\frac{b}{4} c^4
\mylab{en1}
\end{equation}
corresponding to the nondimensional form (Eqs.~(\ref{scales1}))
\begin{equation}
f'(c')\,=\tfrac{1}{4}\,(c'^2-1)^2 \,+\,\text{const}
\mylab{en1b}
\end{equation}
with $E=b C^4$ and $C=\sqrt{a/b}$.
For the surface energies of the two interfaces we use the respective quadratic expressions
\begin{equation}
f^\pm(c)\,=\,\gamma_0^\pm \,+\, \tilde{a}^\pm c \,+\, \frac{\tilde{b}^\pm}{2} c^2.
\mylab{en2}
\end{equation}
Note that in the framework of model-H for a film of binary mixture
the surface energies $f^-(c)$ and $f^+(c)$ correspond to the
concentration dependent surface tensions of the liquid-solid and the
liquid-gas interface, respectively. This implies that $f^+(c)$ is
responsible for a linear ($\tilde{b}^+=0$) or nonlinear
($\tilde{b}^+\neq0$) Marangoni effect. The surface energies
$\gamma_0^+$ and $\gamma_0^-$ are the respective reference surface tensions at
$c=0$. 

Using the reference surface tension of the free surface as a scale for
both interfaces, i.e.\ $E^+=E^-= \gamma_0^+= \gamma_0$, we arrive at
the nondimensional expressions
\begin{equation}
f'^\pm(c')\,=\, \frac{\gamma_0^\pm}{\gamma_0^+} \,+\, a^\pm c' \,+\, \frac{b^\pm}{2} c'^2
\mylab{en2b}
\end{equation}
and identify $\gamma'\,=\,f'^+(c')$. The dimensionless parameters 
\begin{equation}
a^\pm\,=\,\frac{\tilde{a}^\pm C}{\gamma_0^+}\qquad\text{and}\qquad
b^\pm\,=\,\frac{\tilde{b}^\pm C^2}{\gamma_0^+}
\mylab{en2c}
\end{equation}
describe preferential adsorption of one of the species at the
interface and changes in the interaction between the species at the
respective interfaces. Inspecting Eq.\,(\ref{bc-sca2}) it becomes
clear that the 'classical' Marangoni number for a linear Marangoni
effect is $\text{Ma}=a^+\text{S}$. The corresponding number for a
quadratic Marangoni effect is $\text{Ma}_2=b^+\text{S}$ (compare, for
instance, Ref.~\onlinecite{OrRo92,OrRo94}). Furthermore, we can now specify
\begin{equation}
\text{En}^+=\text{En}^-=\frac{\gamma_0}{lE}=\text{S}.
\mylab{en2d}
\end{equation}
This implies that the boundary conditions for concentration and
momentum transport at the free surface are intrinsically coupled.
Note, finally that these considerations only apply for
$\text{Ts}^\pm\rightarrow\infty$ and Ko$^\pm=0$. See discussion in
Sections~\ref{sec-bc} and~\ref{sec-nondim}. In the following we only
work with dimensionless quantities and drop all primes.

We are now equipped with a complete model to investigate a wide
variety of systems involving decomposing mixtures with free
surfaces. Although, the boundary conditions in Section~\ref{sec-bc}
are written for a film on a solid substrate they can be easily adapted
for free standing films, i.e.\ for a film with two free surfaces. Also
droplets of a mixture on a solid substrate can be studied if the given
model is supplemented by a condition at the contact line such as a
concentration dependent equilibrium contact angle. This will be the
scope of future work. 

To understand the evolution of the surface and concentration profiles
of a decomposing film we next analyse (i) the homogeneous and
vertically stratified base state solutions, (ii) the transversal
instability of the base state solutions that lead to the
experimentally observed film profiles and concentration patterns, and
(iii) the full non-linear time evolution. Part (i) will be studied in
the remainder of the present paper, part (ii) forms the content of the
accompanying paper,\cite{MaTh07} and part (iii) will be presented in
a planned sequel.
%
\section{Base states} \mylab{sec-base}
%
The understanding of the behaviour of a thin film of a mixture on a
solid substrate has to be based on an analysis of the base state
solutions. For a film on a horizontal substrate without further
driving forces parallel to the substrate the base states are
quiescent, i.e.\ the velocity of the fluid mixture is zero. We
distinguish two types of quiescent base states: (a) completely
homogeneous flat film and (b) horizontally (transversally) homogeneous
but vertically stratified film.  
%
\subsection{Completely homogeneous film} \mylab{sec-base-chf}
%
A completely homogeneous film of arbitrary thickness $h(x,y)=h_0$,
with arbitrary concentration $c(\vec{x})=c_0$ and with quiescent fluid
$\vec{v}_0=0$ corresponds to a base state solution of the system
(\ref{mh-eqc-sca})-(\ref{en2b}) only if there exists no energetic
bias at the solid-liquid interface or the free surface, 
i.e.\ without any linear or nonlinear Marangoni
effect: $a^\pm=b^\pm=0$. The corresponding effective pressure is
$p_{\mathrm{eff}}=0$

For energetically biased interfaces the boundary conditions for the
concentration field are only fulfilled if $\partial_c f^\pm(c_0)=0$,
i.e.\ for $c_0=-a^+/b^+=-a^-/b^-$. For finite $a^+, a^-$ this is from
the experimental point of view a very unlikely case. Here we will not
pursue it further. However, for $a^+=a^-=0$ a homogeneous film of a
symmetric mixture ($c_0=0$) represents a base state for any $b^+$ and
$b^-$. This case corresponds to a purely quadratic Marangoni
effect. Experimentally, it is not a very common case but was studied in
hydrodynamics for films of alcohol solutions \cite{OrRo94} and also as
a problem of purely diffusive demixing in a gap.  The latter case was
analysed in detail in Refs.~\onlinecite{FMD98,Kenz01} and will serve as a
benchmark for our linear stability analysis in Ref.~\onlinecite{MaTh07}.
%
\subsection{Vertically stratified, horizontally homogeneous film}
\mylab{sec-base-vsf}
%
Depositing a thin film of a mixture on a solid substrate it is to
expect that processes that lead to a vertical stratification are much
faster than processes that lead to a horizontal structuring if the
film thickness is similar or below the length scale of bulk
decomposition. The vertically stratified films may on a larger time
scale undergo a further horizontal structuring. The finally emerging 
horizontal length scales and structures can be understood from the 
'short-time' vertical layering. Therefore we focus
next on a systematic investigation of steady layered films.  

A flat layer
($h=h_0$) of a quiescent fluid mixture ($\vec{v}_0=0$) represents a
base state if the vertical concentration profile $c=c_0(z)$ is a
steady solution of the classical one-dimensional non-convective
Cahn-Hilliard equation
\begin{equation}
\partial_t c \,=\, \partial_{zz} \left[\partial_{zz} c - \partial_c f(c)
\right].
\mylab{ch-1d}
\end{equation}
and the boundary conditions (i) $0= \partial_z \left[\partial_{zz} c_0 -
\partial_c f(c_0)\right]$ (at $z=0$ and $z=h$) and (ii) $0= [\pm\partial_z
c_0 + \text{S} \partial_c f^\pm(c_0)]$ (`$-$' at $z=0$ and `$+$' at
$z=h$). Taking into account (i) one has to solve the bulk equation
$\partial_{zz} c_0 - \partial_c f(c_0)+K_1=0$ with  boundary conditions (ii).
The constant of integration $K_1$ represents the dimensionless chemical potential
for an inhomogeneous equilibrium as discussed after (Eq.~\ref{mh-And1}).

The remaining equations and boundary conditions are
fulfilled with $p_{\mathrm{eff}}=p_{\mathrm{eff}}(z)=-(\partial_z
c_0)^2 + \text{const}$, i.e.\ the layers are completely characterized by 
$c_0(z)$. 
\begin{figure}[tbh]
\centering
(a)\includegraphics[width=0.6\hsize]{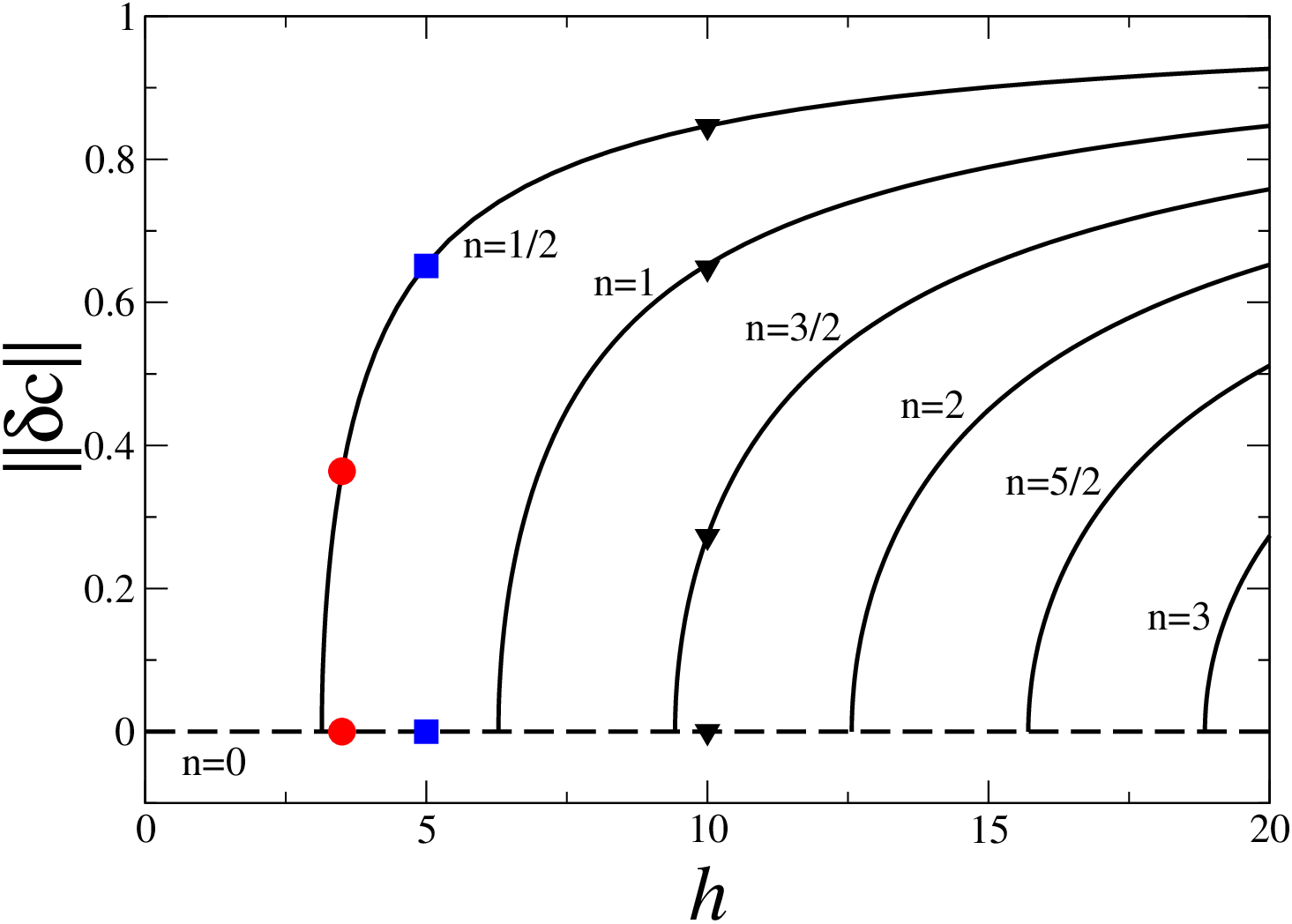}
(b)\includegraphics[width=0.6\hsize]{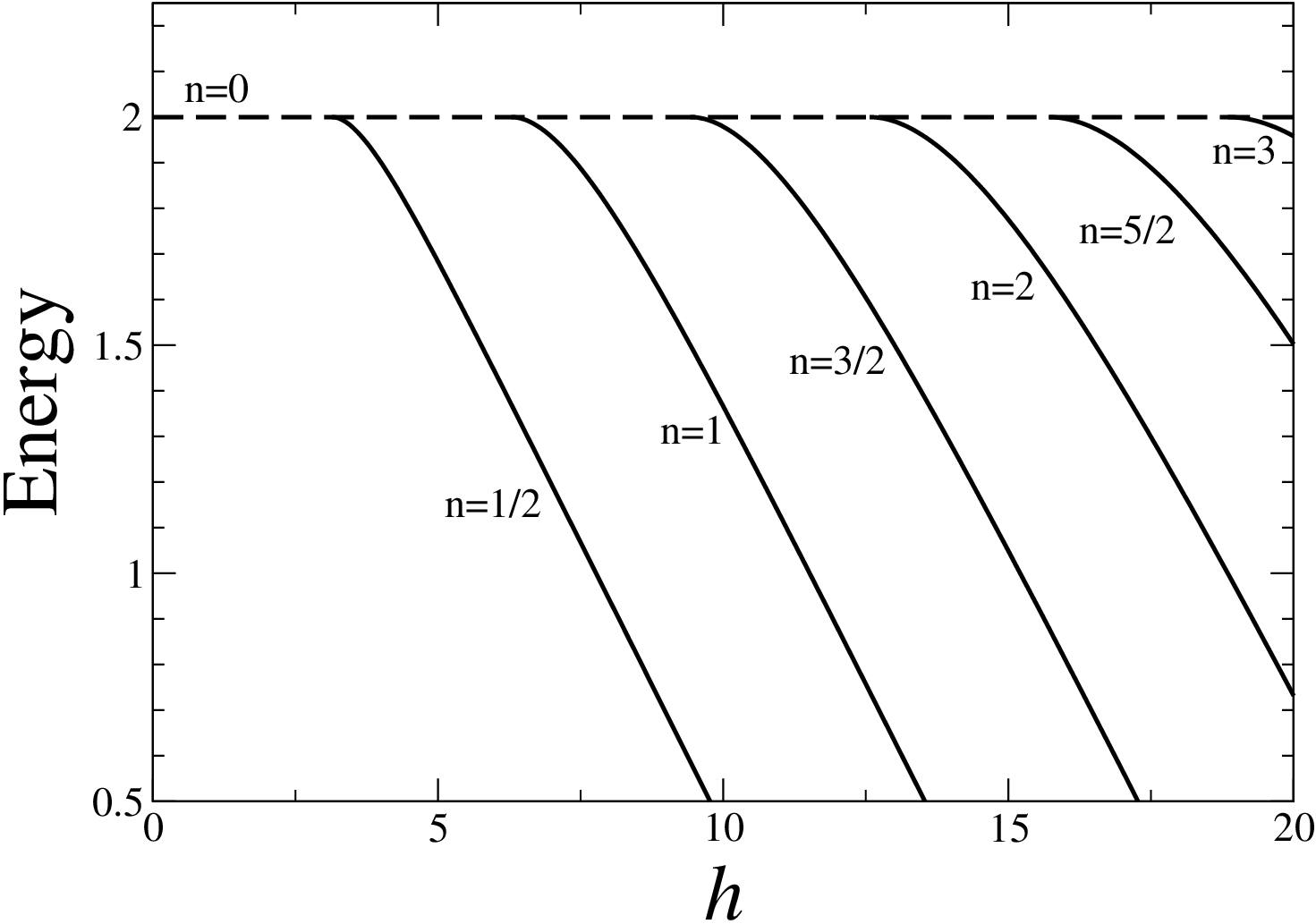}
\caption{Branches of steady vertical concentration profiles 
for energetically non-biased (neutral) surfaces ($a^\pm=b^\pm=0$)
and a symmetric mixture ($\bar{c}=0$) in dependence
on the film thickness $h$. 
Shown are (a) the $L_2$-norm and (b) the energy $E$ of $c_0(z)$. 
Selected corresponding profiles are given in Fig.~\ref{ss2prof}.
$S=1$, and the symbols are explained in the main text.
}
\mylab{ss2}
\end{figure}
In the following we determine families of solutions in terms of
concentration profiles for (i) energetically neutral or non-biased surfaces, (ii)
symmetrically biased surfaces, (iii) antisymmetrically biased
surfaces, and (iv) asymmetrically biased surfaces. Thereby we
characterize the concentration profiles by their energy
\begin{equation}
E\,=\, f^+ \,+\, f^- \,+\,\int_0^h [(\partial_z c)^2 \,+\,f(c)]\,dz
\mylab{ss-en}
\end{equation}
and the $L_2$-norm
\begin{equation}
||\delta c||\,=\,\sqrt{\frac{1}{h}\int_0^h [c(z)-\bar{c}]^2\,dz}
\mylab{ss-l2}
\end{equation}
where $\bar{c}$ is the mean concentration. Note that $E$ should only be used to compare
films of identical $h$ and $\bar{c}$. The profiles are determined using numerical 
continuation techniques detailed in Ref.~\onlinecite{MaTh07}. 
\subsubsection{Energetically neutral surfaces}
\mylab{sec-ens}

\begin{figure}[tbh]
\centering
\includegraphics[width=0.65\hsize]{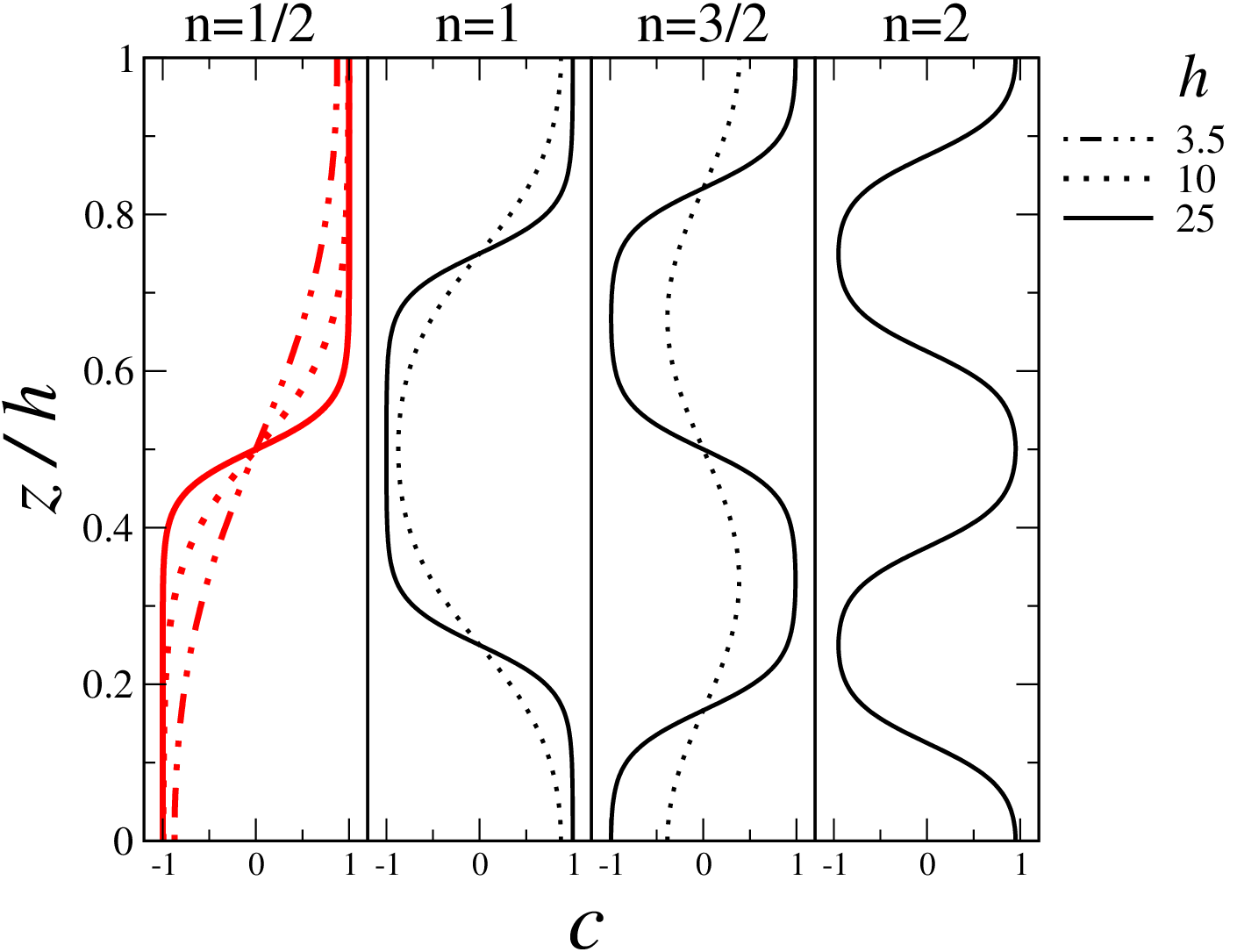}
\caption{Selected concentration profiles corresponding to Fig.~\ref{ss2}.
Note that each of the solutions has a 'twin' obtained by $c\rightarrow
-c$ that also corresponds to an allowed profile. This degeneracy may
be lifted by energetically biased surfaces (depending on the symmetry
$z\rightarrow h_0-z$, see below). The $n=1/2$ profiles corresponding
to the energy minimum for the respective film thickness are shown as
heavy (red online) lines.}
\mylab{ss2prof}
\end{figure}

The results for the trivial case of energetically neutral solid
substrate and free film surface are given in Fig.~\ref{ss2} for a
symmetric mixture, i.e.\ the case of zero mean concentration
$\bar{c}=0$. Shown are the $L_2$-norm and the energy $E$ per film
area. Fig.~\ref{ss2prof} presents selected concentration profiles.
The base states for a film correspond to selected solutions of the
one-dimensional bulk Cahn-Hilliard equation. A multiple of the period
has to be equal to the film thickness.  At both, the substrate and the
free surface, the profile has a minimum or a maximum.  This allows to
classify the obtained solution branches by the number of periods
$n$. The simplest stratified films correspond to half a period
($n=1/2$), one period ($n=1$), one and a half period ($n=3/2$) and so
on.

Note that the solutions with an integer $n$ are symmetric with respect
to a reflection at the plane $z=h_0/2$, i.e.\ $c_0(z)=c_0(h_0-z)$.  We
call them in the following '$z$-reflection-symmetric'. They are
accompanied by a twin solution with identical $L_2$-norm and energy
obtained by an inversion of concentration:
$c_0(z)\rightarrow-c_0(z)$. On the contrary, the solutions with a
non-integer $n$ are antisymmetric with respect to a reflection at the
plane $z=h_0/2$, i.e.\ $c_0(z)=-c_0(h_0-z)$
('$z$-reflection-antisymmetric'). The resulting second solution has
naturally identical $L_2$-norm and energy, and can also be obtained by
an inversion of concentration. We will also call it the twin solution.

\begin{figure}[tbh]
\centering
(a)\includegraphics[width=0.65\hsize]{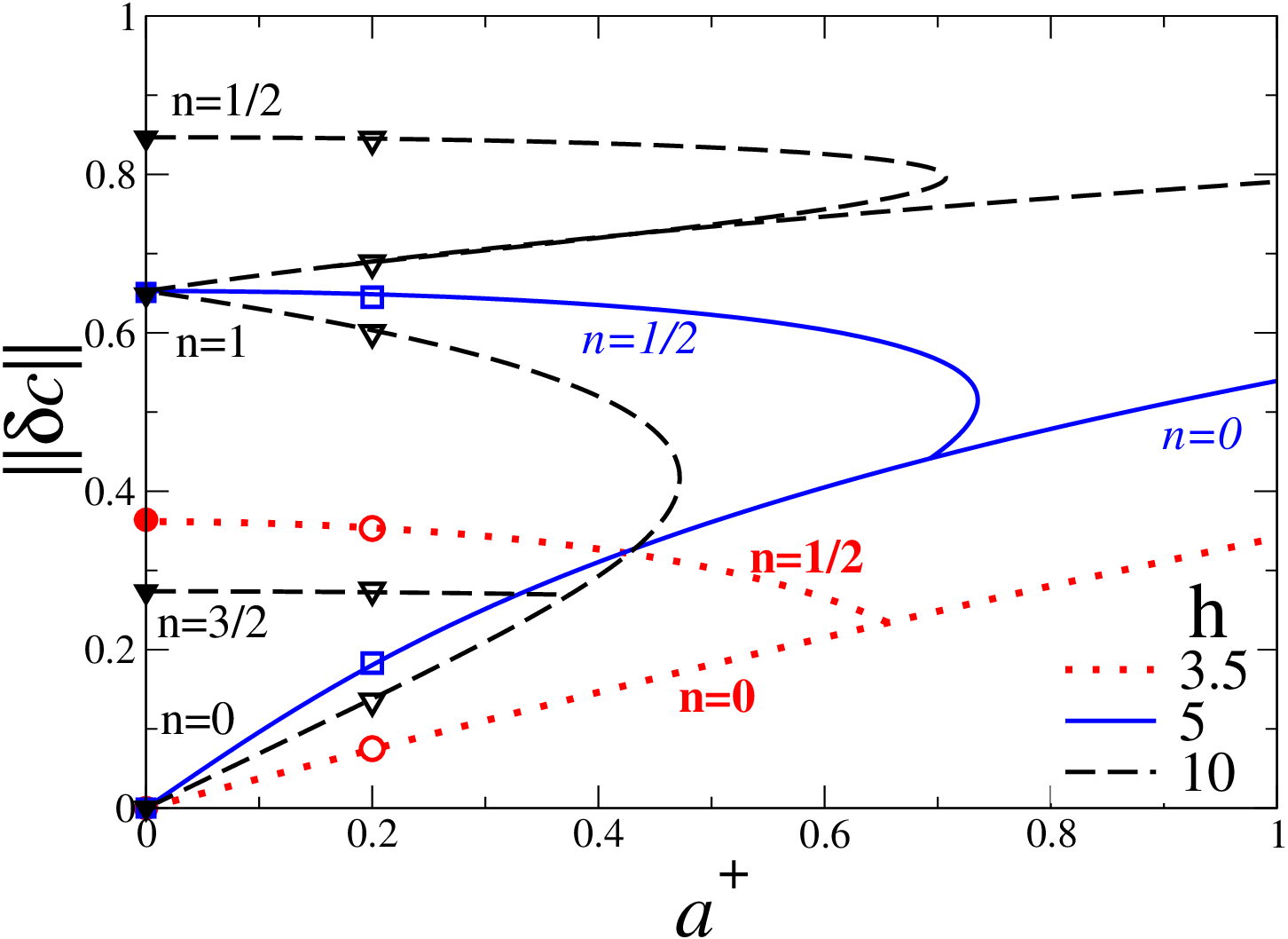}
(b)\includegraphics[width=0.65\hsize]{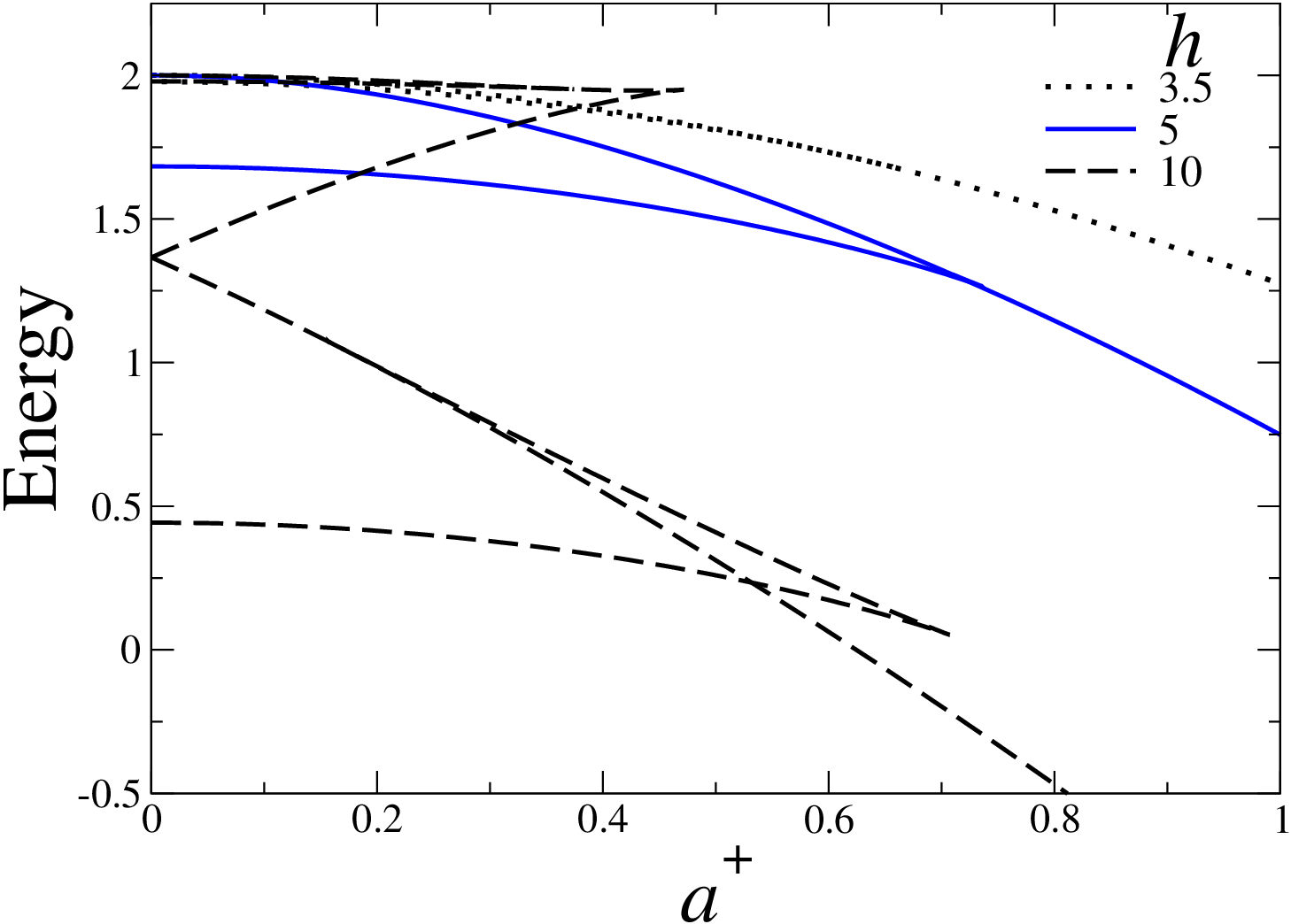}
\caption{Branches of steady vertical concentration profiles 
for linearly symmetrically biased surfaces ($a^+=a^-$, $b^\pm=0$) and
a symmetric mixture ($\bar{c}=0$) in dependence of bias $a$ for film
thicknesses as given in the legend. Shown are (a) the $L_2$-norm and
(b) the energy $E$ of $c_0(z)$. Selected corresponding profiles are
given in Fig.~\ref{ss4prof}. The branch numbers $n$ for the different
film thicknesses are given using different fonts: $h=3.5$ bold, $h=5$
italic, $h=10$ normal. Lines and labels are of corresponding colors
(online).  $S=1$, and the symbols are explained in the main text.}
\mylab{ss3}
\end{figure}

The bifurcation diagram is not very involved. For all film thicknesses
there exists the trivial homogeneous solution with $||\delta c||=E=0$
(broken line in Fig.~\ref{ss2}). Non-trivial branches bifurcate
successively at $h_i=i\,\pi/k_c$ where $i=1,2,3\dots$ and
$k_c=\sqrt{-\partial_{cc}f(c_0)}=\sqrt{1-3c_0^2}$ corresponds to the
critical wavenumber for the linear instability of the homogeneous
solution $c=c_0$ of Eq.~(\ref{ch-1d}). For the symmetric mixture
considered in Fig.~\ref{ss2} one finds $h_i=i\,\pi$. Furthermore, all
characteristics like $E_i[h]$ or $||\delta c||_i[h]$ of all branches
$n=i$ with $i\ge1$ can be mapped onto the characteristics of the $n=1/2$
branch. For example, for the energy one has
$E_i[h]=E_{1/2}[h/(2i)]$. Note, however, that the bifurcations are
degenerate because as discussed above two twin solutions related by symmetry
bifurcate at once.

A thin film in an experiment will tend towards the constellation with
the minimal energy (see Figs.~\ref{ss2}\,b and~\ref{ss2prof}), i.e.\
for $h<\pi$ the homogeneous layer and for $h>\pi$ the stratified layer
with $n=1/2$. The multilayer constellations with $n\ge1$ may, however,
appear as transients as they are saddle fixed points in phase space
that attract time evolutions from a certain basin of attraction and
repel them consecutively into the few unstable directions (for a more
extensive discussion of that concept in connection with dewetting on
heterogeneous substrates see Ref.~\onlinecite{TBBB03}).
%
\subsubsection{Symmetrically biased surfaces}
\mylab{sec-sbs}
%
The presented rather detailed description of the steady states for
energetically neutral surfaces will help us to understand the involved
behaviour for biased surfaces. Allowing for arbitrary linear ($a^-,
a^+$) and quadratic ($b^-, b^+$) energetic biases opens a four
dimensional parameter space additionally to the parameter 'film
thickness'. We give an overview of the system behaviour by focusing on
a linear bias ($b^-=b^+=0$), and by using several special ratios
$a^+/a^-$. In this way we obtain a 2d parameter space spanned by $a^+$
and $h$.

In the present section we assume that the two surfaces energetically
prefer the same component with equal strength ($a^+=a^-$), i.e.\ we
have symmetrically biased surfaces. Figs.~\ref{ss3} and~\ref{ss4} show
characteristics of solution branches in dependence of the bias for
fixed film thickness and in dependence of the film thickness for fixed
bias, respectively.
Corresponding solutions between Figs.~\ref{ss2} and~\ref{ss3} are
marked by filled symbols in the $L2$-norm plots. Hollow symbols
indicate correspondences between Figs.~\ref{ss3} and~\ref{ss4}.
Concentration profiles for $a^+=0.2$ corresponding to the hollow symbols are given
in Fig.~\ref{ss4prof}~(a) whereas panel (b) gives profiles for  a large bias of 
$a^+=0.6$.
\begin{figure}[tbh]
\centering
(a)\includegraphics[width=0.65\hsize]{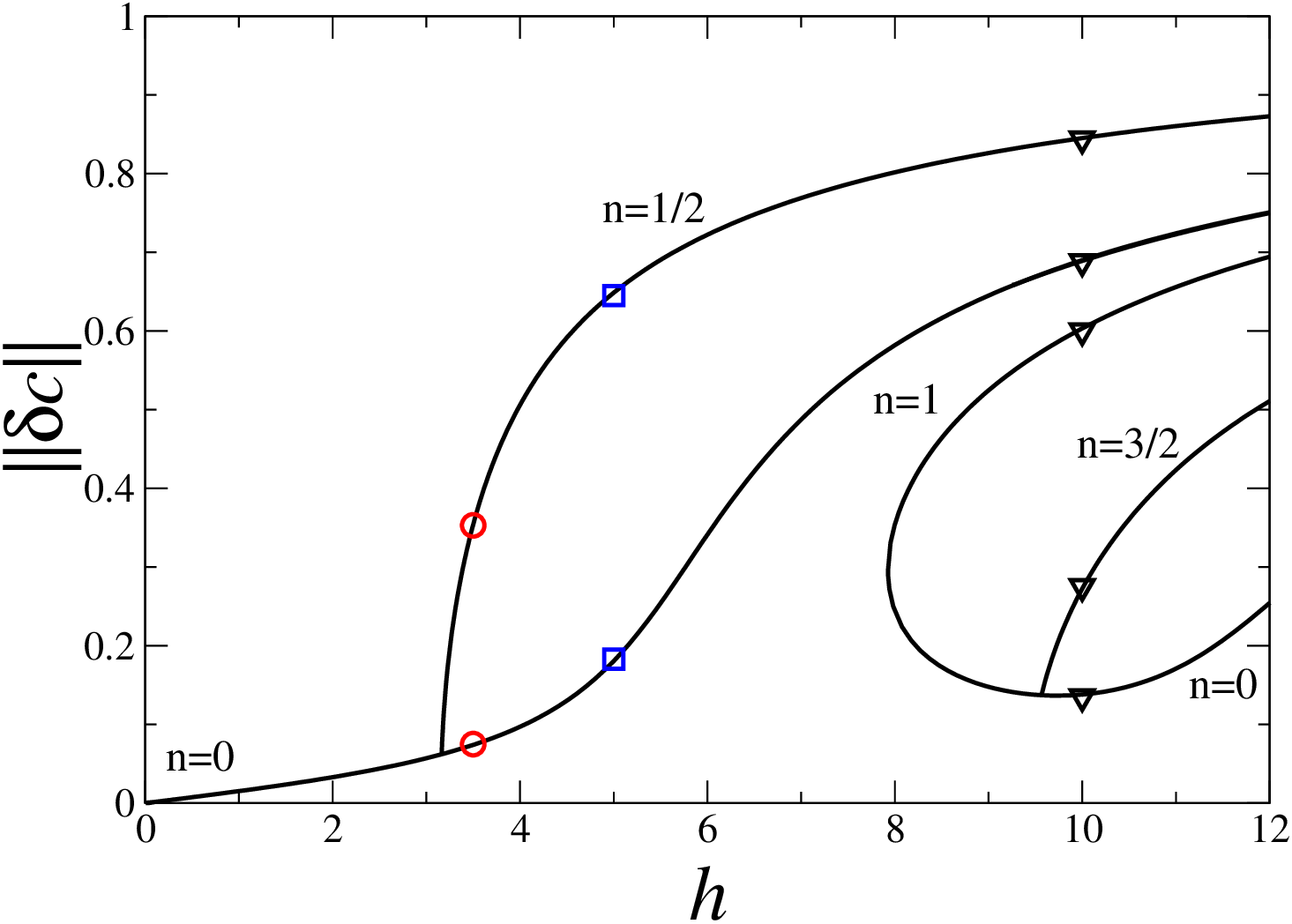}
(b)\includegraphics[width=0.65\hsize]{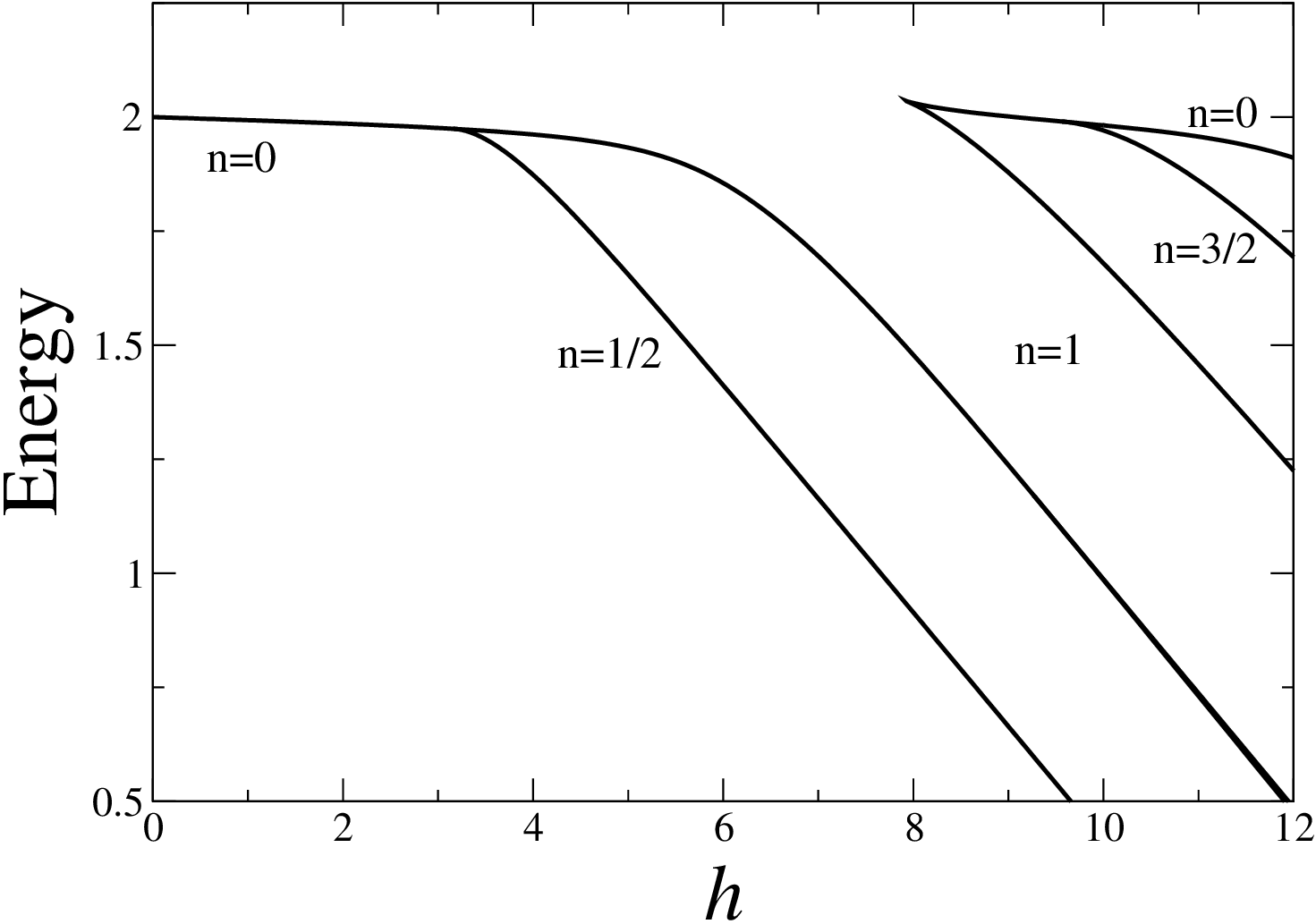}
\caption{Branches of steady vertical concentration 
profiles in dependence of film thicknesses for linearly symmetrically
biased surfaces ($a^+=a^-=0.2$, $b^\pm=0$) and a symmetric mixture
($\bar{c}=0$). Shown are (a) the $L_2$-norm and (b) the energy $E$ of
$c_0(z)$. Selected corresponding profiles are given in
Fig.~\ref{ss4prof}. $S=1$, and the symbols are explained in the main
text.}
\mylab{ss4}
\end{figure}

\begin{figure}[tbh]
\centering
(a)\includegraphics[width=0.65\hsize]{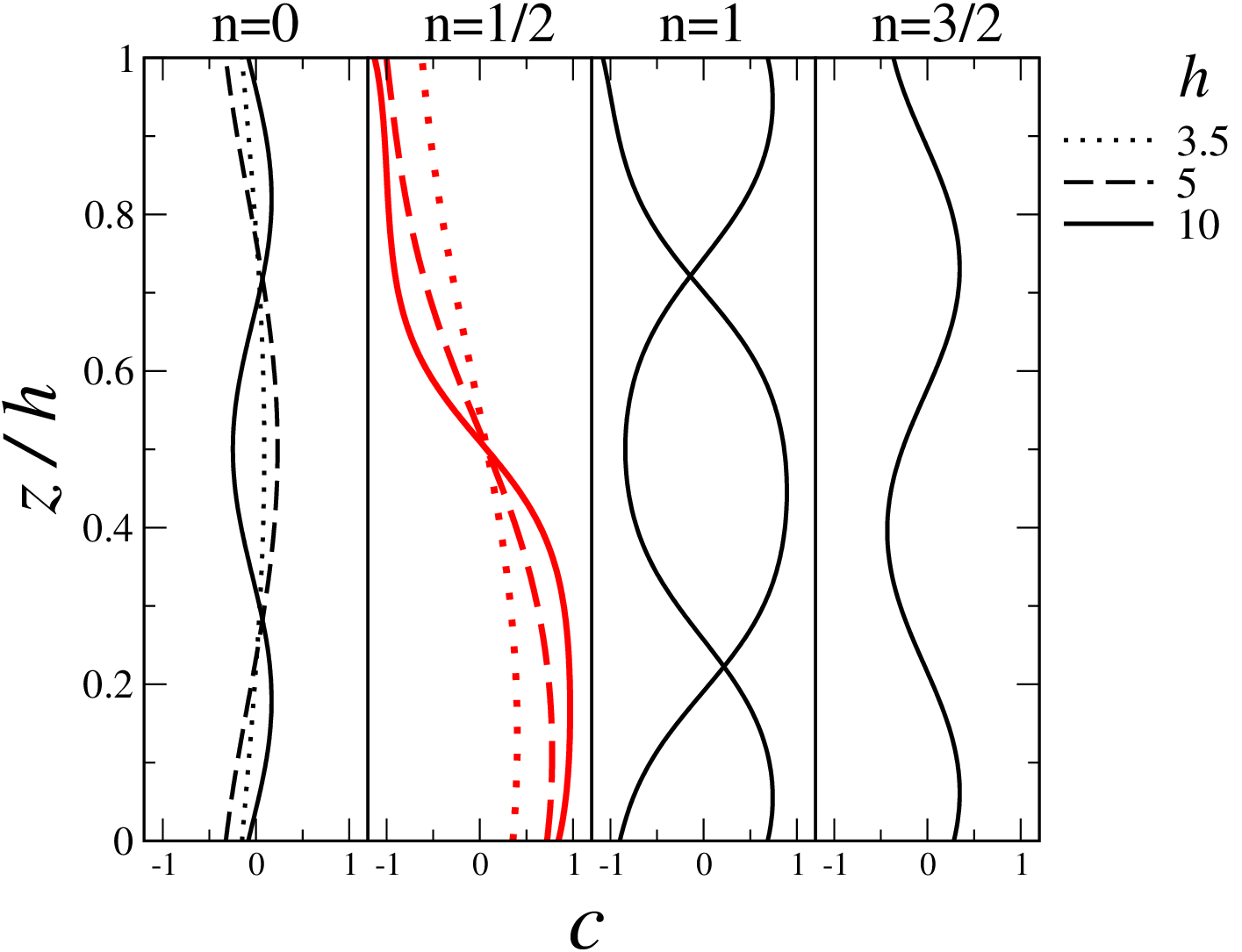}\\
(b)\includegraphics[width=0.65\hsize]{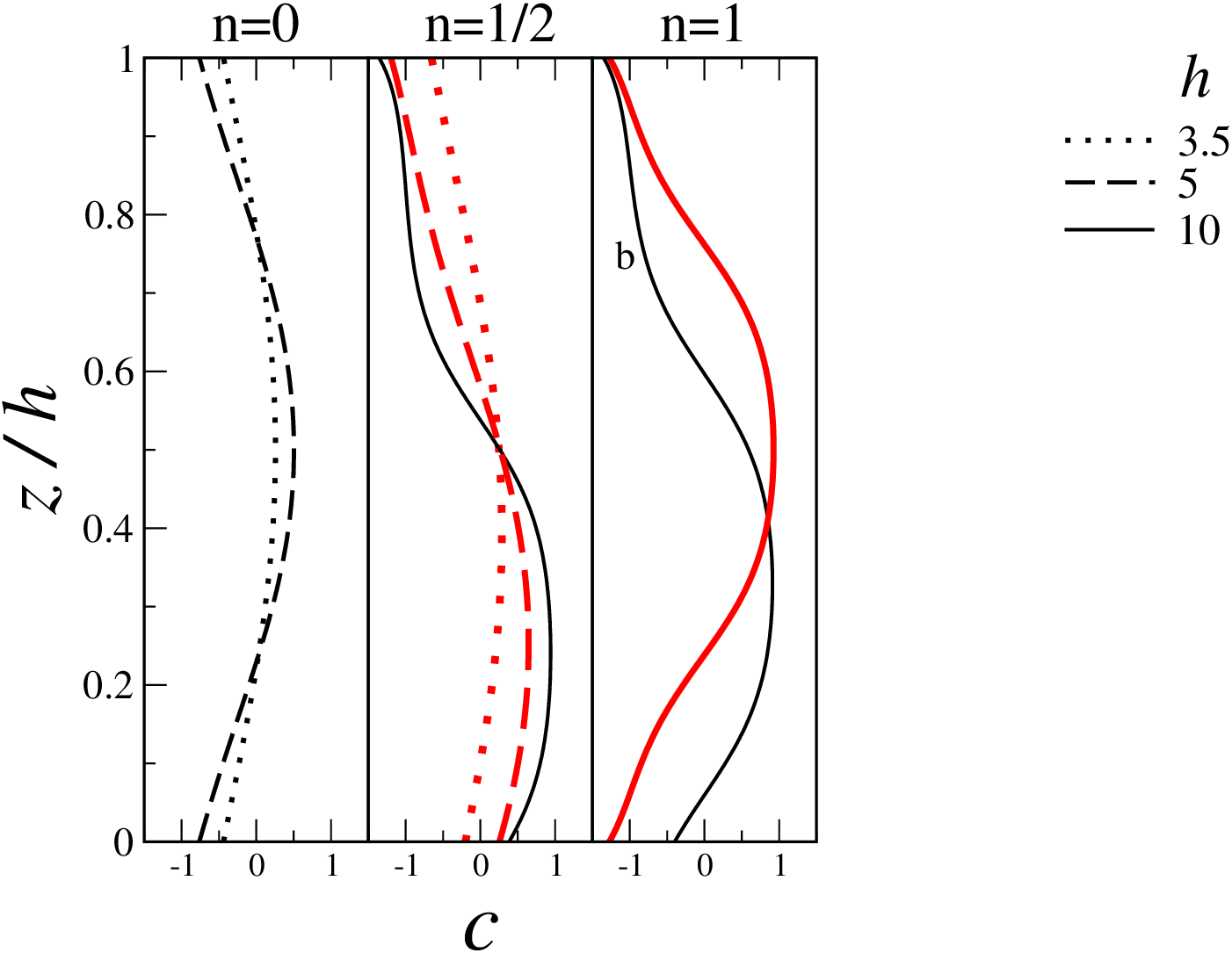}
\caption{Selected concentration profiles corresponding to Figs.~\ref{ss3} 
and~\ref{ss4} for (a) $a^+=0.2$ and (b) $a^+=0.6$ sorted by branch
number as indicated in Fig.~\ref{ss4}.  Film thicknesses are indicated
in the legends. The profiles corresponding to the energy minimum for
the respective film thickness and bias are shown as heavy (red online)
lines. The minuscule 'b' in the $n=1$ panel for  $a^+=0.6$ denotes the 
profile on the side branch of the $n=1$ branch (cp.~Fig.~\ref{ss3}).
}
\mylab{ss4prof}
\end{figure}

First, we focus on Fig.~\ref{ss3}. We introduce branch names
indicating the 'non-biased branch' (Fig.~\ref{ss2}) they are emerging
from. This convention does not correspond to actual properties of the
concentration profile. For instance, the profiles on the $n=0$ branch
in Fig.~\ref{ss3} are not homogeneous any more.  For small film
thicknesses $\pi<h<2\pi$ only two solutions exist at $a^+=0$
corresponding to two branches for increasing $a^+>0$. Thereby the
$n=0$ [$n=1/2$] branch is unstable [stable]. Further increasing the
bias the two branches approach each other. For $h=3.5$ the $n=1/2$
branch terminates in a supercritical bifurcation on the $n=0$
branch. For $h=5$ the stable $n=1/2$ branch first undergoes a
saddle-node bifurcation turning unstable before it finally terminates
in a subcritical bifurcation on the $n=0$ branch.  Beyond the
bifurcation the $n=0$ branch is stable in both cases.

For $i\pi<h<(i+1)\pi$ one finds $i$ solutions at $a^+=0$ exemplified
in Fig.~\ref{ss3} for $h=10$, where 4 solutions exist.  Increasing
$a^+>0$ one finds, however, 5 emerging branches because the degeneracy 
of the $n=1$ solution at $a^+=0$ is lifted by the energetic bias. See
the discussion of symmetries above in Section~\ref{sec-ens}.

The $z$-reflection-symmetric solutions (integer $n$) have at $a^+=0$ a
'twin'-solution obtained by $c(z)\rightarrow-c(z)$ that reacts
differently when imposing a symmetric energetic bias, i.e.\ the bias
lifts the degeneracy and two distinct branches are generated like, for
instance, in Fig.~\ref{ss3} for $h=10$ and $n=1$ (see also profiles in
Fig.~\ref{ss4prof}). Note that the $n=0$ branch is a special case with
out degeneracy at $a^+=0$ (trivial solution at $a^+=0$).  On the
contrary, the 'twin'-solutions of the $z$-reflection-antisymmetric
solutions (non-integer $n$) do not react in a different way to a
symmetric bias, i.e.\ their degeneracy is not lifted.

\begin{figure}[tbh]
\centering
(a)\includegraphics[width=0.65\hsize]{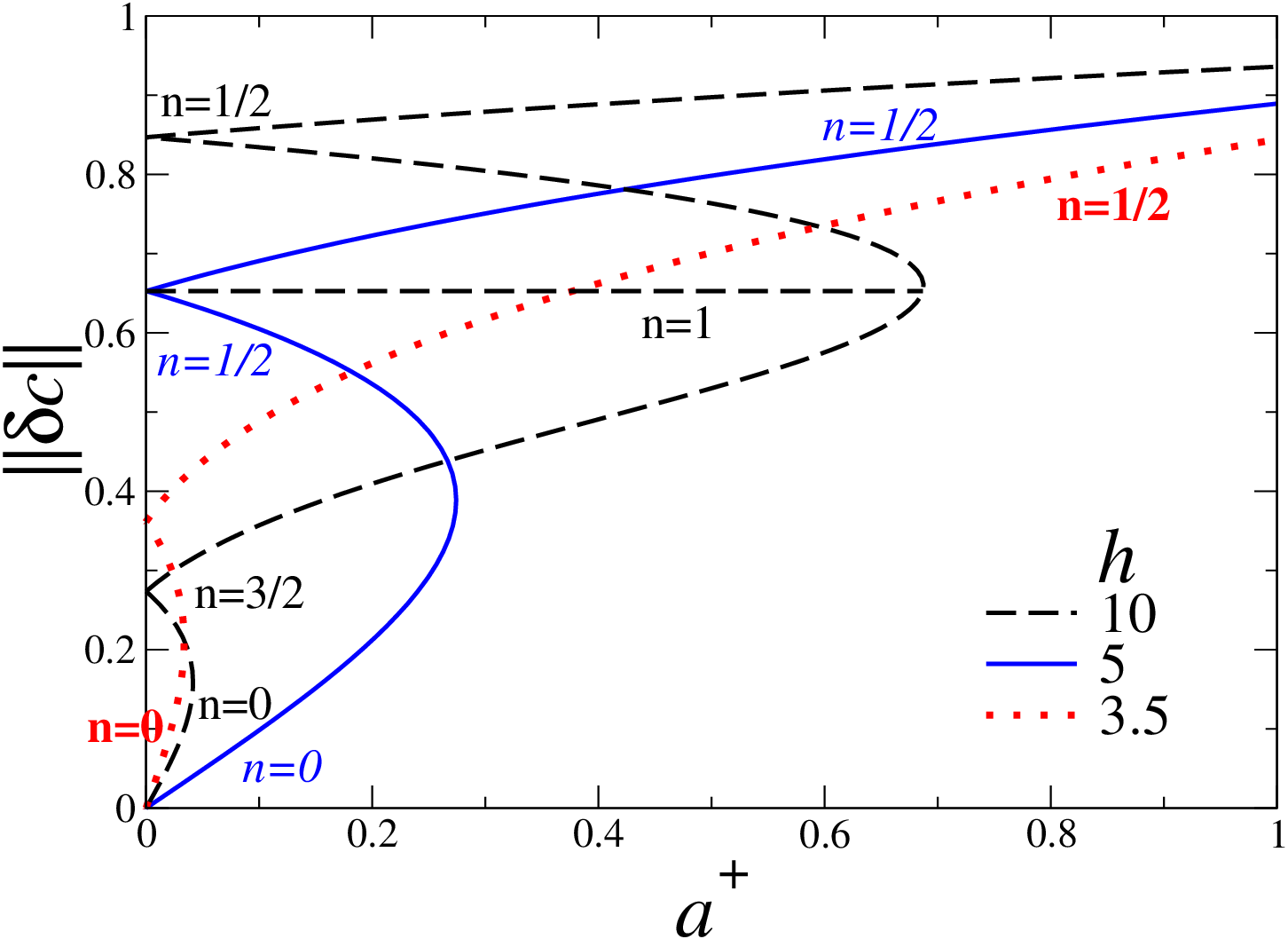}
(b)\includegraphics[width=0.65\hsize]{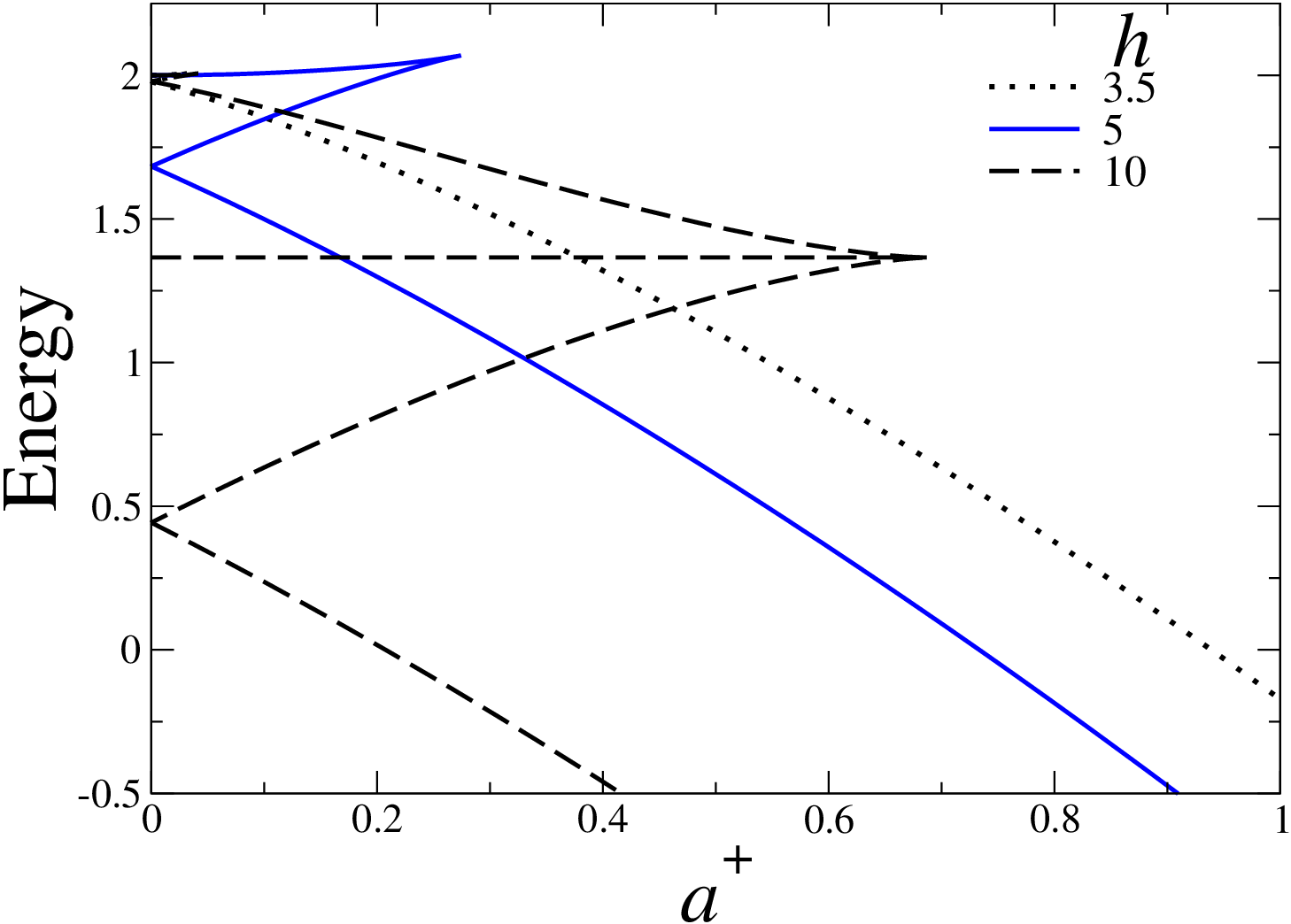}
\caption{Branches of steady vertical concentration profiles 
for linearly antisymmetrically biased surfaces ($a^+=-a^-$, $b^\pm=0$)
and a symmetric mixture ($\bar{c}=0$) in dependence of bias $a$ for
film thicknesses as given in the legend. Shown are (a) the $L_2$-norm
and (b) the energy $E$ of $c_0(z)$. Selected corresponding profiles
are given in Fig.~\ref{ss5prof}. $S=1$. The branch numbers $n$ for the
different film thicknesses are given in different fonts: $h=3.5$ bold,
$h=5$ italic, $h=10$ normal. Lines and labels are of
corresponding colors (online).}
\mylab{ss5}
\end{figure}

Coming back to the case $h=10$ we see that when further increasing
$a^+$ most branches end in bifurcations. One branch finally survives
for large bias corresponding to a layer of 'liquid $+$' enclosed by
two layers of 'liquid $-$' that is preferred by both surfaces for
$a^+>0$. This implies that depending on the strength of bias the
energetic minimum corresponds to qualitatively different
stratifications -- bilayer ($n=1/2$) structure for small $a^+$ and a
sandwich trilayer ($n=1$) structure for large $a^+$ (see heavy (red
online) lines in Fig.~\ref{ss4prof}.

Note finally that the diagram is symmetric w.r.t.\
$a^+\rightarrow-a^+$.  Focusing on the branch of lowest energy that
represents the solutions selected by the system we see that the role
is taken for small [large] $a^+$ by the $n=1/2$ [$n=1$ or $n=0$]
branch. It is intuitively clear that a strong symmetric bias will
suppress the $z$-reflection-antisymmetric solutions.
The alternative view of fixing $a^+$ and changing $h$ is given for
$a^+=0.2$ in Fig.~\ref{ss4} allowing for a better comparison with
Fig.~\ref{ss2}.  From this representation it becomes clear that for
$a^+>0$ the 2 branches emerging from the $n=1$ solution 'break off'
the $n=0$ branch at $h=2\pi$. A similar process occurs at all
$h=2i\pi$ for integer $i$.
%
\subsubsection{Antisymmetrically biased surfaces}
\mylab{sec-antibs}
%
In contrast to the preceding section, here we assume $a^+=-a^-$,
i.e.\ the two surfaces energetically prefer different components. The
preference is, however, equally strong. We focus on $a^+>0$, i.e.\ the
free surface prefers the $c<0$ component. The case $a^+<0$ is related
by symmetry.  Figs.~\ref{ss5} and~\ref{ss6} show solution branches in
dependence of the bias and of film thickness, respectively (in analogy
to Figs.~\ref{ss3} and ~\ref{ss4}). Selected corresponding profiles are
given in Fig.~\ref{ss5prof}.

\begin{figure}[tbh]
\centering
\includegraphics[width=0.65\hsize]{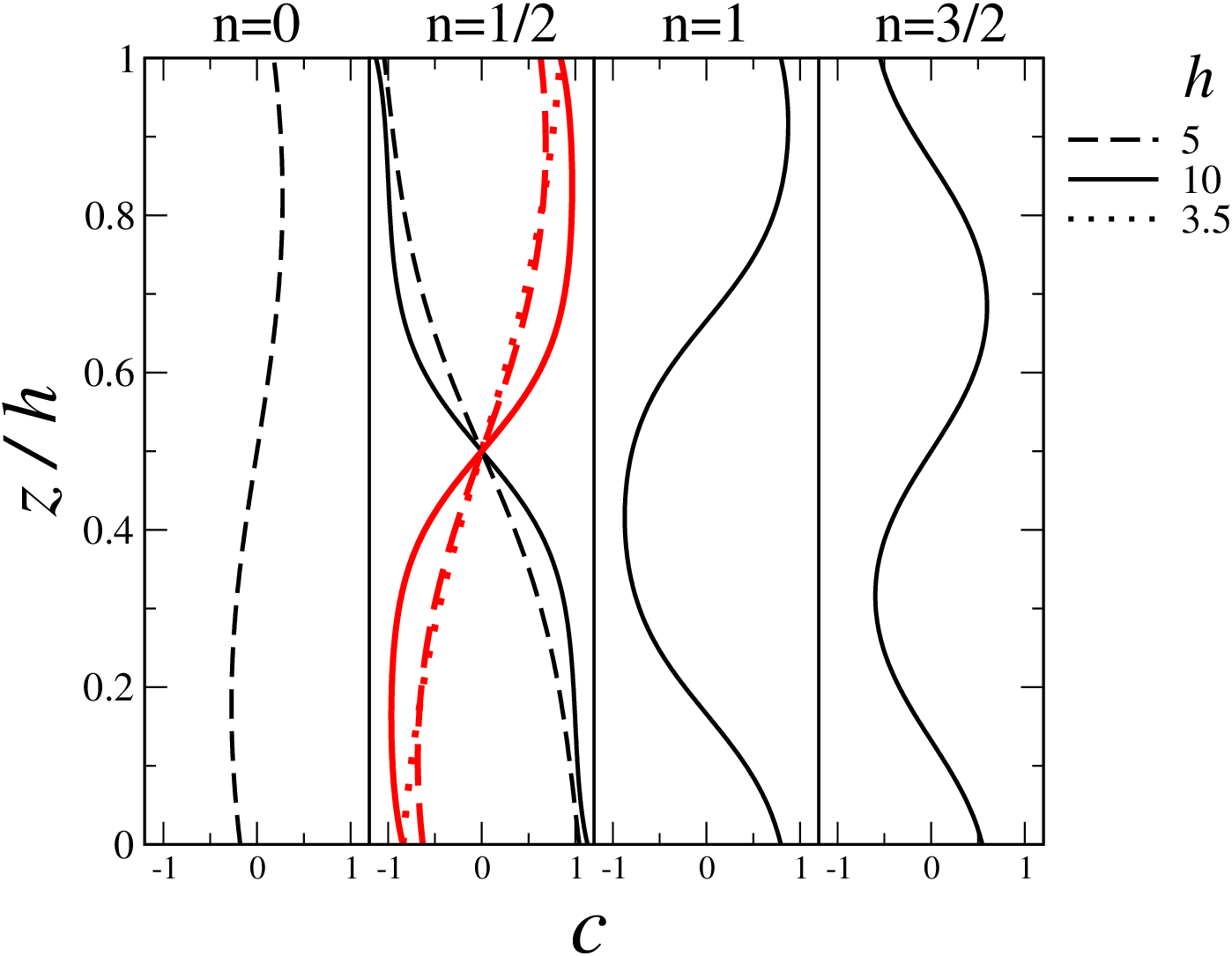}
\caption{Selected concentration profiles corresponding to Fig.~\ref{ss5}
for $a^+=0.2$ sorted by branch number as indicated in Fig.~\ref{ss5}.
Film thicknesses are indicated in the legend. The profiles
corresponding to the energy minimum for the respective film thickness
and bias are shown as heavy (red online) lines.}
\mylab{ss5prof}
\end{figure}

In contrast to the case of symmetrically biased surfaces we find that
for small $a^+>0$ two branches emerge from the $n=1/2$ and $n=3/2$
solutions but only one from the $n=1$ solution. Here, the degeneracy
of the solutions at $a^+=0$ is only lifted for the
$z$-reflection-antisymmetric solutions (non-integer $n$), but not for
the $z$-reflection-symmetric solutions (integer $n$).  In the former
case one of the twin solutions at $a^+=0$ is favored by the
antisymmetric bias whereas the other one is disfavored, i.e.\ they
decrease and increase their energy with $a^+$, respectively
(Fig.~\ref{ss5}\,b). One of the $n=1/2$ solutions is the only one that
'survives' for large bias $a^+$. It is furthermore this solution that
corresponds to the energy minimum for all $a^+$ (see heavy (red
online) lines in Fig.~\ref{ss5prof}).

The alternative view of fixing $a^+$ and changing $h$ is given for
$a^+=0.2$ in Fig.~\ref{ss6}. Contrary to section~\ref{sec-sbs} for
$a^+>0$ the $n=0$ branch is 'broken off' at $h=2n\pi$ by the
respective 2 branches emerging from the non-integer $n$ solutions.

\begin{figure}[tbh]
\centering
(a)\includegraphics[width=0.65\hsize]{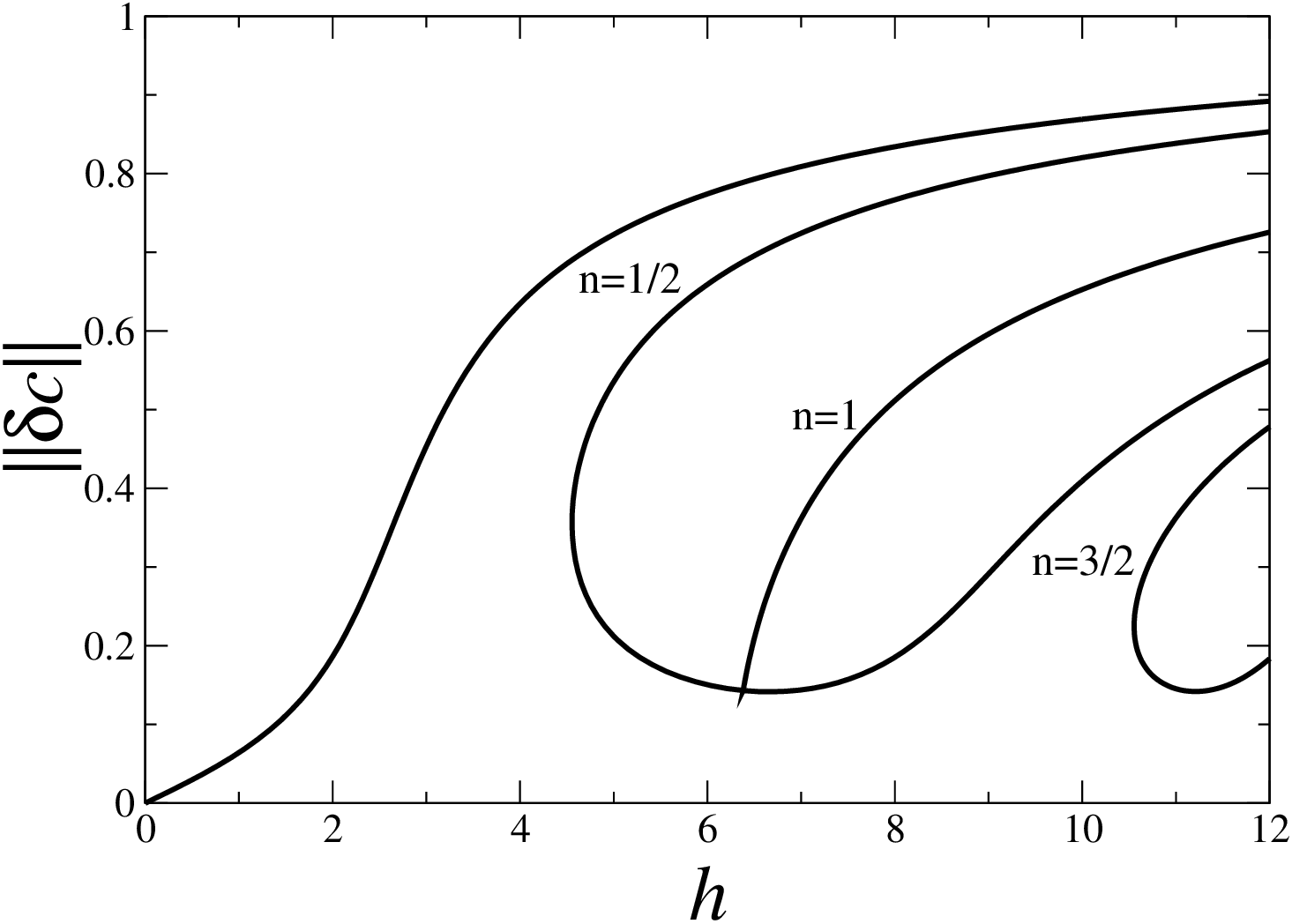}
(b)\includegraphics[width=0.65\hsize]{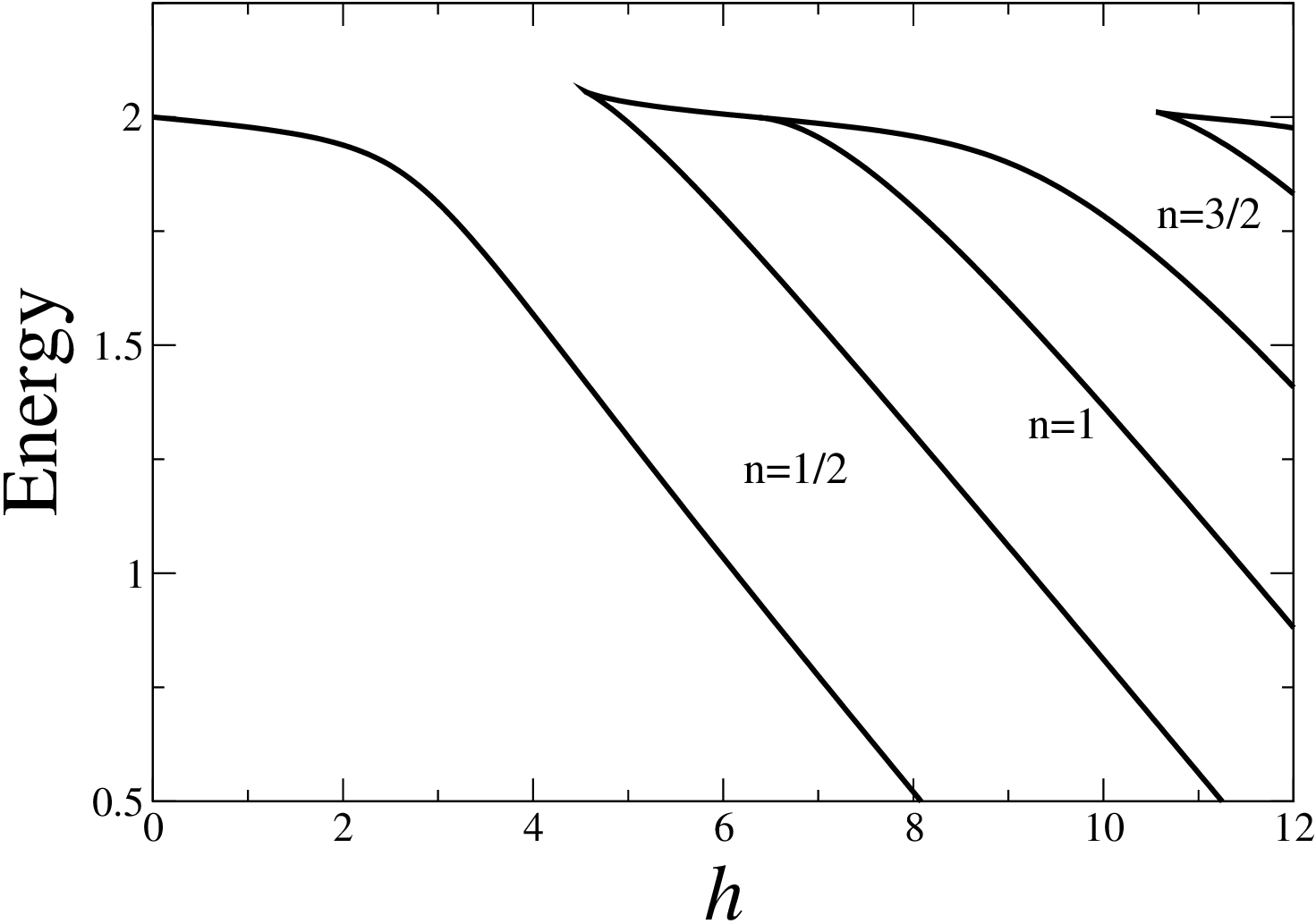}
\caption{Branches of steady vertical concentration profiles 
for linearly antisymmetrically biased surfaces ($a^+=-a^-=0.2$,
$b^\pm=0$) and a symmetric mixture ($\bar{c}=0$) in dependence of film
thickness. Shown are (a) the $L_2$-norm and (b) the energy $E$ of
$c_0(z)$. Selected corresponding profiles are given in
Fig.~\ref{ss5prof}. $S=1$.}
\mylab{ss6}
\end{figure}
%
\subsubsection{Asymmetrically biased surfaces}
\mylab{sec-abs}
%
%
\begin{figure}[tbh]
\centering
(a)\includegraphics[width=0.65\hsize]{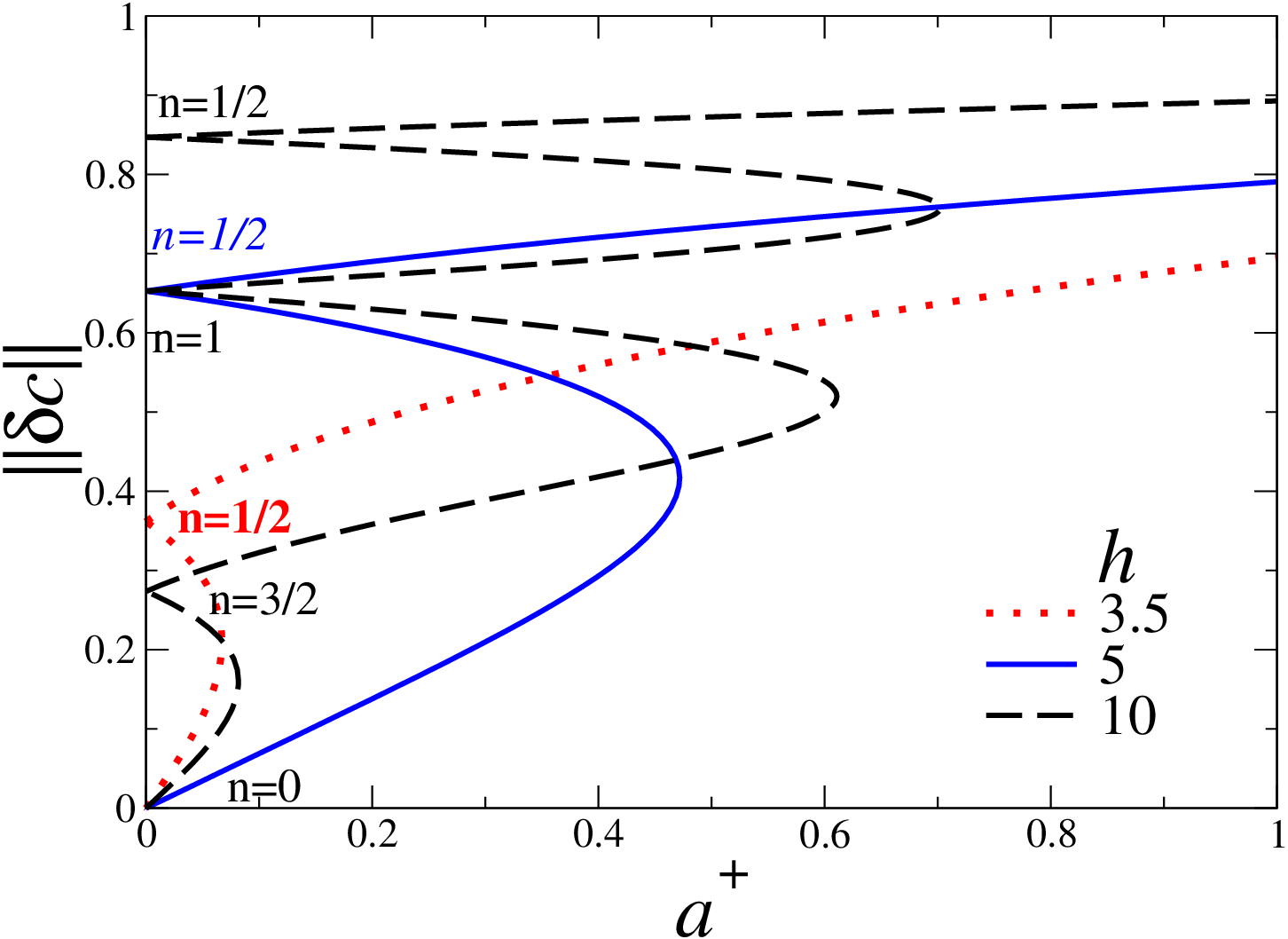}
(b)\includegraphics[width=0.65\hsize]{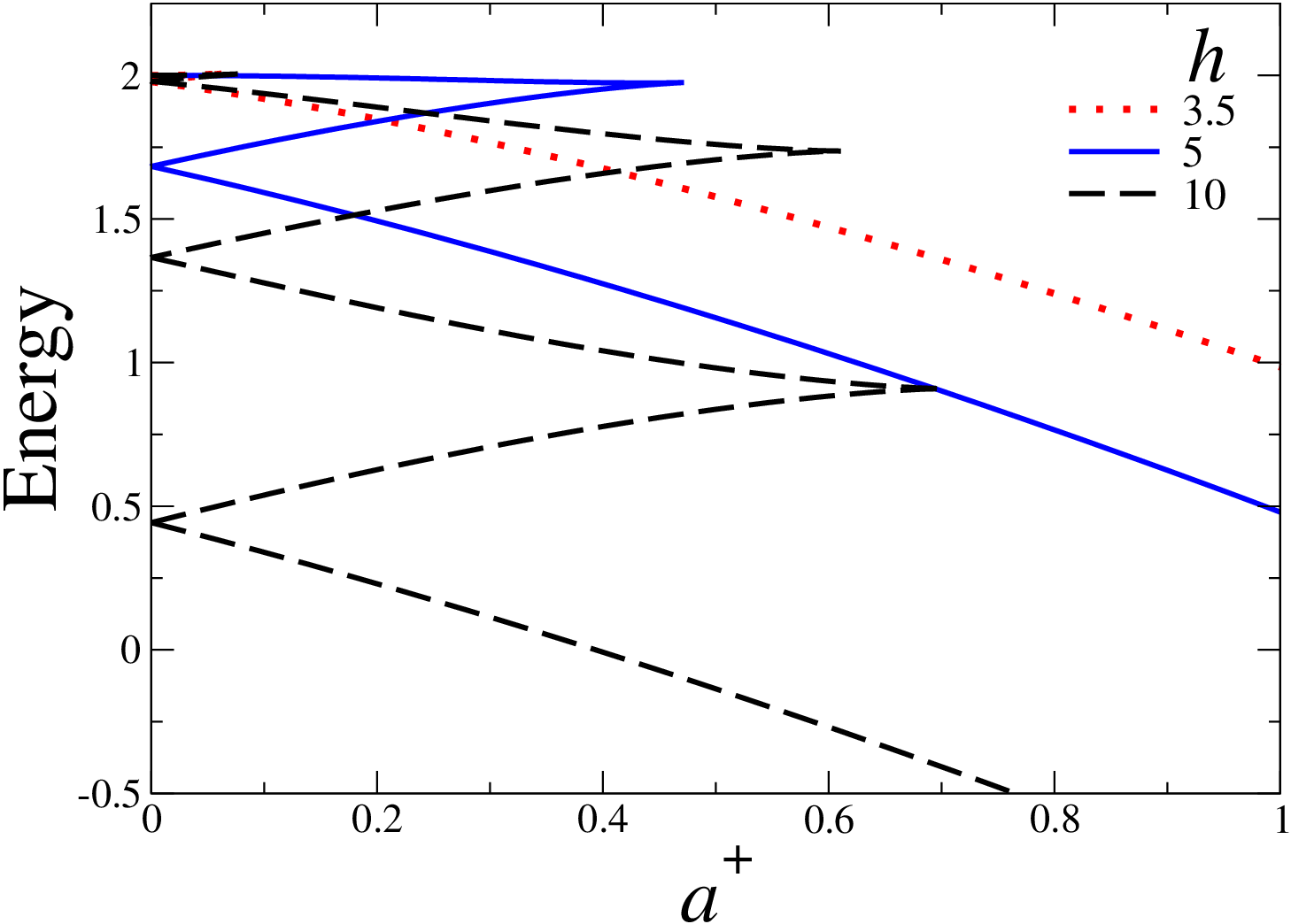}
\caption{Branches of steady vertical concentration profiles 
for linearly asymmetrically biased surfaces ($a^+>0$, $a^-=b^\pm=0$)
and a symmetric mixture ($\bar{c}=0$) in dependence of bias $a^+$ for
film thicknesses as given in the legend. Shown are (a) the $L_2$-norm
and (b) the energy $E$ of $c_0(z)$. $S=1$.
The branch numbers $n$ for the different film thicknesses are given in different
fonts: $h=3.5$ bold, $h=5$ italic, $h=10$ normal. 
Lines and labels are of corresponding colors (online).
}
\mylab{ss7}
\end{figure}
As an intermediate case compared to the two
preceding sections, we focus next on $a^-=0$ and $a^+>0$, i.e.\ the
substrate is energetically neutral whereas the free surface prefers 
the $c<0$ component. Note that the cases  $a^-=0$ and $a^+<0$, 
$a^->0$ and $a^+=0$, $a^-<0$ and $a^+=0$ are related by symmetry.

Figs.~\ref{ss7} and~\ref{ss8} show solution branches in
dependence of the bias for fixed film thickness and in dependence of
the film thickness for fixed bias, respectively.
Here the bias lifts all degeneracies existing for $a^+=0$, i.e.\ 
from each solution at $a^+=0$ emerge two branches (beside the $n=0$
branch). Correspondingly, Fig.~\ref{ss8} shows that for 
$a^+>0$ the 2 branches 
emerging from every integer and  non-integer $n$ solution 
'break off' the $n=0$ branch at all $h=2n\pi$.

\begin{figure}[tbh]
\centering
(a)\includegraphics[width=0.65\hsize]{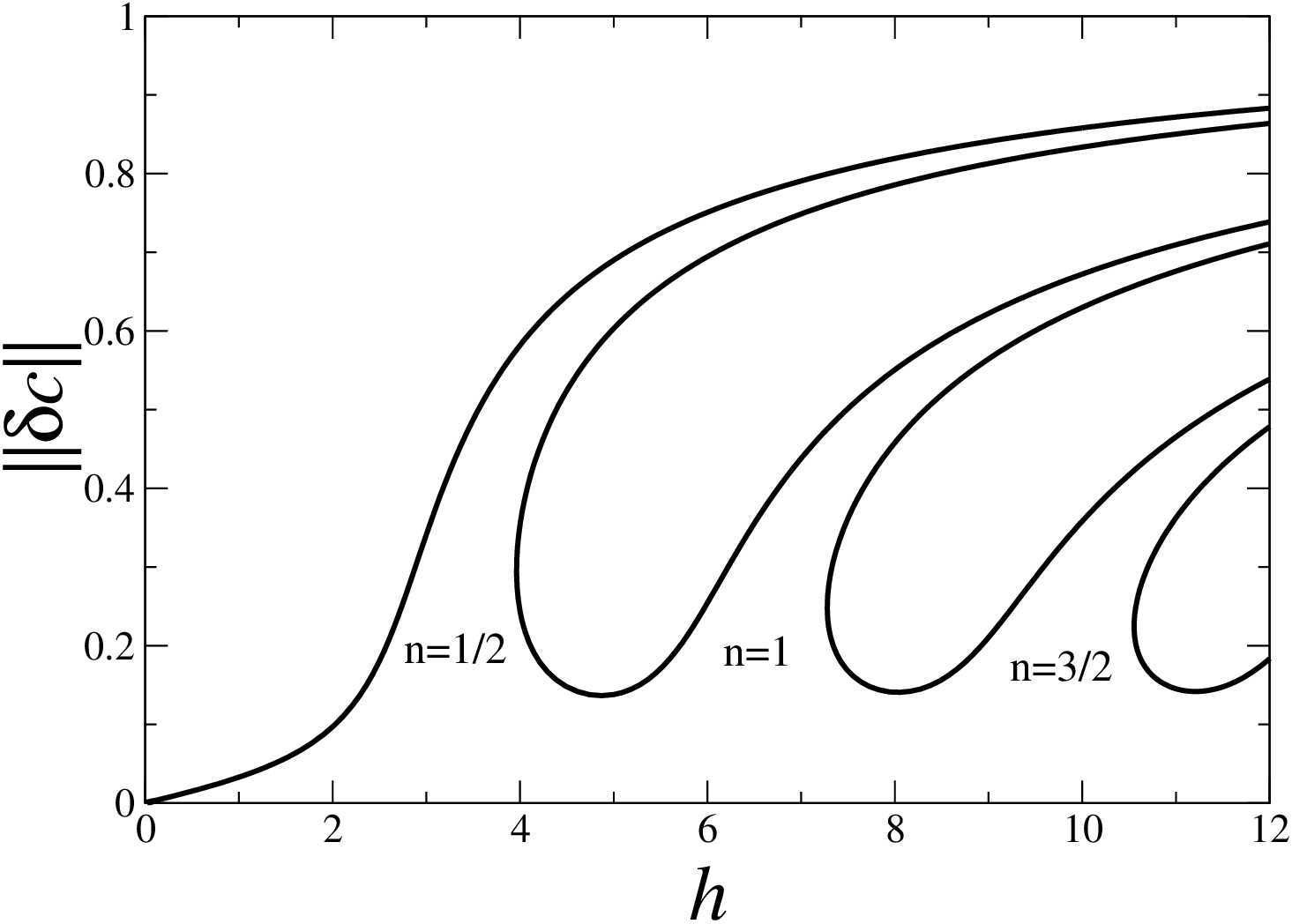}
(b)\includegraphics[width=0.65\hsize]{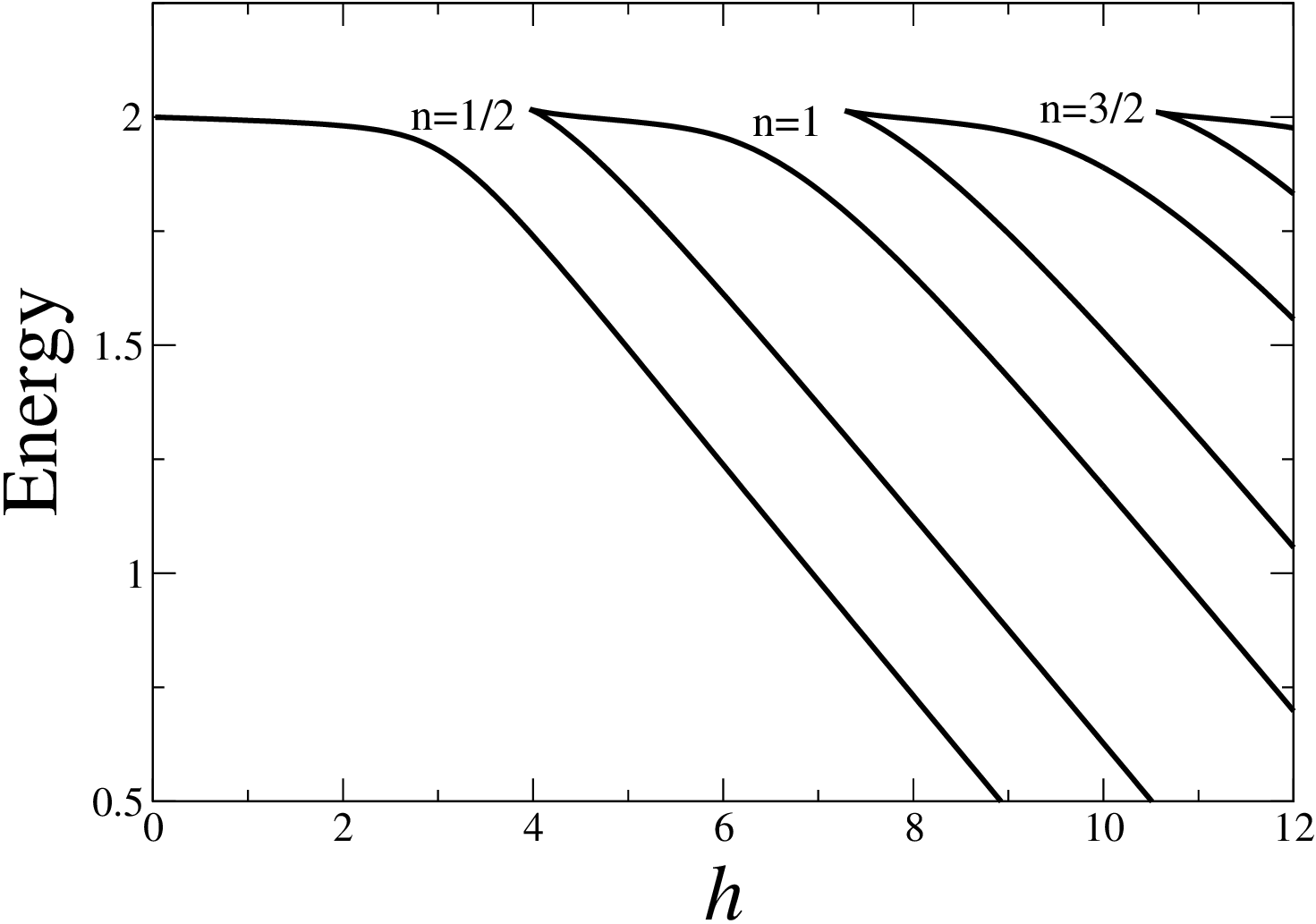}
\caption{Branches of steady vertical concentration profiles 
for linearly asymmetrically biased surfaces ($a^+=0.2$, $a^-=0$, $b^\pm=0$)
and a symmetric mixture ($\bar{c}=0$) in dependence of 
film thicknesses. Shown are (a) the
$L_2$-norm and (b) the energy $E$ of $c_0(z)$. $S=1$.}
\mylab{ss8}
\end{figure}

The branch of lowest energy is here for all $a^+$ and $h$ the $n=1/2$
branch, i.e.\ a simple two layer structure. This is, however, by no
means a general result but depends on the specific asymmetry
chosen. For a strong bias that is only slightly asymmetric (like, for
instance, $a^-=a^++\Delta$ with $\Delta\ll a^+$ the branch of lowest
energy will still be the sandwich structure discussed in
Section~\ref{sec-sbs}.

\subsection{Non-symmetric mixtures}
\mylab{sec-nsc}
%
\begin{figure}[tbh]
\centering
(a)\includegraphics[width=0.65\hsize]{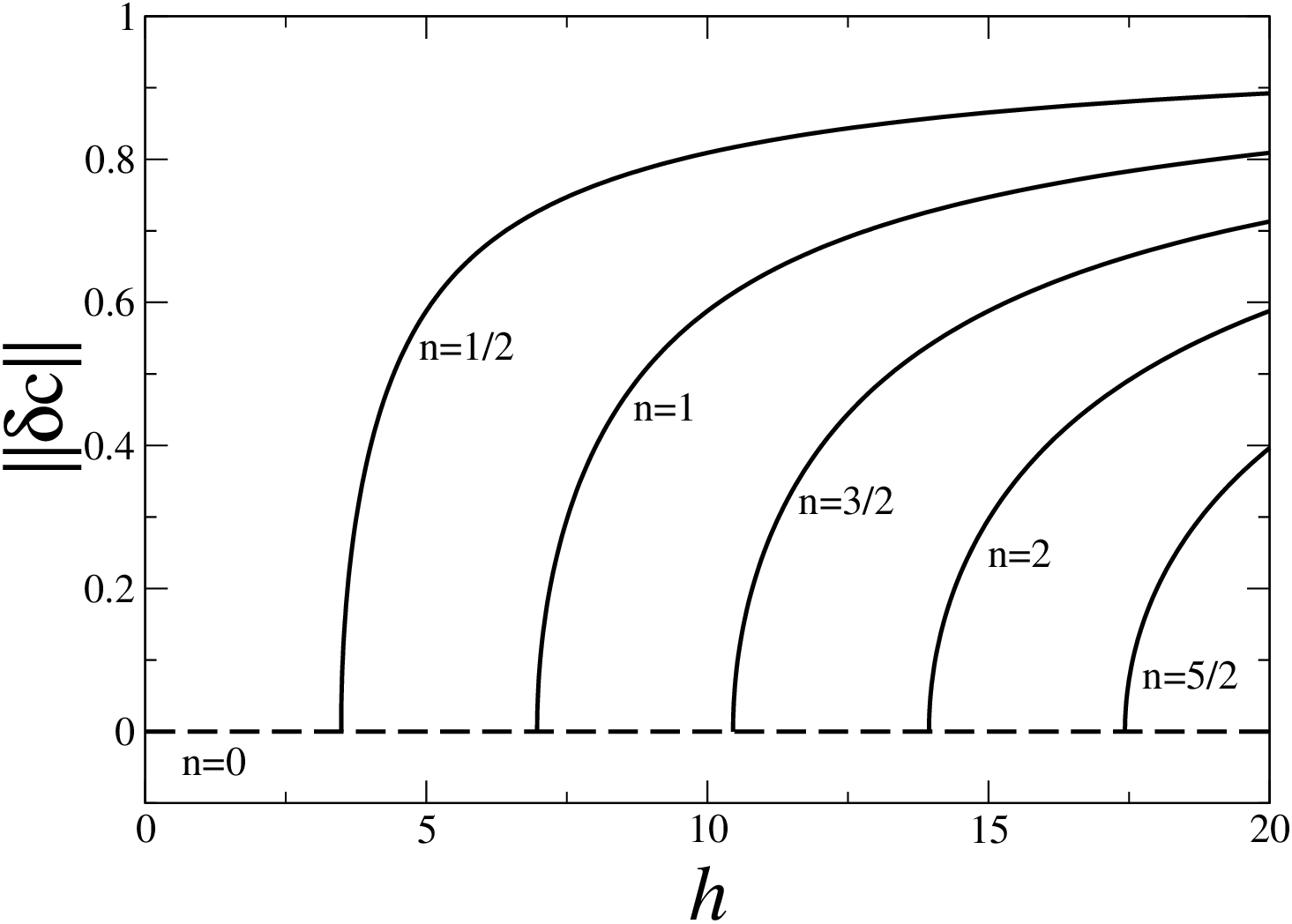}
(b)\includegraphics[width=0.65\hsize]{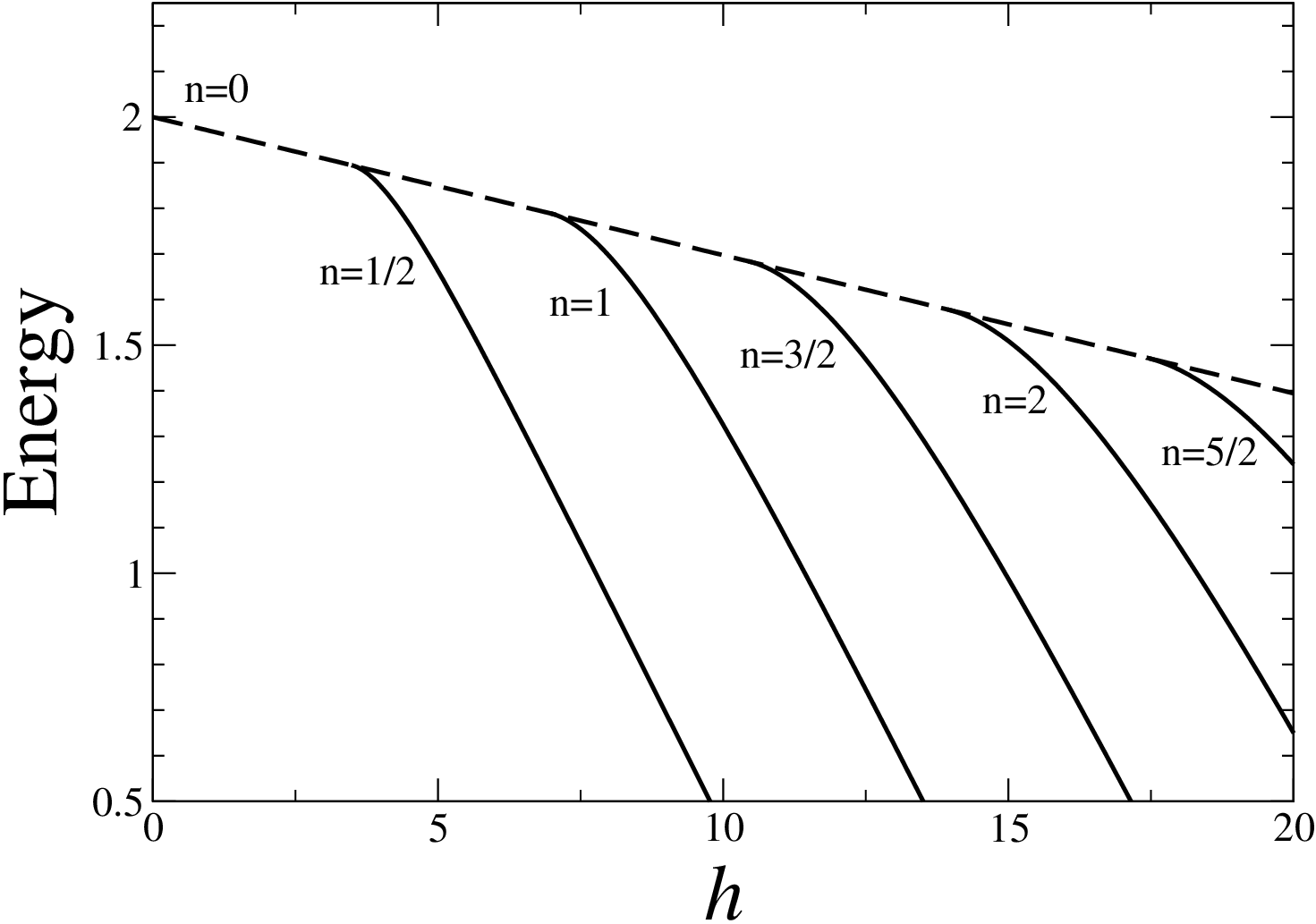}
\caption{Branches of steady vertical concentration profiles 
for non-biased surfaces ($a^\pm=b^\pm=0$) and non-symmetric mixtures
($\bar{c}=0.25$) in dependence of film thickness $h$. Shown are (a)
the $L_2$-norm and (b) the energy $E$ of $c_0(z)$. $S=1$.  }
\mylab{ss9}
\end{figure}

We have seen that the case of a symmetric mixture $\bar{c}=0$ can well
be used to understand the basic solution structure for stratified
films. However, it has to be kept in mind that a symmetric mixture
represents a rather special case. Experimental systems will normally
consist of non-symmetric mixtures with $\bar{c}\neq0$. In the present
section we give selected results for the general case.

For small $\bar{c}\neq0$ the branch structure for non-biased surfaces
is given in Fig.~\ref{ss9} for $\bar{c}=0.25$.  It is qualitatively
equivalent to the one for a symmetric mixture
(cp.~Fig.~\ref{ss2}). Note, however, that in our normalization the
energy for the trivial homogeneous film now depends linearly on film
thickness because the bulk energy density for $\bar{c}$ is not zero
any more. For larger $\bar{c}$ the primary bifurcations become
subcritical.  Eventually the trivial solution becomes linearly
stable. It is, however, metastable, because finite perturbations may
trigger a nonlinear instability. For those $\bar{c}$ no primary
bifurcations exist. All branches of stratified solutions continue
towards infinite thickness.  The bifurcation diagrams for biased
surfaces become quite involved for the subcritical and metastable case
and will be discussed elsewhere.

\begin{figure}[tbh]
\centering
(a)\includegraphics[width=0.65\hsize]{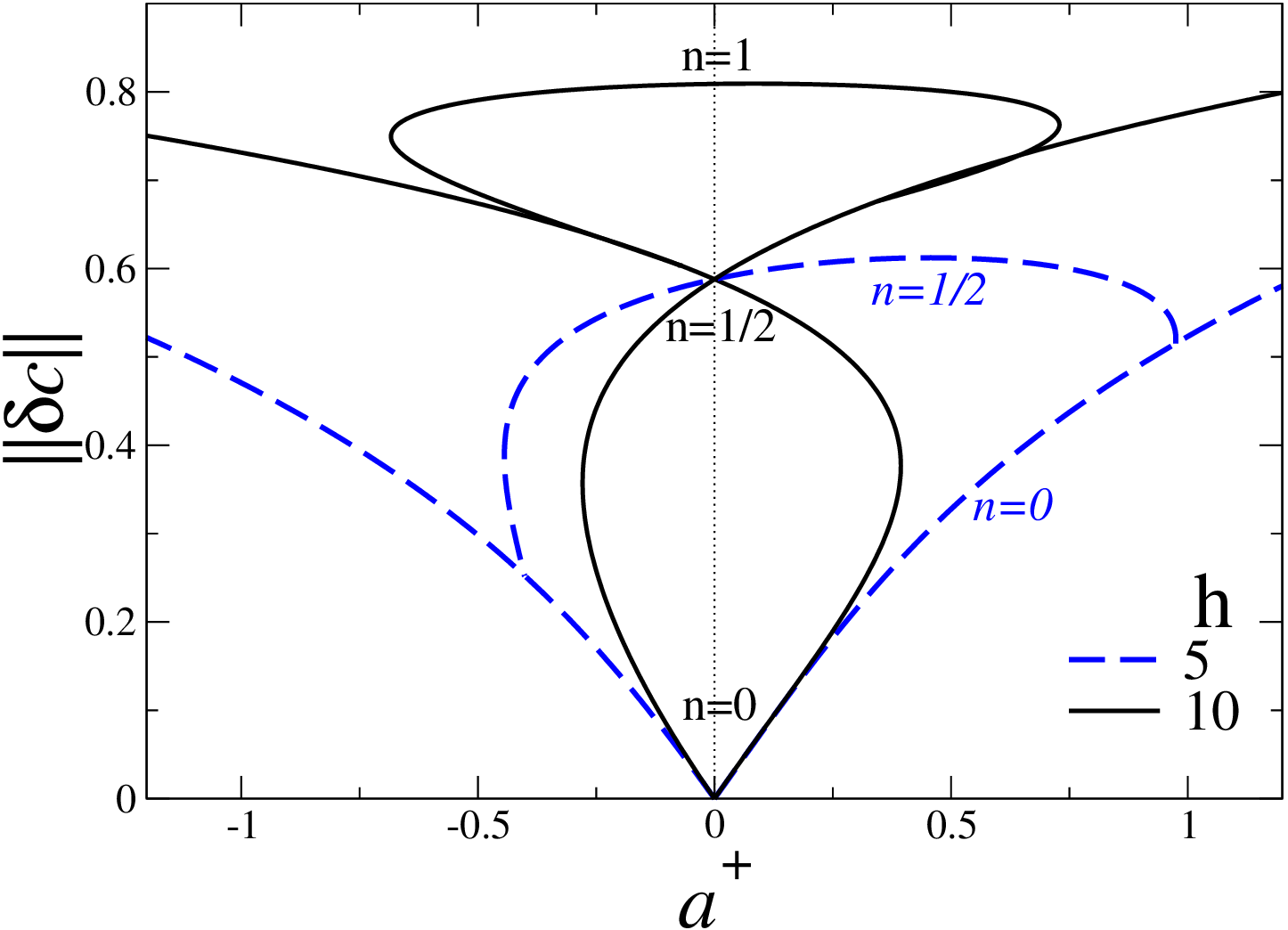}
(b)\includegraphics[width=0.65\hsize]{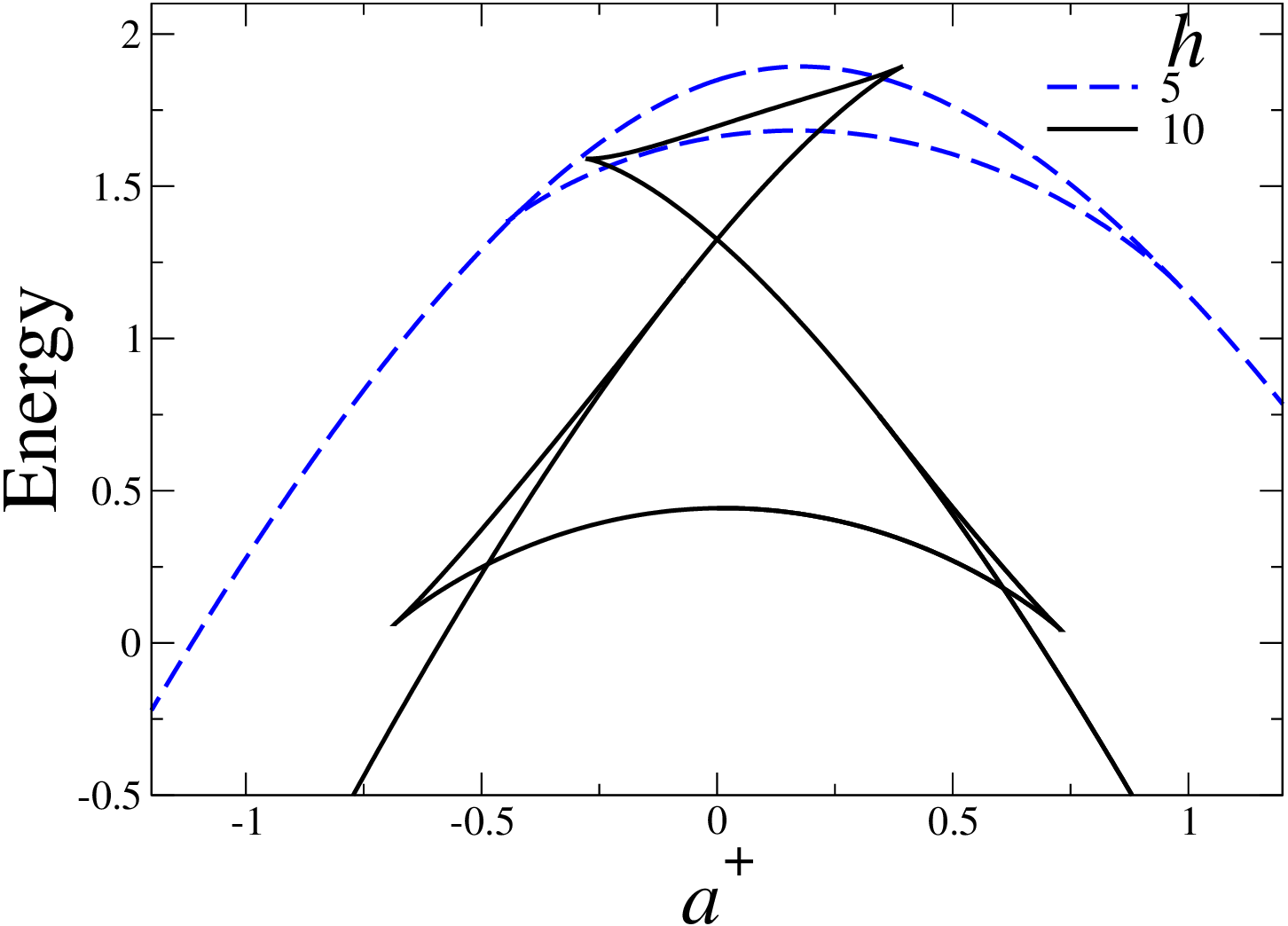}
\caption{Branches of steady vertical concentration profiles 
for linearly symmetrically biased surfaces ($a^+=a^-$, $b^\pm=0$) and
a non-symmetric mixture ($\bar{c}=0.25$) in dependence of bias $a^+$
for film thicknesses as given in the legend. Shown are (a) the
$L_2$-norm and (b) the energy $E$ of $c_0(z)$. $S=1$. The branch
numbers $n$ for the different film thicknesses are given in different
fonts: $h=5$ italic, $h=10$ normal. Lines and labels are of
corresponding colors (online).  }
\mylab{ss10}
\end{figure}

Focusing on the case of supercritical primary bifurcations shown in
Fig.~\ref{ss9} next we discuss the influence of symmetrically biased
surfaces presented in Fig.~\ref{ss10}. The general form of the
bifurcation diagram for $a^+>0$ is qualitatively very similar to the
case of a symmetric mixture (Fig.~\ref{ss3}). However, the symmetry
w.r.t.\ $a^+\rightarrow -a^+$ does not hold anymore. It is replaced by
a symmetry w.r.t.\ $(\bar{c}, a^+)\rightarrow (-\bar{c}, -a^+)$.

\begin{figure}[tbh]
\centering
(a)\includegraphics[width=0.65\hsize]{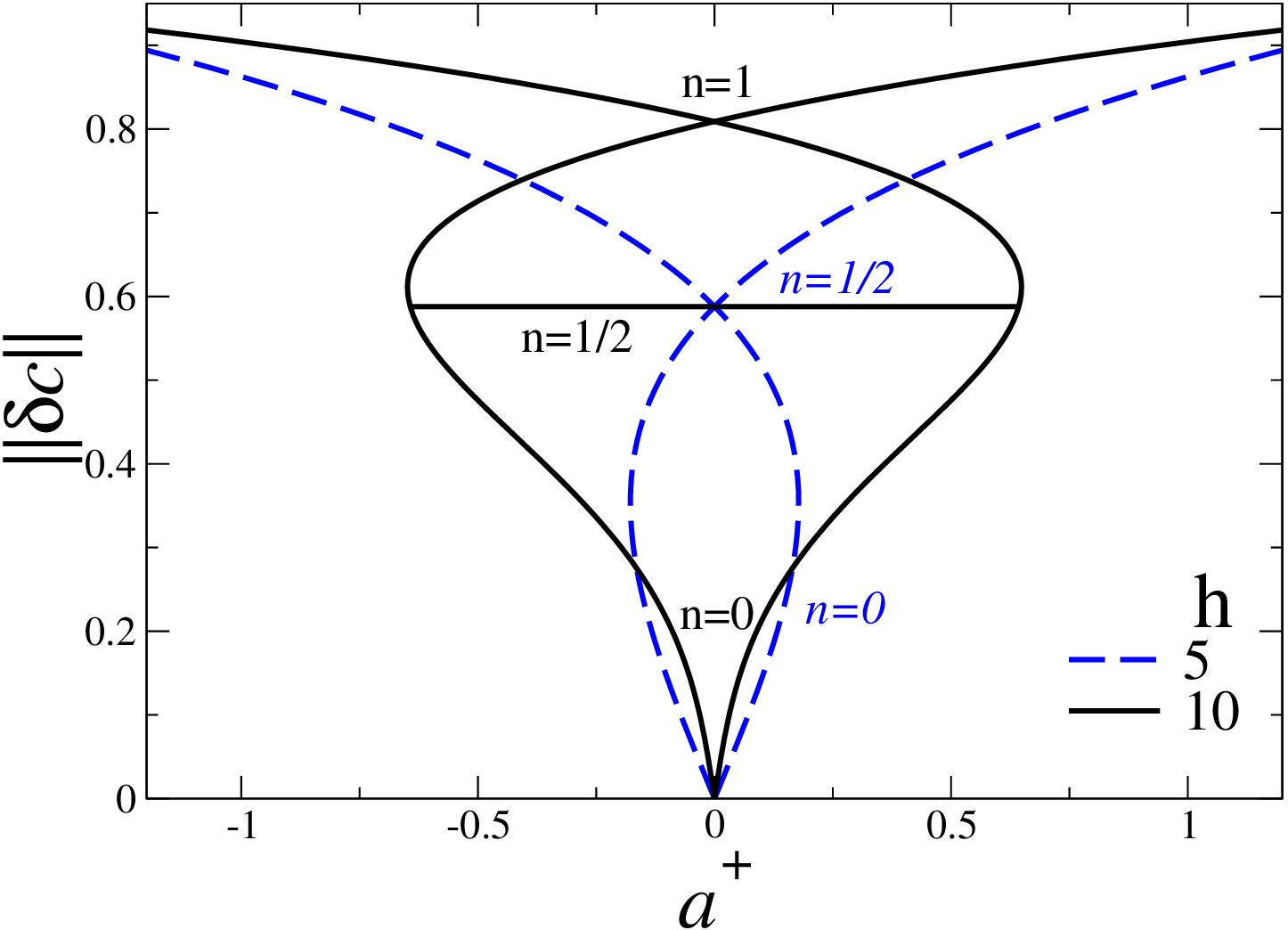}
(b)\includegraphics[width=0.65\hsize]{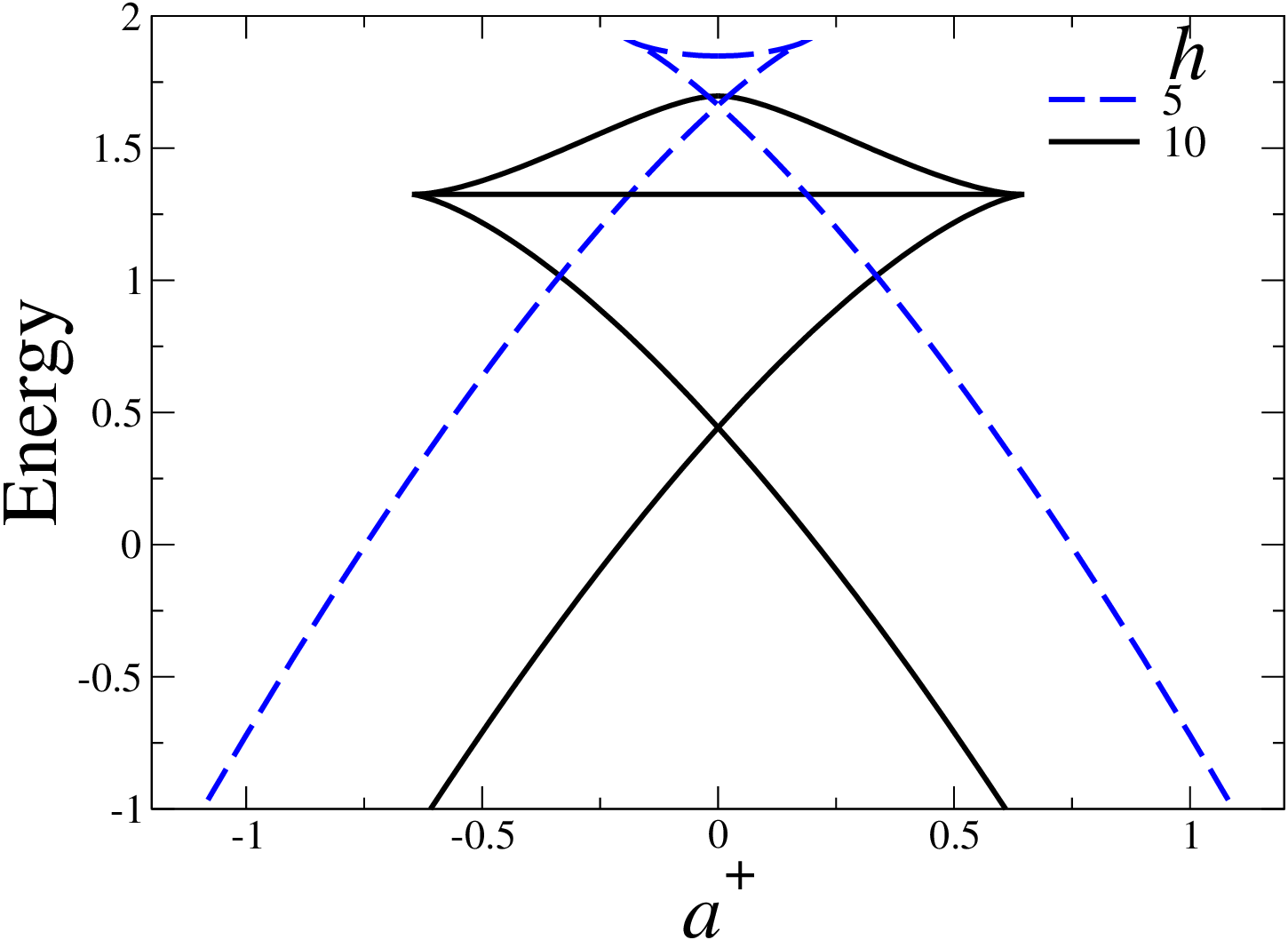}
\caption{Branches of steady vertical concentration profiles 
for linearly antisymmetrically biased surfaces ($a^+=-a^-$, $b^\pm=0$)
and a non-symmetric mixture ($\bar{c}=0.25$) in dependence of bias
$a^+$ for film thicknesses as given in the legend. Shown are (a) the
$L_2$-norm and (b) the energy $E$ of $c_0(z)$. $S=1$. The branch
numbers $n$ for the different film thicknesses are given in different
fonts: $h=5$ italic, $h=10$ normal. Lines and labels are of
corresponding colors (online).
}
\mylab{ss11}
\end{figure}
\begin{figure}[tbh]
\centering
(a)\includegraphics[width=0.65\hsize]{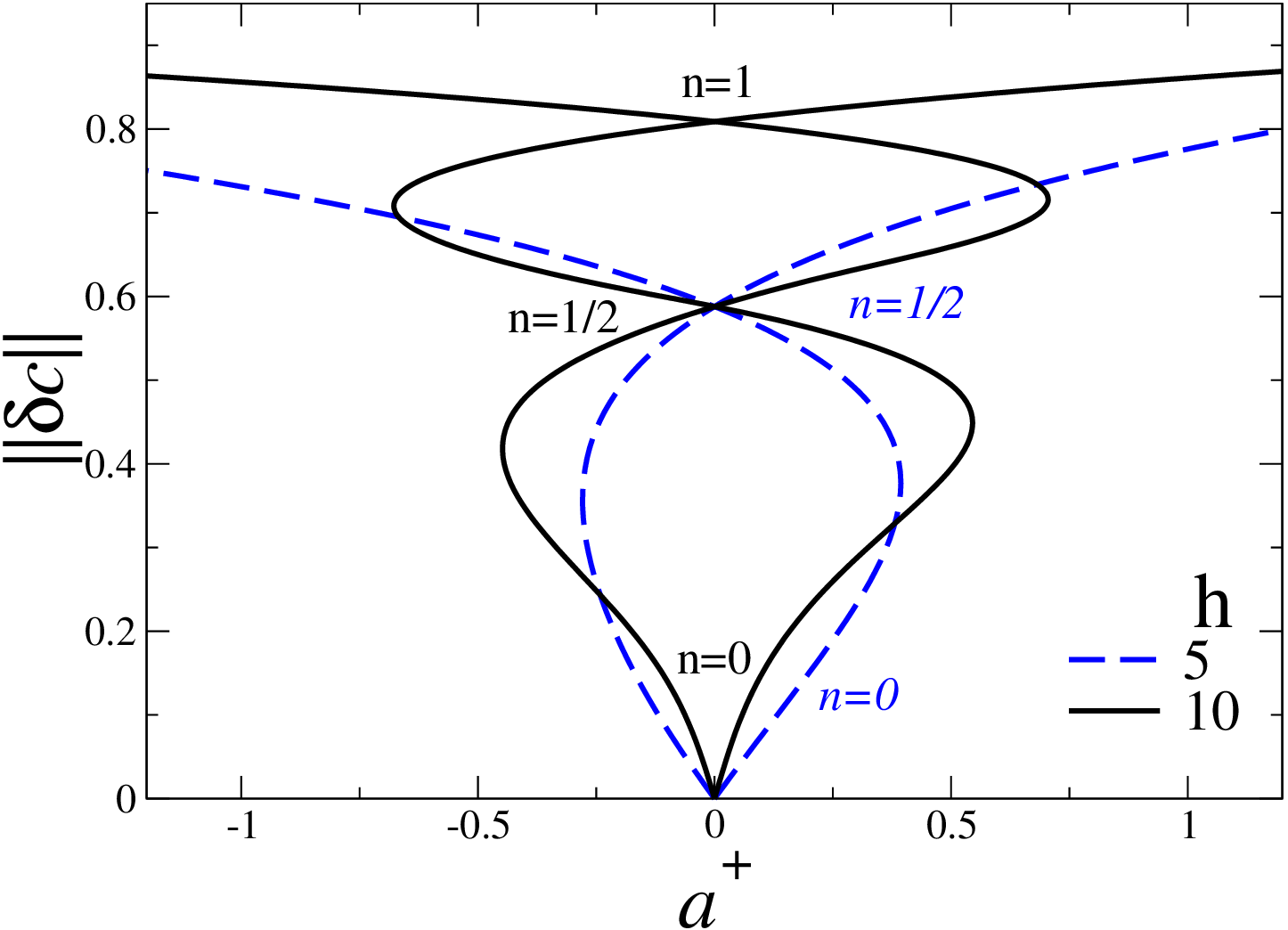}
(b)\includegraphics[width=0.65\hsize]{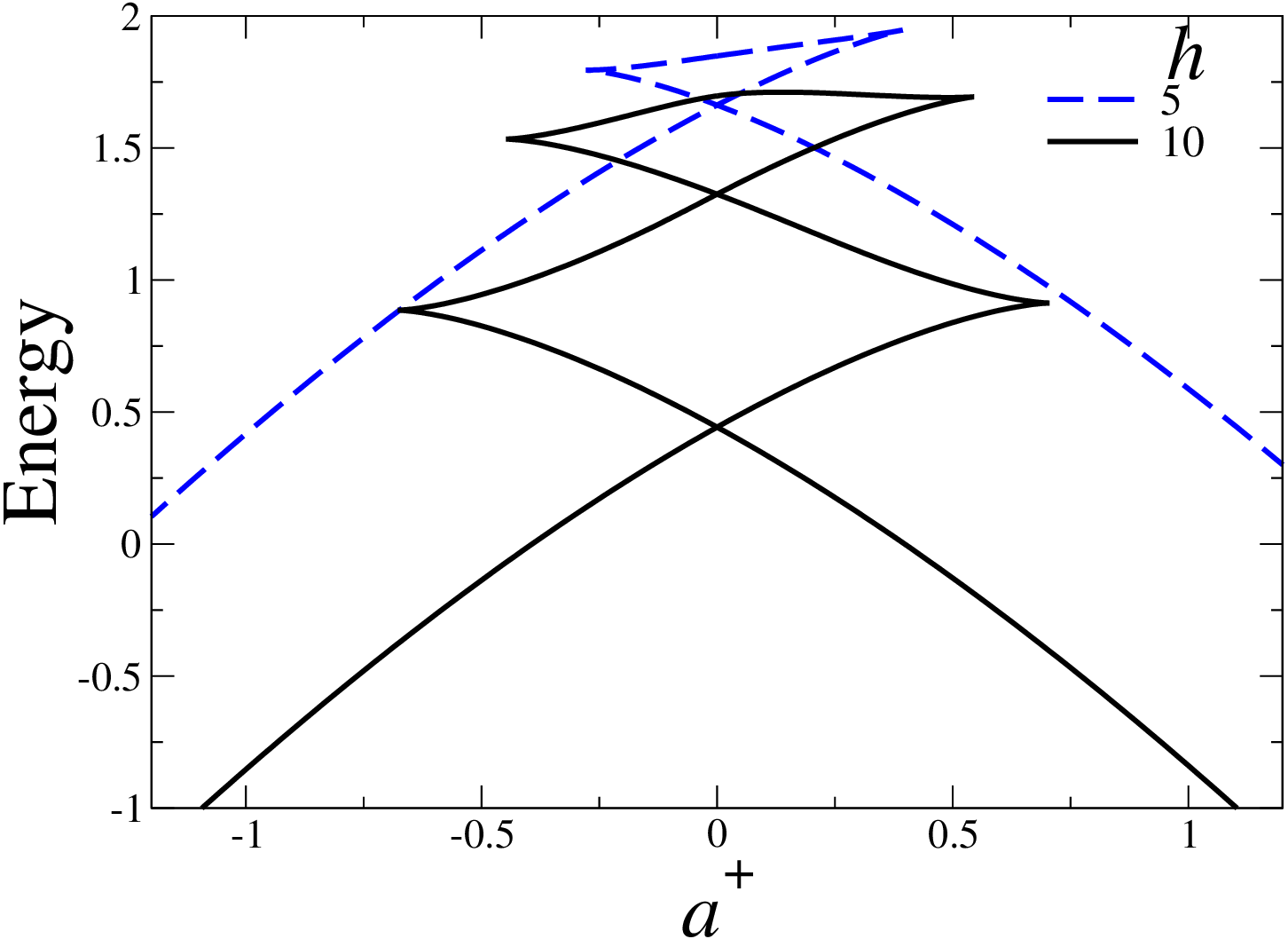}
\caption{Branches of steady vertical concentration profiles 
for linearly asymmetrically biased surfaces ($a^+\neq0$, $a^-=0$, $b^\pm=0$)
and a non-symmetric mixture ($\bar{c}=0.25$) in dependence of 
bias $a^+$ for film thicknesses as given in the legend. 
Shown are (a) the $L_2$-norm and (b) the energy $E$ of $c_0(z)$. 
$S=1$. The branch
numbers $n$ for the different film thicknesses are given in different
fonts: $h=5$ italic, $h=10$ normal. Lines and labels are of
corresponding colors (online).
}
\mylab{ss12}
\end{figure}

Also for antisymmetrically biased surfaces one finds similar
bifurcation diagrams for non-symmetric (Fig.~\ref{ss11}) and symmetric
(Fig.~\ref{ss5}) mixtures. The antisymmetry of the boundary conditions
implies that the symmetry w.r.t.\ $a^+\rightarrow -a^+$ also holds for
the non-symmetric mixture.
Finally, in the asymmetrically biased case shown in Fig.~\ref{ss12}
all degeneracies at $a^+=0$ are broken as in the case of a symmetric
mixture (Fig.~\ref{ss7}). Furthermore, now also the symmetry w.r.t.\
$a^+\rightarrow -a^+$ is broken.

%
\section{Conclusion} \mylab{sec-conc}
%
We have proposed a dynamical model that describes the coupled
decomposition and profile evolution of a free surface film of a binary
mixture, a process frequently encountered in coating and structuring
processes. An example is a thin film of a polymer blend on a solid
substrate undergoing simultaneous phase separation and dewetting. We
have based our approach on model-H coupling transport of the mass of
one component (convective Cahn-Hilliard equation) and momentum
(Navier-Stokes-Korteweg equations). We have used the framework of
phenomenological non-equilibrium thermodynamics to derive a
generalized model-H coupling transport equations for momentum, density
and entropy in the framework of phenomenological non-equilibrium
thermodynamics. Then we have discussed the individual contributions
representing, for instance, an internal Soret effect and interface
viscosity. The model has been simplified for isothermal setting,
vanishing interface viscosity and internal energies resulting from a
setting close to the critical point of demixing. A comparison with
literature results has clarified the issue of defining pressure and
chemical potential.

To facilitate the description of a free surface profile we have
introduced boundary conditions at the solid substrate and the free
interface. It has been explained that the incorporation of
hydrodynamic flow even in the case of extremely slow creeping flow is a
necessary precondition for the description of evolving surface
deflections.
It has been shown that the dimensionless numbers entering
the boundary conditions for the Cahn-Hilliard and the
Korteweg-Navier-Stokes are closely related. Therefore they can not
by any means be chosen independently of each other.
After non-dimensionalization we have analysed possible steady base
state solutions for laterally homogeneous films of decomposing
mixtures. In doing so we have distinguished vertically homogeneous and
vertically stratified films. It has been shown that a plethora of
stratified solutions exist that can be mapped, ordered and understood
using continuation techniques and symmetry arguments. The obtained
systematics will form the basis for the analysis of the lateral
stability of the base states undertaken in an accompanying paper.
\cite{MaTh07}
In an Appendix~\ref{deribc}
 we have used variational calculus to independently
confirm the boundary conditions for the static limiting case.

Our results on vertical stratifications can be compared to a broad variety
of experimental data on static film structures. However, they can 
also be used to interpret transitions observed in slow time evolutions.
Most of the results on vertical layering reviewed in 
\cite{GeKr03} can be explained at least qualitatively. 
Most static vertical layerings observed in thin films of polymer
blends either correspond to two-layer 
or sandwich-like three-layer 
structure that we have found to be the only structures of lowest
energy depending on the energetic bias of the surfaces.

The interesting case of the evolution of a relatively thick (500\,nm)
decomposing d-PMMA/SAN blend film \cite{WaCo00} is presented in Fig.~16 of
Ref.~\onlinecite{GeKr03}. The vertical profile develops from a homogeneous film to
a two-layer structure, with d-PMMA collecting at the solid
substrate. However, the path to equilibrium passes through a
sandwich-like three-layer structure. This can be easily understood
from the solution structure presented for antisymmetrically biased
surfaces in Fig.~\ref{ss5}. There the $n=1$ (sandwich-like three-layer
structure) has a lower energy than the $n=0$ or $n=3/2$ solution but a
higher energy than the final $n=1/2$ solution. The $n=1$ solution
corresponds to a saddle in phase space, i.e.\ it attracts time
evolutions of a broad range of initial conditions and then expels
evolutions in its only unstable direction directing the evolution
towards the final two-layer structure.  Which 'saddle solutions' will
be involved in a time evolution depends on the wavelength of the
fastest linear mode. For a similar discussion for dewetting films on a
heterogeneous substrate see Ref.~\onlinecite{TBBB03} section 3.

In another experiment using a PEP/d-PEP blend film \cite{GeKr98}
presented in Fig.~21 of
Ref.~\onlinecite{GeKr03} it is shown that the equilibrium layer structure changes from
d-PEP/PEP/d-PEP to d-PEP/PEP by varying the substrate surface
energy. This corresponds in our idealized setting to a transition from
symmetrically biased (or unbiased) surfaces to anti- or asymmetrically
biased surfaces. The here observed change in the layering of lowest
energy corresponds well to the experiments.

To conclude, the present work has presented a complete model to
investigate a wide variety of systems involving the coupled evolution
of surface and concentration profiles of free surface films of a
decomposing mixture. It can be used to analyse vertically stratified
but horizontally homogeneous films and their evolution in time. This
includes layer inversions of two-layer systems with diffuse or 
sharp interfaces
that can not be described by two-layer models for immiscible liquids.
\cite{PBMT04,FiGo05,PBMT05,BGS05,PBMT06}
The dynamics of such an inversion is quite complex. For instance, for
a substrate/liquid~1/liquid~2/air two-layer structure it involves the
transient formation of drops of liquid~1 at the liquid~2/air interface.
\cite{SKF94}  The typical distance of those droplets can now be
calculated using a linear transversal stability analysis of the
unstable two-layer profile. For antisymmetrically biased surfaces it
exists, for instance, for $h=5$ up to $a^+\approx0.3$
(Fig.~\ref{ss5}).

Our model also allows to study the emergence of permanent lateral
structures using a transversal stability analysis of the stratified
layers
\cite{MaTh07} or a simulation in time.  Especially, it allows to
quantify the influence of hydrodynamic transport on the dynamics.

Note that the model can be adapted for several related
problems. Although, the boundary conditions in Section~\ref{sec-bc}
are posed for a film on a solid substrate they can easily be adapted
for free standing films, i.e.\ for a film with two free surfaces. Also
drops of a mixture on a solid substrate can be studied if the presented
model is supplemented by a condition at the contact line. 
The latter will be the scope of future work. Also the used model of local bulk 
and interface energies can easily be replaced by more realistic functions
as discussed in Ref.~\onlinecite{GeKr03}. Choosing parameters that correspond to
a stable mixture also the dynamics of mixing in a free surface film can be 
studied.

\appendix
\section{Variational approach}
\mylab{deribc}
%
This appendix uses variational calculus to derive the static limit of
the bulk equations and boundary conditions from the underlying Helmholtz
free energy functional of Cahn-Hilliard \cite{CaHi58} type
for a two-dimensional film of binary mixture.
The three dimensional case will be presented elsewhere for a more 
general setting.
The free energy functional
\begin{equation}
F[c(x,z),h(x)]=\Fb[c(x,z),h(x)] + \Fs[c(x,h(x)),h(x)].
\end{equation}
consists of a bulk part $\Fb$ and a surface part $\Fs$
defined as
\begin{align}
\Fb &= \int_{-\infty}^\infty\int_0^{h(x)}\left[
   \frac{\sigma_c}{2}(\nabla c)^2+f(c)\right]\diff z\diff x
 -\lambdadif\left[\int_{\varOmega}c_1\,\diff A - C_1A\right]
 -\lambda\left[\int_{\varOmega}\diff A - A\right],\label{e:Fb}\\
\Fs &= \oint_{\partial\varOmega} \fsurf(c)\diff s.
\end{align}
The second and third integral in $\Fb$ are taken over the same area as the
first one, the symbol $\varOmega$ is introduced for convenience.
The symbol $\partial\varOmega$ denotes the boundary of the domain of
integration $\varOmega$.
We assume that the surface free energy does not depend on $(\nabla c)^2$, i.e., 
the bulk free energy gradient term is not amended at the free surface. 
Such a contribution and its physical consequences will be discussed elsewhere.
The Lagrange multipliers $\lambdadif$ and $\lambda$ enforce mean
concentration of component 1, $c_1=(c+1)/2$, and total area of the domain to have the
prescribed values $C_1$ and $A$, respectively. Later, we will discuss
their relation to the local chemical potentials $\mu_2, \mudif$ and
mechanical pressure $p$. 

To vary $F$ with respect to all possible degrees of freedom in a
transparent way we define small changes of the functions $c$ and $h$
as
\begin{align}
h(x;\alpha) &= h(x) + \alpha\zeta(x),\label{e:hxa}\\
c(x,z;\alpha)=c(\vec x;\alpha) &=c(\vec x) + \alpha\eta(\vec x),\label{e:cxa}
\end{align}
where $\zeta$ and $\eta$ are arbitrary admissible functions and $\vec{x}
 = (x,z)$. The variation of $h$ and $c$ corresponds to the operation
$\partial_\alpha|_{\alpha=0}$, i.e., $\delta h(x) = \zeta(x)$, $\delta
c(\vec{x}) = \eta(\vec x)$. However, not only the local concentration
and the interface position are varied. Varying the latter also implies
that fluid elements have to vary their position due to convective
motion.

The variation of the bulk
contribution to the free energy functional can be written as
\begin{align}
\delta\Fb&=\int_{-\infty}^\infty\int_0^{h(x)}\left[
   \sigma_c\nabla c\cdot\nabla\eta
   + \left(\partial_cf-\frac{\lambdadif}{2}\right)\eta
   \right]\diff z\diff x
 -\delta\lambdadif\left[\int_{\varOmega}c_1\diff A - C_1A\right]
\nonumber\\
&{} -\delta\lambda\left[\int_{\varOmega}\diff A - A\right]
  +\int_{-\infty}^\infty\left[
   \frac{\sigma_c}{2}(\nabla c)^2+f(c)-\lambdadif c_1 -
   \lambda\right]\zeta(x)\diff x,
\end{align}
where we used $\delta\nabla c=\nabla\delta c$.
Integrating by parts the term containing $\nabla\eta$, we get
\begin{align}
\delta\Fb&=\int_{-\infty}^\infty\int_0^{h(x)}\left[
   - \sigma_c(\Delta c)\eta
   +\left(\partial_cf-\frac{\lambdadif}{2}\right)\eta
   \right]\diff z\diff x
 -\delta\lambdadif\left[\int_{\varOmega}c_1\diff A - C_1A\right]
\nonumber\\
&{} -\delta\lambda\left[\int_{\varOmega}\diff A - A\right]
+\oint_{\partial\varOmega}\sigma_c(\nabla c)\cdot\vec{n}\,\eta\diff s
+\int_{-\infty}^\infty\left[
   \frac{\sigma_c}{2}(\nabla c)^2+f(c)-\lambdadif c_1 -
   \lambda\right]\zeta(x)\diff x.
\label{e:deltaFb}
\end{align}

Next, we turn our attention to the surface contribution $\Fs$.
concentrating on the most interesting
top part of the boundary $\partial\varOmega$, i.e.\ the free surface.
We denote the corresponding part of $\Fs$ as $\Fstop$ and write it as
\begin{equation}
\Fstop=\int_{-\infty}^\infty
   \fsurf[c(\xsurf)]
\frac{\diff s}{\diff x}\diff x,\label{e:Fstop}
\end{equation}
where
\begin{equation} 
\frac{\diff s}{\diff x} = \sqrt{1+(\partial_x
h(x))^2}.
\label{e:dsdx} 
\end{equation}
For the point $\xsurf(x) = (x,h(x))$ at the free surface and its variation we have
\begin{align}
\xsurf(x;\alpha) &= (x,h(x)+\alpha\zeta(x)),\nonumber\\
\delta\xsurf(x) &= (0,\zeta(x)).\label{e:deltaxs}
\end{align} 
Then $\delta c(\xsurf(x)) = \nabla c(\xsurf)\cdot\delta\xsurf +
\eta(\xsurf)$. Using eqs.\ (\ref{e:dsdx}) and $\xsurf(x)$ we have expressed
the integrand of (\ref{e:Fstop}) as a function of $x$.
The variation of $\Fstop$ is
\begin{align}
\delta\Fstop&=\int_{-\infty}^\infty\bigg\{
  \partial_c\fsurf\,\frac{\diff s}{\diff x}\,\delta c(\xsurf) 
+ \fsurf(c) \delta\frac{\diff s}{\diff x}\bigg\}
\diff x\nonumber\\
&{}=\int_{-\infty}^\infty\bigg\{
\partial_c\fsurf(\nabla c\cdot\delta\xsurf+\eta)
   \frac{\diff s}{\diff x} + \fsurf(c)
   \partial_x h\,\vec{t}\cdot\nabla\zeta\bigg\}
\diff x,
\label{e:deltaFstopx}
\end{align}
where we used Eq.~(\ref{e:dsdx}) and applied
\begin{equation}
\frac{\partial_x\zeta(x)}{(\diff s/\diff x)}
= \frac{\diff\zeta(x)}{\diff x}\frac{\diff x}{\diff s}
 = \frac{\diff\zeta(x)}{\diff s}\,=\,\vec{t}\cdot\nabla\zeta.
\end{equation}
The last step is correct if fields are only defined at the surface (here $\zeta$, 
but valid also for $h$, $\vec{n}$, $\vec{t}$, etc.) are interpreted as being defined 
everywhere with values independent of $z$.

Next, integration by parts has to be applied to eliminate derivatives of the
variations. One uses
\begin{equation}
\int a\,(\vec{t}\cdot\nabla b) \, \diff s
\,=\,\int a\,(\vec{t}\cdot\nabla b) \, \frac{\diff s}{\diff x}\diff x
\,=\,- \int b\,\nabla\cdot\left[\vec{t} \frac{\diff s}{\diff x} a\right]\, \diff x
\,=\,- \int b\,(\vec{t}\cdot\nabla a) \, \diff s.
\label{e:dtibp}
\end{equation}
Assuming laterally periodic or localized structures, surface terms resulting from the 
integration by parts are zero here.
We obtain
\begin{equation}
\delta\Fstop=\int_{-\infty}^\infty\left\{
\partial_c\fsurf(\nabla c\cdot\delta\xsurf+\eta)
\,-\, \vec{t}\cdot\nabla\left[
\fsurf(c)\,\frac{\diff x}{\diff s}\partial_x h\right]\zeta\right\}\diff s.
\label{e:fstop_terms_ds}
\end{equation}
Performing derivatives and substituting from (\ref{e:deltaxs}) we arrive at
\begin{align}
\delta\Fstop=\int_{-\infty}^\infty\left\{
\partial_c\fsurf\,[(\partial_zc)\zeta+\eta]
\,-\, \left[(\vec{t}\cdot\nabla\fsurf)\,\partial_x h 
\,-\,\fsurf\kappa\right]
\,\frac{\diff x}{\diff s}\,\zeta\right\}\diff s,
\label{e:fstop_terms_ds2}
\end{align}
where
\begin{equation}
\kappa = -\frac{\partial_{xx}h}{[1+(\partial_x h)^2]^{3/2}}
\label{e:curv}
\end{equation}
is the curvature. Note that $\kappa$ is positive for a
convex surface of the fluid.
Terms with $(\partial_c\fsurf)\zeta$ can be simplified as follows
\begin{equation}
(\partial_c\fsurf)\,\left[(\partial_z c)\,
\frac{\diff s}{\diff x} \,-\, (\vec{t}\cdot\nabla c)\,\partial_x h\right]
\frac{dx}{ds}\zeta
\,=\,(\partial_c\fsurf)\,(\vec{n}\cdot\nabla c)\,\frac{dx}{ds}\zeta
\end{equation}
resulting in the final expression
\begin{equation}
\delta\Fstop=\int_{-\infty}^\infty\left\{
\left[(\vec{n}\cdot\nabla c)\,
\partial_c\fsurf 
\,+\,\fsurf\,\kappa\right]\,\frac{dx}{ds}\zeta
\,+\,(\partial_c\fsurf)\eta\right\}\diff s.
\label{e:fstop_terms_ds4}
\end{equation}

The variation of the contribution of the free energy at the bottom part
of the boundary $\partial\varOmega$, denoted by $\delta\Fsbot$, can be
obtained as a special case of $\delta\Fstop$. In (\ref{e:deltaFstopx})
we consider $\xsurf(x)=(x,0),\delta\xsurf=0, ds/dx = 1$
resulting in
\begin{equation}
\delta\Fsbot=\int_{-\infty}^\infty\partial_c\fsurf\,\eta\diff x.
\label{e:fsbot}
\end{equation}

Next, writing $\delta F=\delta\Fb+\delta\Fstop+\delta\Fsbot=0$,
one is in principle
ready to extract governing equations and natural
boundary conditions of the problem. Inspecting the form of 
(\ref{e:deltaFb}), (\ref{e:fstop_terms_ds4}), and (\ref{e:fsbot})
one notes that we obtained two scalar boundary conditions on the free
surface as the prefactors of arbitrary admissible
functions $\eta$, $\zeta$ in the boundary integral of the stationarity
condition $\delta F=0$
\begin{align}
\sigma_c\vec{n}\cdot\nabla c + \partial_c\fsurf &= 0,\\
\frac{\sigma_c}{2}(\nabla c)^2+f(c)-\lambdadif c_1-\lambda + (\vec{n}\cdot\nabla
c)\,\partial_c\fsurf + \fsurf\kappa &=0,
\end{align}
respectively. In order to obtain force boundary conditions,
we need to express our variations $\zeta,\eta$ in terms of the virtual
displacements because mechanical forces are energetically conjugated to them.

For this purpose, we introduce the variation of the position of
a fluid element due to convective motion. The
varied Euler coordinates $\vec{x}$ of a fluid element specified
by its material (Lagrange) coordinates $\vec X$ can be expressed as
\begin{align}
\vec{x}(\vec X;\alpha) &= \vec X+\alpha\vecg{\chi}(\vec
X),\label{e:eulvialagr}\\
\delta\vec{x}(\vec X) &= \vecg{\chi}(\vec X).\label{e:deltaeul}
\end{align}
where $\vecg{\chi}=(\chi_x,\chi_z)$ is an arbitrary admissible
displacement vector. Both coordinate systems coincide for $\alpha=0$.
The free surface has to follow the fluid as is expressed by the kinematic
condition Eq.~(\ref{th-hddf4b}). This introduces a dependency between the
variations $\zeta$ and $\vecg\chi$, i.e.\ 
\begin{equation} 
\delta h(x) = \zeta(x) = -\partial_xh(x)\chi_x(x) +
\chi_z(x),\qquad\mbox{or}\qquad 
\zeta\,\vec{e}_z\cdot\vec n = \zeta\,\frac{dx}{ds} = \vecg\chi\cdot\vec n,
\label{e:kincond} 
\end{equation} 
Next, we consider a fluid element identified by its referential position
$\vec{X}$. The varied concentration at this element can be expressed using
(\ref{e:eulvialagr}) as
\begin{equation}
c(\vec x;\alpha) = c(\vec X+\alpha\vecg\chi(\vec X))
 + \alpha\eta(\vec X+\alpha\vecg\chi(\vec X)),
\end{equation}
being consistent with (\ref{e:cxa}). Consider for a moment 
that no diffusion is active. In that case the concentration
$c(\vec x;\alpha)$ changes only due to convection described by
$\vecg\chi$. In consequence, the concentration at the arbitrary
but fixed fluid element $\vec X$ should remain constant, i.e.\
\begin{equation}
\frac{\diff}{\diff\alpha}\left\{c[\vec X+\alpha\vecg\chi(\vec X)]
 + \alpha\eta[\vec X+\alpha\vecg\chi(\vec X)]\right\}
 = \nabla c[\vec X+\alpha\vecg\chi(\vec X)]\cdot\vecg\chi(\vec X)
 + \eta[\vec X+\alpha\vecg\chi(\vec X)] = 0.
\end{equation}
We denote the variation $\eta$ that satisfies this condition as $\etaco$
(convective) and the remaining part as $\etanc$ (diffusive) variation. In
consequence, we have 
\begin{equation}
\eta = \etaco + \etanc = -\nabla c\cdot\vecg\chi + \etanc
\label{e:etasplit}
\end{equation}
where $\etanc$ is a variation independent of $\vecg\chi$ because it is caused
by a different physical process.

Using Eqs.~(\ref{e:etasplit}) and (\ref{e:kincond}), we write the stationarity
conditions for the variation $\delta F=\delta\Fb+\delta\Fs$.
Prefactors of the variations in the bulk and at the free and bottom
surface give the Euler-Lagrange equations of the problem.
The prefactor of $\etanc$ in the bulk integral gives
\begin{equation}
-\sigma_c\Delta c + \partial_cf-\frac{\lambdadif}{2} \,=\,0,
\label{e:eqetancbulk}
\end{equation}
i.e.\ the correct static limit of the 2d version of
Eq.~(\ref{mh-eqc-our}). From this equation we deduce that
$-2\sigma_c\Delta c+2\partial_cf = -\sigma_{c_1}\Delta c_1 + \mudift
= \lambdadif$ is the chemical potential for a heterogeneous equilibrium
discussed in Section~\ref{mhf}.
The prefactor of $\vecg\chi$ gives the same equation as
(\ref{e:eqetancbulk}). The surface integrals yield as the prefactor
of $\etanc$
\begin{equation}
\sigma_c\,\vec{n\cdot\nabla} c \,+\,\partial_c \fsurf
\,=\,0,
\label{e:eqetancbdtop}
\end{equation}
i.e., the static limit of the 2d version of
Eq.~(\ref{bc-conc2b}) with $\sigma^-=0$ and of
Eq.~(\ref{bc-conc2}) with $\sigma^+=0$. 
On the free surface the vectorial prefactor of $\vecg\chi$
\begin{align}
(\vec n\cdot\nabla c)\,(\partial_c\fsurf) \,\vec{n}
\,-\,(\nabla c)\,\partial_c\fsurf 
\,+\,\fsurf\kappa\,\vec{n}
\,-\,\sigma_c\,\vec n\cdot(\nabla c)(\nabla c)&\nonumber\\
+\,\left[
   \frac{\sigma_c}{2}(\nabla c)^2+f(c)-\lambdadif c_1 -
   \lambda\right] \vec n
&\,=\,0
\label{e:eqchibdtop}
\end{align}
gives, using $\tens I - \vec{nn} = \vec{tt}$ and reordering,
\begin{align}
-\sigma_c\,\vec n\cdot(\nabla c)(\nabla c)
\,-\,p_{\rm eff} \vec n
\,=\,\vec{tt}\cdot\nabla\fsurf\,-\,\fsurf\kappa\,\vec{n}
\label{e:eqchibdtop3}
\end{align}
with
\begin{equation}
p_{\rm eff}= (\partial_c f-\sigma_c\Delta c)(c+1)
-\frac{\sigma_c}{2}(\nabla c)^2 \,-\, f(c)\,+\,\lambda,
\mylab{e:peff}
\end{equation}
where we used Eq.~(\ref{e:eqetancbulk}) for $\lambdadif$
We proceed to identify the quantities in
$p_{\rm eff}$. Using (\ref{gibbsduhemf}), we substitute the free
energy density $f(c)$ in (\ref{e:peff}) and simplify to
\begin{equation}
p_{\rm eff}= p - \mu_2\rho-\sigma_c(c+1)\Delta c
\,-\,\frac{\sigma_c}{2}(\nabla c)^2\,+\,\lambda,
\mylab{e:peff2}
\end{equation}
taking into account that $2(\partial_c f)c_1 = \mudift c_1 = \mudif\rho_1$.
The pressure 
$p_\mathrm{eff}$ in (\ref{e:peff2}) coincides with the one defined
in (\ref{pco-our}) in case that
\begin{equation}
\lambda = \mu_2\rho.\mylab{e:peff3}
\end{equation}
We conclude that the Eq.~(\ref{e:eqchibdtop3}) gives the tangential
and the normal force equilibrium conditions at
the free surface and corresponds to the static limit of the boundary
conditions (\ref{th-hddf5}) with (\ref{mh-tau-our}).

\acknowledgments

This work was supported by the European Union and Deutsche
Forschungsgemeinschaft under grants MRTN-CT-2004-005728 and SFB 486
B13, respectively. UT thanks MPIPKS and Peter H\"anggi for support at
various stages of the project.

%

%
\end{document}